\documentclass[12pt,a4paper]{article}

\usepackage{amsmath}
\usepackage{amssymb}
\usepackage{epsfig}
\usepackage{graphics}

\reversemarginpar

\def\nostrocostrutto#1\over#2{\mathrel{\mathop{\kern 0pt \rlap 
  {\raise.2ex\hbox{$#1$}}}
  \lower.9ex\hbox{\kern-.190em $#2$}}}

\catcode`@=11
\newcount\@tempcntc
\def\@citex[#1]#2{\if@filesw\immediate\write\@auxout{\string\citation{#2}}\fi
  \@tempcnta\z@\@tempcntb\m@ne\def\@citea{}\@cite{\@for\@citeb:=#2\do
    {\@ifundefined
       {b@\@citeb}{\@citeo\@tempcntb\m@ne\@citea\def\@citea{,}{\bf ?}\@warning
       {Citation `\@citeb' on page \thepage \space undefined}}%
    {\setbox\z@\hbox{\global\@tempcntc0\csname b@\@citeb\endcsname\relax}%
     \ifnum\@tempcntc=\z@ \@citeo\@tempcntb\m@ne
       \@citea\def\@citea{,}\hbox{\csname b@\@citeb\endcsname}%
     \else
      \advance\@tempcntb\@ne
      \ifnum\@tempcntb=\@tempcntc
      \else\advance\@tempcntb\m@ne\@citeo
      \@tempcnta\@tempcntc\@tempcntb\@tempcntc\fi\fi}}\@citeo}{#1}}
\def\@citeo{\ifnum\@tempcnta>\@tempcntb\else\@citea\def\@citea{,}%
  \ifnum\@tempcnta=\@tempcntb\the\@tempcnta\else
   {\advance\@tempcnta\@ne\ifnum\@tempcnta=\@tempcntb \else \def\@citea{--}\fi
    \advance\@tempcnta\m@ne\the\@tempcnta\@citea\the\@tempcntb}\fi\fi}
\catcode`@=12

\begin{document}
\ifx\href\undefined\else\hypersetup{linktocpage=true}\fi
 
\begin{titlepage}
\begin{center}
\hfill 15.09.2003

\hfill + 3.05.2004
\vspace*{0.5cm}

\section*{Central hadron production}
\section*{in crossing of dedicated hadronic beams \footnote{
This material served as base for the contributions of the author
to the 9th Adriatic Meeting on
Particle Physics and the Universe,
Dubrovnik, Croatia, 4 - 14 September 2003,
and to the 310th Heraeus Seminar,
Quarks in Hadrons and Nuclei II, September 15 - 20, 2003,
Rothenfels Castle, Oberwoelz, Austria.
}
}
\vspace*{0.5cm}

{\bf Peter Minkowski} \\
Institute for Theoretical Physics, University of Bern, \\
CH-3012 Bern, Switzerland
\vspace*{1.5cm}

{\large\bf Abstract\\[10pt]} \parbox[t]{\textwidth}{
The original aim of this work, was to give a {\it brief} review
of gluonic mesons, to be searched for in an experiment dedicated
to central production of a relatively low mass hadronic system,
whereby rapidity gaps are possible to impose, requiring initial
hadron beams of sufficient energy and intensity. Sections 1 - 4
are devoted to this aim. The various theoretical ingrediants,
covering several decades of thinking by many, including the author,
are contained in 5 appendices, dedicated to specifically
gluonic binaries within QCD and their underlying Yang-Mills base structure
}

\end{center}
\end{titlepage}



\tableofcontents

\newpage

\listoffigures

\newpage

\subsection*{The physics potential}
\subsection*{of an in depth experimental investigation}

\noindent
\section{Introduction}
\vspace*{0.2cm}

\noindent
I shall begin a historical survey, quoting a recent article
\cite{KaKhoMaRy}
entitled  

\noindent
"Central exclusive diffractive production as a spin--parity
analyser: \\
from hadrons to Higgs" ,
written by four authors :
A.B. Kaidalov, V.A. Khoze, A.D. Martin and M.G. Ryskin.

\noindent
'Pour fixer les id\'{e}es' , let me reproduce the first figure of the
above paper

\begin{figure}[htb]
\vskip 1.cm 
\begin{center}
\vskip -1.5cm 
\mbox{\epsfig{file=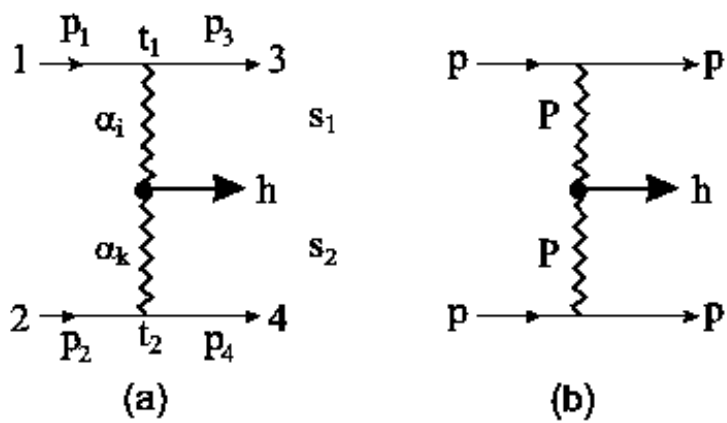,%
width=14cm}}
\end{center}
\caption{(a) The central production of a state h by double Reggeon-exchange.
(b) The double-Pomeron exchange contributions to 
$pp \ \rightarrow \ p \ + \ h \ + \ p$, which dominates
at high energies, where the $+$ signs are used to indicate the presence
of Pomeron-induced rapidity gaps.}
\label{fig1}
\end{figure}

\newpage

\noindent
In figure \ref{fig1} the reaction of central type

\begin{equation}
\label{eq:1}
\begin{array}{l}
\left \lbrace
\begin{array}{c}
\hspace*{0.3cm} H_{\  1} \ ( \ p_{\ 1} \ ; \ q.\# 1 \ )
\vspace*{0.3cm} \\
 +
\ H_{\  2} \ ( \ p_{\ 2} \ ; \ q.\# 2 \ )
\end{array}
\ \right \rbrace
\ \rightarrow
\left \lbrace
\begin{array}{c}
\hspace*{0.3cm} H_{\  3} \ ( \ p_{\ 3} \ ; \ q.\# 3 \ )
\vspace*{0.3cm} \\
 +
\ H_{\  4} \ ( \ p_{\ 4} \ ; \ q.\# 4 \ )
\vspace*{0.3cm} \\
 +
 \ \left \lbrack
\ h_{\  c} \ ( \ p_{\ c} \ ; \ q.\# c \ )
\ + \ X_{\ c}
\ \right \rbrack
\end{array}
\right \rbrace
\end{array}
\end{equation}

\noindent
is represented with the following identifications  

\begin{description}
\item 1) initial and tagged final hadron pairs
inducing central production

\begin{equation}
\label{eq:2}
\begin{array}{l}
H_{\  1 \ , \ 2} \ : 
\ \begin{array}{ll}
 \mbox{initial hadron pair} & \hspace*{1.2cm} \mbox{with}
 \end{array}
\ \left \lbrace
\begin{array}{cr}
 \mbox{momenta} & p_{\ 1 \ , \ 2}
\vspace*{0.3cm} \\
\mbox{and $q. \#$} & q_{\ 1 \ , \ 2}
\end{array}
\right \rbrace
\vspace*{0.3cm} \\
H_{\  3 \ ,\ 4} \ : 
\ \begin{array}{ll}
 \begin{array}{l}
\ \mbox{$3 \ (\leftarrow \ 1)$ and $4 \ (\leftarrow \ 2)$} 
\vspace*{0.3cm} \\
\mbox{associated hadron pair}
\end{array}
& \mbox{with}
\end{array}
\ \left \lbrace
\begin{array}{cr}
 \mbox{momenta} & p_{\ 3 \ , \ 4}
\vspace*{0.3cm} \\
\mbox{and $q. \#$} & q_{\ 3 \ , \ 4}
\end{array}
\right \rbrace
\end{array}
\end{equation}

\item 2) centrally produced (hadronic) system $h_{\ c}$
{\it conditioned} by $h_{\ c} \ | \ X_{\ c}$

\begin{equation}
\label{eq:3}
\begin{array}{l}
\begin{array}{lll ll}
h_{\  c} & : &
\begin{array}{l}
 \mbox{centrally produced}
\vspace*{0.3cm} \\
\mbox{system of interest} 
\end{array}
& \mbox{with}
& \hspace*{-0.1cm} \left \lbrace
\begin{array}{cl}
 \mbox{momentum} & p_{\ c}
\vspace*{0.3cm} \\
\mbox{mass} & M_{\ c} =  \sqrt{p_{\ c}^{\ 2}}
\vspace*{0.3cm} \\
\mbox{and $q. \#$} & q_{\ c}
\end{array}
\right \rbrace
\vspace*{0.3cm} \\
X_{\  c} & : &
\mbox{{\it specified conditions}} 
& &
 \left \lbrace
\begin{array}{cl}
 \mbox{optimized to isolate} \ h_{\ c} 
\vspace*{0.3cm} \\
\mbox{from background}  
\end{array}
\right \rbrace
\end{array}
\end{array}
\end{equation}

\end{description}

\noindent
As is illustrated by the range of topics discussed in ref.
\cite{KaKhoMaRy} , the general issue of central production
is not restricted to strong interactions limited as far
as  quark- and antiquark flavors are concerned to the three light ones 
u,d and s, denoted $ QCD_{\ 3}$ hereafter. 
\vspace*{0.1cm} 

\noindent
{\it This is our main focus here.}
\vspace*{0.1cm} 

\noindent
Rather at sufficiently high c.m. energy 
strong and electrweak synthesis of the central system '$h_{\ c}$'
well includes the following processes, becoming
dominantly reducible to fusion of virtual
gauge boson pairs formed out of the sequence
gluon (g) , photon ($\gamma$) , W , Z. We list only the combinations
where $h_{\ c} \ = \ Q \overline{Q}$
$g \ g \ ( \ g \ \gamma \ , \ \gamma \gamma \ )
\ \rightarrow \ Q \ \overline{Q}$ , 
$g \ \gamma \ \rightarrow \ Q \ \overline{Q}$ for heavy flavors
$Q \ = \ c \ , \ b \ , \ t$ and the top quark induced hadronic
production of Higgs boson(s), where $h_{\ c} \ = h^{\ JP}$
$g \ g \ \rightarrow \ t \overline{t} \ \rightarrow \ h^{\ JP}$.

\begin{description}
\item i) hadronic production of (single) heavy quark-antiquark pairs,
both bound and open 

\begin{equation}
\label{eq:4}
\begin{array}{l}
h_{\ c} \ ( \ QCD_{\ 6} \ ) \ =
\left \lbrace
\begin{array}{lll ll}
c \overline{c} & : &
J/ \Psi, \chi \ , \cdots & , & D \overline{D} \ , \cdots
\vspace*{0.3cm} \\
b \overline{b} & : &
Y \ , \chi_{\ b} \ , \cdots & , & B \overline{B} \ , \cdots
\vspace*{0.3cm} \\
t \overline{t} &  &
\end{array}
\right \rbrace
\vspace*{0.3cm} \\
g \ g \ \rightarrow \ Q \overline{Q}
\end{array}
\end{equation}

\item ii) hadronic production of (single) Higgs bosons $h^{\ PC}$

\begin{equation}
\label{eq:5}
\begin{array}{l}
h_{\ c} \ ( \ QCD_{\ 6} \ , \ {\cal{Y}}_{\ h^{\ PC} \ t \overline{t}} \ ) \ =
\left \lbrace
\begin{array}{l}
h^{\ ++} \ , \ h^{\ -+}
\end{array}
\right \rbrace
\vspace*{0.3cm} \\
g \ g \ \rightarrow \ t \overline{t} \ \rightarrow \ h^{\ JP}
\end{array}
\end{equation}

In eq. (\ref{eq:5}) ${\cal{Y}}_{\ h^{\ PC} \ t \overline{t}}$
denotes the Yukawa coupling between the Higgs boson(s) and
the top quark.

\end{description}

\noindent
The association of central production with {\it perturbatively preconceived}
gauge boson fusion is not fortuitous. It goes back to seminal work
on multiparticle production mainly of electrons and positrons
in QED, by Landau, Lifschitz, Pomeranchuk and others. I only wish to cite
a selcted subset for historical accuracy \cite{LandauPom} .

\noindent
The perturbative approach to QED governed 
high energy elastic scattering amplitudes for initial
particle pairs 
$e^{-} e^{\pm}$ , $e^{-} p$ ,  $e^{-} \gamma$ , $\gamma p$
and  $\gamma \gamma$
was pioneered by Cheng and Wu \cite{ChengWu} . The proton can be
replaced by a nucleus (A, Z), where the nuclear charge 
$Q \ = \ Z e$ serves to represent 'strong' coupling, for
large Z.

\newpage

\noindent
\section{Theoretical expectations for primary gluonic binary (gb) Regge
trajectories}
\vspace*{0.2cm}

\noindent
We follow the identification of the gluonic binary states
lowest in mass discussed in ref. \cite{PMWO}.
\vspace*{0.5cm}

\begin{figure}[htb]
\vskip 1.cm 
\begin{center}
\vskip -1.5cm 
\mbox{\epsfig{file=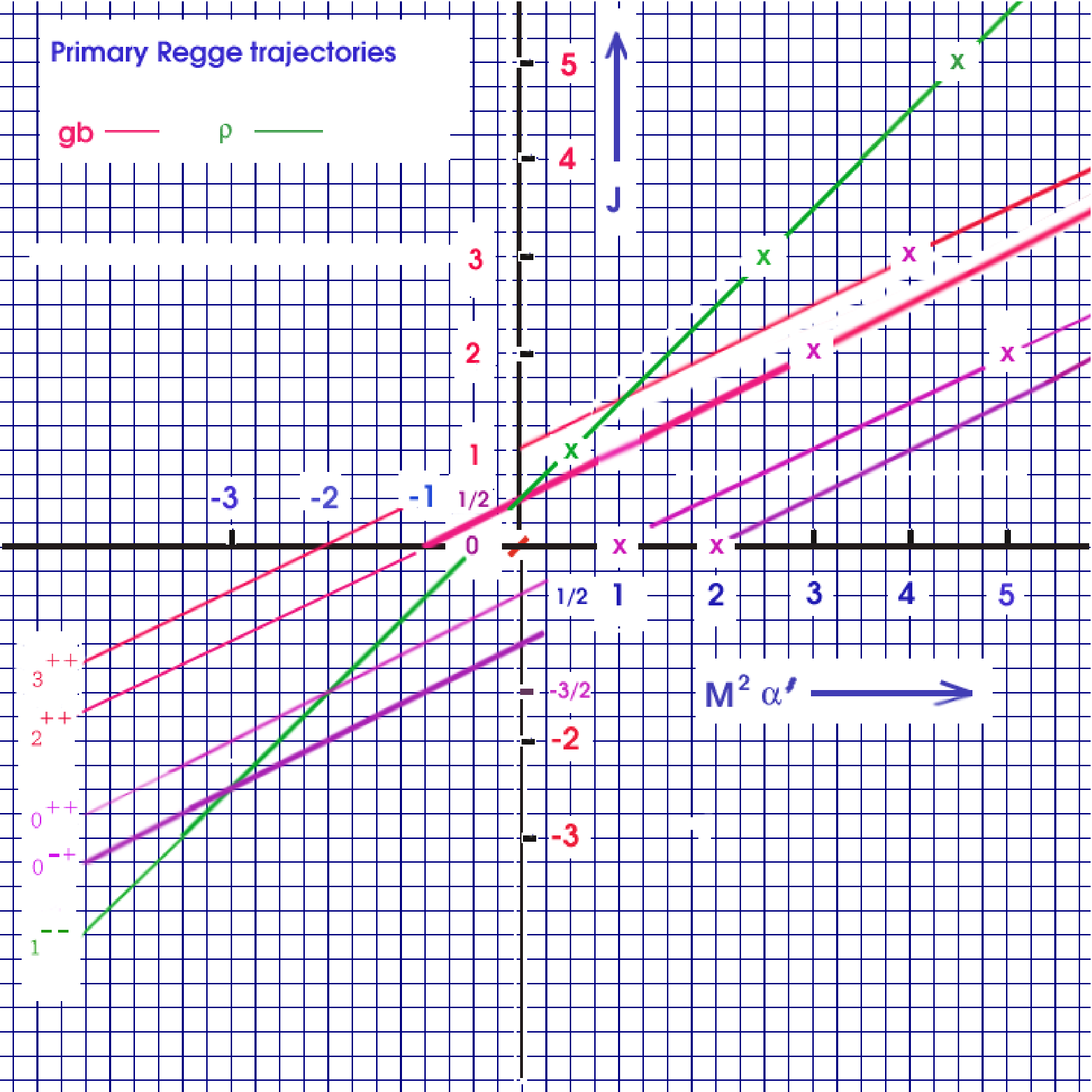,%
width=14cm}}
\end{center}
\caption{Primary gluonic binary (gb) Regge trajectories in comparison
with the $\varrho-(q \overline{q})$ trajectory.
}
\label{fig2}
\end{figure}

\noindent
It shall be clear, that here we follow a combination
of {\it hypotheses and theoretical expectations}.
We will comment on alternatives below.   

\newpage

\noindent
We begin with the spectrosopic classification of gluonic binaries
\cite{HFPM},
which, apart from the confined nature of binary gluons, is
identical to the classification of photon binaries \cite{Land}, \cite{Yang}.

\noindent
It will prove useful to obtain a 'Richtwert'
for the inverse slope
of the rho-Regge trajectory at 0 momentum transfer, since
it is this quantity which sets the unit of mass square with
respect to which hadronic resonances, gb and others are to be
placed in the simplified harmonic {\it and zero width-} approximation.

\noindent
In this objective we define 
the quantity $m_{\ \varrho}^{\ 2} \ ( \ 0 \ = \ t \ )$,
which represents the $\varrho$ mass square as seen in the
limit of spacelike momentum transfer $t \ \rightarrow \ 0$
through the electromagnetic form factor of charged pions
$F_{\ \gamma}^{\ \pi^{\ +}} \ ( \ t \ )$

\begin{equation}
\label{eq:6}
\begin{array}{l}
m_{\ \varrho}^{\ - 2} \ ( \ 0 \ ) \ =  
\ \left .
d \  F_{\ \gamma}^{\ \pi^{\ +}} \ ( \ t \ ) \ / \ ( \ dt \ )
\ \right |_{\ t \ = \ 0}
\ = \ \frac{1}{6} \ \left \langle \ r^{\ 2} 
\ \right \rangle^{\ \pi^{\ +}}_{\ \gamma}
\end{array}
\end{equation}

\noindent
In eq. (\ref{eq:6}) 
$\left \langle \ r^{\ 2} \ \right \rangle^{\ \pi^{\ +}}_{\ \gamma}$
denotes the e.m. mean square charge radius of charged pions.

\noindent
This quantity is presently beeing investigated by
Caprini, Colangelo and \\ 
Leutwyler \cite{CaCoLeu},
from where I quote the preliminary result

\begin{equation}
\label{eq:7}
\begin{array}{l}
\left \langle \ r^{\ 2} 
\ \right \rangle^{\ \pi^{\ +}}_{\ \gamma}
\ = 
\ \left \lbrace
\begin{array}{l}
 0.4332 \ \pm \ 0.005 \ (stat.) \ \pm
\ 0.0004 \ (syst.) \ \pm \ 0.0004 \ (P)
\vspace*{0.3cm} \\
\ \rightarrow \ 0.4332 \ \pm \ 1.3 \ \% \ \mbox{fm}^{\ 2}
\end{array}
\right .
\end{array}
\end{equation}

\noindent
Converting to GeV units we obtain

\begin{equation}
\label{eq:8}
\begin{array}{lll}
m_{\ \varrho}^{\ 2} \ ( \ 0 \ ) & = & 
 0.5393 \ \pm \ 1.3 \ \% \ \mbox{GeV}^{\ 2}
\vspace*{0.3cm} \\
m_{\ \varrho}^{\ 2} \ ( \ 0 \ ) \ - \ m_{\ \pi^{0}}^{\ 2}
& = & 
 0.5211 \ \pm \ 1.3 \ \% \ \mbox{GeV}^{\ 2}
\vspace*{0.3cm} \\
& = & 
\left ( \ 721.9  \ \pm \ 0.65 \ \% \ \mbox{MeV} \ \right )^{\ 2}
\end{array}
\end{equation}

\noindent
The quantity $m_{\ \varrho} \ ( \ 0 \ )$ in eq. (\ref{eq:8})
deviates substantially from the resonance parameters of the rho,
whether obtained from the pole position in the complex energy plane
or other parametrizations of physical cross sections.
For comparison I quote a recent determination by the Kloe collaboration
\cite{Kloe}

\begin{equation}
\label{eq:8a}
\begin{array}{l}
m_{\ \varrho} \ = \ 775.9 \ \pm \ 0.5 \ \pm \ 0.3 \ \mbox{MeV}
\vspace*{0.3cm} \\
\Gamma_{\ \varrho} \ = \ 143.9 \ \pm \ 1.3 \ \pm \ 1.1 \ \mbox{MeV}
\end{array}
\end{equation}

\noindent
The relation to the inverse Regge slope parameter 
$ ( \ \alpha^{'} \ )^{\ -1}$ is

\begin{equation}
\label{eq:9}
\begin{array}{l}
\begin{array}{lll}
( \ \alpha^{'} \ )^{\ -1}
& = & 
2 \ ( \ m_{\ \varrho}^{\ 2} \ ( \ 0 \ ) \ - \ m_{\ \pi^{0}}^{\ 2} \ )
\vspace*{0.3cm} \\
& = & 
 1.0422 \ \pm \ 1.3 \ \% \ \mbox{GeV}^{\ 2}
\vspace*{0.3cm} \\
& = & ( \ 1.0209 \ \pm \ 0.65 \ \% \ \mbox{GeV} \ )^{\ 2}
\end{array}
\end{array}
\end{equation}

\noindent
We remark here, that the relation in eq. (\ref{eq:9}) is
{\it not} a rigorous one. 
We can compare with the direct $t \ > \ 0$ spectroscopic
masses along the $\Lambda$ baryon trajectory, assumed
unperturbed 

\begin{equation}
\label{eq:10}
\begin{array}{l}
\frac{1}{2}  \left ( \ m^{2} \ ( \ \Lambda^{\ 5/2 +} \ ) \ - 
m^{2} \  ( \ \Lambda^{\ 1/2 +} \ ) \ \right )
\ = \ ( \ \alpha^{'}_{\ \Lambda} \ )^{\ -1}
 =  1.034 \ \pm \ 0.010 \ \mbox{GeV}^{\ 2}
\vspace*{0.3cm} \\
\frac{1}{4}  \left ( \ m^{2}  \  ( \ \Lambda^{\ 9/2 +} ) \ - 
m^{2}  \ ( \ \Lambda^{\ 1/2 +} ) 
\ \right ) \ =
\ ( \ \alpha^{'}_{\ \Lambda} \ )^{\ -1}
\ =  
\ 1.069 \ \pm \ 0.024 \ \mbox{GeV}^{\ 2}
\end{array}
\end{equation}

\noindent
\section{Quantum numbers of binary gluonic mesons}
\vspace*{0.1cm} 

\noindent
The binary gluon system is only singled out
in the present discussion, because it is expected to contain
gluonic meson resonances lower in mass, than ternary or more complex
multi-gluonic mesons.

\noindent
Let us first consider the finite dimensional (nonunitary) representations of
$SL2C \ \times \ SL2C$ restricted and unrestricted to the 
covering group of the real Lorentz group. Details are presented in 
appendices A.1 and A.2 .

\noindent
To this end we associate with a bosonic resonance a free, massive state
or collection of spin states. 
Let the total spin be J. 
The spinor wave functions are obtained by direct products
of full and chiral spin 1/2 spinors, and the four-momentumi p,
neglecting here the width of the associated resonance.

\begin{equation}
\label{eq:11}
\begin{array}{l}
t_{\ \alpha_{1}  \alpha_{ 2} \cdots \alpha_{N}} 
\ ( \ p \ ; \ \left \lbrace spin \right \rbrace \ ) \ e^{\ - i p x}
\ = 
\ \left \langle \ \Omega \ \right |
\ \phi_{\ \alpha_{1}  \alpha_{ 2} \cdots \alpha_{N}} \ ( \ x \ )
\ \left | \  p \ ; \ \left \lbrace  spin \right \rbrace 
\ \right \rangle
\vspace*{0.3cm} \\
N \ = \ 2 \ J
\hspace*{0.3cm} , \hspace*{0.3cm} 
p^{\ 2} \ = \ M^{\ 2}
\hspace*{0.3cm} , \hspace*{0.3cm} 
p^{\ 0} \ = \ E \ \ge \ M
\vspace*{0.3cm} \\
t_{\ \alpha_{1}  \alpha_{ 2} \cdots \alpha_{N}} 
\ = \ t_{\ \underline{\alpha}}
\hspace*{0.3cm} : \hspace*{0.3cm} 
\begin{array}{l}
\mbox{totally symmetric under}
\vspace*{0.3cm} \\
\mbox{permutations of the indices}
\end{array}
\ \alpha_{1} \ \cdots \ \alpha_{N}
\vspace*{0.3cm} \\
\alpha_{j} \ = \ 1,2 
\hspace*{0.3cm} , \hspace*{0.3cm} 
j \ = \ 1 \ \cdots \  N 
\vspace*{0.2cm} \\
\hline
\vspace*{-0.2cm} \\
\widetilde{t}^{\ \dot{\gamma}_{1}  \dot{\gamma}_{ 2} \cdots 
\dot{\gamma}_{N}} 
\ ( \ p \ ; \ \left \lbrace spin \right \rbrace \ ) \ e^{\ - i p x}
\ = 
\ \left \langle \ \Omega \ \right |
\ \psi^{\ \dot{\gamma}_{1}  \dot{\gamma}_{ 2} \cdots 
\dot{\gamma}_{N}} \ ( \ x \ )
\ \left | \  p \ ; \ \left \lbrace  spin \right \rbrace 
\ \right \rangle
\vspace*{0.3cm} \\
\widetilde{t}^{\ \dot{\gamma}_{1}  \dot{\gamma}_{ 2} \cdots 
\dot{\gamma}_{N}} 
\ = \ \widetilde{t}^{\ \underline{\dot{\gamma}}}
\hspace*{0.3cm} : \hspace*{0.3cm} 
\begin{array}{l}
\mbox{totally symmetric under}
\vspace*{0.3cm} \\
\mbox{permutations of the indices}
\end{array}
\ \dot{\gamma}_{1} \ \cdots \ \dot{\gamma}_{N}
\vspace*{0.3cm} \\
\dot{\gamma}_{j} \ = \ 1,2 
\hspace*{0.3cm} , \hspace*{0.3cm} 
j \ = \ 1 \ \cdots \  N 
\end{array}
\end{equation}

\noindent
In eq. (\ref{eq:11}) $\left \lbrace spin \right \rbrace$
denotes the spin state, to be specified in a general frame of motion,
and 
$( \ \phi_{\ \underline{\alpha}} \ , 
\ \psi^{\ \underline{\dot{\gamma}}} \ )$ a pair of free fields,
( right chiral , left chiral ) , associated with the particle in question.


\noindent
The transformation rules of the spinor wave functions 
$( \ t_{\ \underline{\alpha}}
\ , \ \widetilde{t}^{\ \underline{\dot{\gamma}}} \ )$ are

\begin{equation}
\label{eq:12}
\begin{array}{l}
\left \lbrace spin \right \rbrace \ \rightarrow \ s
\hspace*{0.3cm} , \hspace*{0.3cm} 
\# \ s \ = \ N \ + \ 1
\vspace*{0.3cm} \\
t_{\ \underline{\alpha}} 
\ ( \ \Lambda p \ ; \ s \ ) 
\ = 
\ S_{\ \underline{\alpha}}^{\hspace*{0.3cm} \underline{\beta}} \ ( \ a \ )
\ t_{\ \underline{\beta}} 
\ ( \ p \ ; \ s^{'} \ ) 
\ D_{\ s  \ s^{'} }
\vspace*{0.3cm} \\
\widetilde{t}^{\ \underline{\dot{\gamma}}} 
\ ( \ \Lambda p \ ; \ s \ ) 
\ =
\ \widetilde{S}^{\ \underline{\dot{\gamma}}}_{\hspace*{0.4cm} 
\underline{\dot{\delta}}} \ ( \ b \ )
\ \widetilde{t}^{\ \underline{\dot{\delta}}} 
\ ( \ p \ ; \ s^{'} \ ) 
\ D_{\ s  \ s^{'} }
\vspace*{0.3cm} \\
D_{\ s  \ s^{'} } \ = \ D_{\ s  \ s^{'} }^{\ J} \ ( \ \Lambda \ , \ p \ )
\hspace*{0.3cm} ; \hspace*{0.3cm}
b \ = \overline{a}
\end{array}
\end{equation}

\noindent
The sought representations of the Lorentz group are obtained as
symmetric products of the spin 1/2 chiral spinors. They are presented in
appendix A.1 . 

\noindent
There is a {\it small} step from binary photon to binary gluon compounds,
even though their classification with respect to quantum numbers
$J^{\ PC}$ is identical.

\noindent
To see this let me first discuss the $SU3_{\ c}$ gauge invariant
binary gauge boson operator

\begin{equation}
\label{eq:101}
\begin{array}{l}
B_{\ \left \lbrack \ \mu_{1} \ \nu_{1} \  \right \rbrack
\ , \ \left \lbrack  \ \mu_{2} \ \nu_{2}  \ \right \rbrack}
\ ( \ x_{\ 1} \ , \ x_{\ 2} \ )
\ =
\vspace*{0.3cm} \\
\hspace*{0.5cm}
F_{\ \left \lbrack \ \mu_{1} \ \nu_{1} \ \right \rbrack}
\ ( \ x_{\ 1} \ ; \ A \ )
\ U \ ( \ x_{\ 1} \ , \ A \ ; \ x_{\ 2} \ , \ B \ )
\ F_{\ \left \lbrack \ \mu_{2} \ \nu_{2} \ \right \rbrack}
\ ( \ x_{\ 2} \ ; \ B \ )
\vspace*{0.3cm} \\
A \ , \ B \ , \cdots \ = \ 1, \cdots , 8
\end{array}
\end{equation}

\noindent
Adjoint representation indices referring to the color gauge group
are denoted by A , B in eq. (\ref{eq:101}). Summation over
repeated such indices is implied.

\noindent
$F_{\ \left \lbrack \ \mu \ \nu \ \right \rbrack} \ ( \ x \ ; \ A \ )$
denote the color octet of field strengths. 

\noindent
The quantity $ U \ ( \ x \ , \ A \ ; \ y \ , \ B \ )$
in eq. (\ref{eq:101}) denotes the octet string operator,
i. e. the path ordered exponential
over a straight line path ${\cal{C}}$ from y to x.

\begin{equation}
\label{eq:102}
\begin{array}{l}
U \ ( \ x \ , \ A \ ; \ y \ , \ B \ )
\ = 
P \ \exp 
\ \left ( 
\ \left . 
{\displaystyle{\int}}_{\ y}^{\ x} 
\ \right |_{\ {\cal{C}}}
\ d \ z^{\ \mu}
\ \frac{1}{i} \ V_{\ \mu} \ ( \ z \ , \ D \ ) \ {\cal{F}}_{\ D}
\ \right )_{\ A \ B}
\vspace*{0.3cm} \\
\left ( \ {\cal{F}}_{\ D} \ \right )_{\ A B} \ = \ i \ f_{\ A \ D \ B}
\end{array}
\end{equation}

\noindent
In eq. (\ref{eq:102}) $f_{\  A \ D \ B}$ denotes the structure
constants of $SU3_{\ c}$ and ${\cal{F}}_{\ D}$ the generators
of its adjoint representation.
$ V_{\ \mu} \ ( \ z \ , \ D \ )$ denote the octet of field potentials. 

\noindent
Properties pertaining to the octet string operators, field strengths
and their potentials are collected in appendix A.3 .
\vspace*{0.2cm}

\noindent
The extensive discussion of Stokes relations in appendix A.3 
serves here to present as clear an argument as possible,
why the octet string operator \\
$\left . U \ ( \ x \ , \ A \ ; \ y \ , \ B \ ) \ \right |_{\ {\cal{C}}}$
taken over a straight line path ${\cal{C}}$ attached to two field
strength operators 
$F_{\ \left \lbrack \ \mu_{k} \ \nu_{k} \ \right \rbrack}
\ ( \ x_{\ k} \ ; \ A_{\ k} \ ) \ ; \ k \ = 1,2$ at the ends
of the string -- as displayed in eqs. (\ref{eq:101}) and (\ref{eq:102} --
form a configuration similar to an $H_{\ 2}$ molecule,
representing an
energetically favored gluonic meson, i.e. a hadronic resonance susceptible
of identification specifically in central production.
\vspace*{0.1cm}

\noindent
The band structure of the $H_{\ 2}$ molecule would then translate 
into the possible quantum numbers of the associated binary gluonic mesons,
{\it so defined}, in appropriately adapted analogy. 
\vspace*{0.1cm}

\noindent
From a purely theoretical point of view it has to be stressed, that this
remains at the present stage {\it a hypothesis}, subject to further
tests, extending the existing analyses in refs. \cite{PMWO} and \cite{HFPM}
as well as related and/or alternative points of view, 
to be substancified below.
\vspace*{0.2cm}

\newpage

\begin{figure}[htb]
\vskip -0.0cm 
\begin{center}
\vskip -0.5cm 
\mbox{\epsfig{file=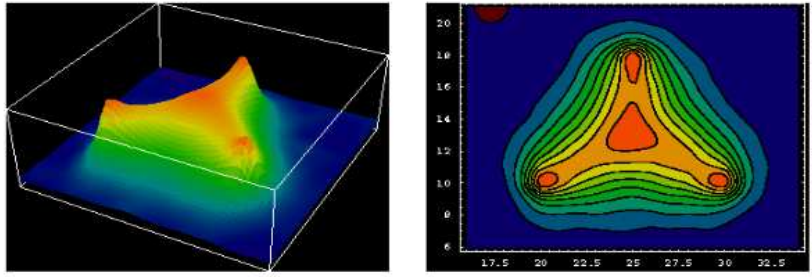,%
width=12cm}}
\end{center}
\vskip -0.5cm 
\caption{Gauge boson action density for a nucleon 
in lattice simulation of QCD.}
\label{fig10}
\end{figure}

\noindent
To illustrate the molecular aspect I reproduce in figure \ref{fig10}
the gauge boson action density in a lattice
calculcation of a nucleon \cite{Schier}

\noindent
The bilinear operator in eq. (\ref{eq:101}) satisfies Bose symmetry

\begin{equation}
\label{eq:103}
\begin{array}{l}
B_{\ \left \lbrack \ \mu_{1} \ \nu_{1} \  \right \rbrack
\ , \ \left \lbrack  \ \mu_{2} \ \nu_{2}  \ \right \rbrack}
\ ( \ x_{\ 1} \ , \ x_{\ 2} \ )
\ =
B_{\ \left \lbrack \ \mu_{2} \ \nu_{2} \  \right \rbrack
\ , \ \left \lbrack  \ \mu_{1} \ \nu_{1}  \ \right \rbrack}
\ ( \ x_{\ 2} \ , \ x_{\ 1} \ )
\ \rightarrow
\vspace*{0.3cm} \\
\widehat{C}^{\ -1} 
B_{\ \left \lbrack \ \mu_{1} \ \nu_{1} \  \right \rbrack
\ , \ \left \lbrack  \ \mu_{2} \ \nu_{2}  \ \right \rbrack}
\ ( \ x_{\ 1} \ , \ x_{\ 2} \ )
\ \widehat{C}
\ = 
B_{\ \left \lbrack \ \mu_{1} \ \nu_{1} \  \right \rbrack
\ , \ \left \lbrack  \ \mu_{2} \ \nu_{2}  \ \right \rbrack}
\ ( \ x_{\ 1} \ , \ x_{\ 2} \ )
\end{array}
\end{equation}

\noindent
In eq. (\ref{eq:103}) $\widehat{C}$ denotes the charge conjugation operator.

\noindent
We shall consider matrix elements of the form

\begin{equation}
\label{eq:104}
\begin{array}{l}
\left \langle 
\ \emptyset
\ \right |
\ B_{\ \left \lbrack \ \mu_{1} \ \nu_{1} \  \right \rbrack
\ , \ \left \lbrack  \ \mu_{2} \ \nu_{2}  \ \right \rbrack}
\ ( \ x_{\ 1} \ , \ x_{\ 2} \ )
\ \left |
\ gb \ ( \ J^{\ P\ C} \ ) \ ; \ p \ , 
\ \left \lbrace spin \right \rbrace
\ \right \rangle \ \rightarrow
\vspace*{0.3cm} \\
\exp^{\ - i p X} 
\ \widetilde{t}_{\ \underline{.}} \ ( \ z \ , \ p \ , \  J^{\ P\ C} \ ; \ . \ ) 
\hspace*{0.3cm} \mbox{with} \hspace*{0.3cm}
\vspace*{0.3cm} \\
\left |
\ \emptyset
\ \right \rangle \ : \ \mbox{ground state}
\hspace*{0.2cm} , \hspace*{0.2cm}
X \ = \ \frac{1}{2} 
\ \left ( \ x_{\ 1} \ + \ x_{\ 2} \ \right )
\hspace*{0.2cm} , \hspace*{0.2cm}
z \ = \ \ \left ( \ x_{\ 1} \ - \ x_{\ 2} \ \right )
\vspace*{0.3cm} \\
 J^{\ P \ C} \ : \ \mbox{total spin , parity , C-parity}
\hspace*{0.2cm} ; \hspace*{0.2cm}
p \ : \ \mbox{c.m. four momentum}
\vspace*{0.3cm} \\
\underline{.} \ : \ \mbox{spinor representation for}
\ \left \lbrack \ \mu_{\ 1} \ \nu_{\ 1} \ \right \rbrack \ , 
\ \left \lbrack \ \mu_{\ 2} \ \nu_{\ 2} \ \right \rbrack
\vspace*{0.3cm} \\
. \ : \ \mbox{spin state}
\end{array}
\end{equation}

\noindent
In eq. (\ref{eq:104}) $J^{\ P \ C}$ , p and  
$. \ = \ \left \lbrace spin \right \rbrace$ refer to 
properties of the gluonic meson
in question (gb) , whereas 
$\underline{.} \ \leftrightarrow
\ \left \lbrack \ \mu_{\ 1} \ \nu_{\ 1} \ \right \rbrack \ , 
\ \left \lbrack \ \mu_{\ 2} \ \nu_{\ 2} \ \right \rbrack$
and z refer to variables initrinsic to the operator 
$B_{\ \left \lbrack \ \mu_{1} \ \nu_{1} \  \right \rbrack
\ , \ \left \lbrack  \ \mu_{2} \ \nu_{2}  \ \right \rbrack}
\ ( \ x_{\ 1} \ , \ x_{\ 2} \ )$ .

\noindent
The four momentum p is introduced, as if 
$gb \ ( \ J^{\ P\ C} \ ) \ ; \ p \ , \ \left \lbrace spin \right \rbrace$
would correspond to a {\it stable} particle. This is at best
approximately justified in the zero width approximation,
which we shall {\it not} a priori assume to be valid.

\noindent
Nevertheless gb-s will manifest themselves as poles
in complex momentum planes, corresponding to analytic continuation
of strong interaction scattering amplitudes.
The latter refer to stable particles, like pions, kaons
and baryons, in the limit where both electromagnetic and weak interactions
are neglected. 

\noindent
Hence, ignoring the above complication for the time beeing, 
the mass of $gb \ ( \ J^{\ P \ C} \ )$ is defined through p

\begin{equation}
\label{eq:105}
\begin{array}{l}
m^{\ 2} \ = \ p_{\ \mu} \ p^{\ \mu} 
\hspace*{0.2cm} ; \hspace*{0.2cm} 
E_{\ p} \ = \ \sqrt{\ p^{\ 2} \ + \ m^{\ 2}}
\vspace*{0.3cm} \\
m \ = \ m \ \left ( \ gb \ ( \ J^{\ P \ C} \ ) \ \right )
\end{array}
\end{equation}

\noindent
As a consequence of eq. (\ref{eq:103}) we have for {\it binary}
gluonic mesons $C \ = \ +$ throughout.

\begin{equation}
\label{eq:106}
\begin{array}{l}
gb \ ( \ J^{\ P \ C} \ ) \ \rightarrow 
gb \ ( \ J^{\ P \ +} \ )
\end{array}
\end{equation}

\noindent
The relativistic spin $\underline{.}$ ,
processed as outlined in appendix A.1 , combines the same way as
in the nonrelativistic case to a total spin $S_{\ 12 }$

\begin{equation}
\label{eq:107}
\begin{array}{l}
S_{\ 12} \ = \ S_{\ 12}^{\ +} \ + \ S_{\ 12}^{\ -}    
\vspace*{0.3cm} \\
\left (
\ S_{\ 12}^{\ +} \ = \ 0 \ , \ 2 
\hspace*{0.2cm} \leftrightarrow \hspace*{0.2cm} 
P \ = \ +
\ \right )
\hspace*{0.2cm} ;  \hspace*{0.2cm} 
\left (
\ S_{\ 12}^{\ -} \ = \ 1   
\hspace*{0.2cm} \leftrightarrow \hspace*{0.2cm} 
P \ = \ -
\ \right )
\end{array}
\end{equation}

\noindent
The total spin states $S_{\ 12}$ in eq. (\ref{eq:107}) are
subject to transversity conditions, to which we will turn below.
But independtly thereof the spectrum of $gb \ ( \ J^{\ P \ +} \ )$
at this stage splits into three

\begin{equation}
\label{eq:108}
\begin{array}{l}
gb \ ( \ J^{\ P \ +} \ )
\ \begin{array}{lll ll}
\nearrow & gb \ ( \ J^{\ + \ +} \ ) & , & S_{\ 12}^{\ +} \ = \ 2 
& \hspace*{0.3cm} : \hspace*{0.3cm} II^{\ +}
\vspace*{0.3cm} \\
\rightarrow & gb \ ( \ J^{\ + \ +} \ ) & , & S_{\ 12}^{\ +} \ = \ 0 
& \hspace*{0.3cm} : \hspace*{0.3cm} I^{\ +}
\vspace*{0.3cm} \\
\searrow & gb \ ( \ J^{\ - \ +} \ ) & , & S_{\ 12}^{\ -} \ = \ 1
& \hspace*{0.3cm} : \hspace*{0.3cm} I^{\ -}
\end{array}
\end{array}
\end{equation}

\noindent
The three spectral types shall be denoted as in eq. (\ref{eq:108}) :
$II^{\ +}$ , $I^{\ +}$ and $I^{\ -}$ , where the superfix stands for parity.

\noindent
To clarify the spin structure we discuss the spectral classes $I^{\ \pm}$
first. To this end we label the amplitudes 
$\widetilde{t}_{\ \underline{.}} 
\ ( \ z \ , \ p \ , \  J^{\ P\ C} \ ; \ . \ )$ in eq. (\ref{eq:104})  

\begin{equation}
\label{eq:122}
\begin{array}{l}
\widetilde{t}_{\ \underline{.}} 
\ ( \ z \ , \ p \ , \  J^{\ P \ +} \ ; \ . \ )   
\ \rightarrow 
\ \widetilde{t}_{ \ 
\underline{.} \ ; \ S_{\ 12}^{\ \pm}} 
\ ( \ z \ , \ p \ , \ J^{\ \pm \ +} \  ; \ . \ )
\vspace*{0.3cm} \\
\hline
\vspace*{-0.2cm} \\
\widetilde{t}_{\ \underline{.} \ ; \ S_{\ 12}^{\ \pm}} 
\ ( \ z \ , \ p \ , \ J^{\ \pm \ +} \  ; \ . \ )
\begin{array}{ll}
\nearrow & 
\widetilde{t}_{\ \underline{.} \ ; \ II^{\ +}} 
\ ( \ z \ , \ p \ , \ J^{\ + \ +} \  ; \ . \ )
\vspace*{0.3cm} \\
\rightarrow & 
\widetilde{t}_{\ \underline{.} \ , \ I^{\ +}} 
\ ( \ z \ , \ p \ , \ J^{\ + \ +} \  ; \ . \ )
\vspace*{0.3cm} \\
\searrow & 
\widetilde{t}_{\ \underline{.} \ ; \ I^{\ -}} 
\ ( \ z \ , \ p \ , \ J^{\ - \ +} \  ; \ . \ )
\end{array}
\vspace*{0.3cm} \\
\underline{.} \ ; S_{\ 12}^{\ \pm} \ \rightarrow \ \underline{.}
\ \rightarrow 
\ \left \lbrack
\ \mu_{\ 1} \ \nu_{\ 1}
\ \right \rbrack
\ , 
\ \left \lbrack
\ \mu_{\ 2} \ \nu_{\ 2}
\ \right \rbrack
\end{array}
\end{equation}

\noindent
The two spectral classes $I^{\ \pm}$ exhibit the
relativistic factorization patterns

\begin{equation}
\label{eq:123}
\begin{array}{l}
\widetilde{t}_{\ \underline{.} \ ; \ I^{\ \pm}} 
\ ( \ z \ , \ p \ , \ J^{\ \pm \ +} \  ; \ . \ )
\ = 
\ \left (
 \begin{array}{l}
\ \left (
\ K^{\ \pm}
\ \right )_{\ 
\ \left \lbrack
\ \mu_{\ 1} \ \nu_{\ 1}
\ \right \rbrack
\ \left \lbrack
\ \mu_{\ 2} \ \nu_{\ 2}
\ \right \rbrack}
\ \times
\vspace*{0.3cm} \\
\hspace*{0.7cm} \times
\ \widetilde{t}_{\ I^{\ \pm}} 
\ ( \ z \ , \ p \ , \ J^{\ \pm \ +} \  ; \ . \ )
\end{array}
\ \right )
\vspace*{0.5cm} \\
\ \left (
\ K^{\ +}
\ \right )_{\ 
\ \left \lbrack
\ \mu_{\ 1} \ \nu_{\ 1}
\ \right \rbrack
\ \left \lbrack
\ \mu_{\ 2} \ \nu_{\ 2}
\ \right \rbrack}
\ =
\ g_{\ \mu_{\ 1} \ \mu_{\ 2}}
\ g_{\ \nu_{\ 1} \ \nu_{\ 2}}
\ -
\ g_{\ \mu_{\ 1} \ \nu_{\ 2}}
\ g_{\ \mu_{\ 2} \ \nu_{\ 1}}
\vspace*{0.5cm} \\
\ \left (
\ K^{\ -}
\ \right )_{\ 
\ \left \lbrack
\ \mu_{\ 1} \ \nu_{\ 1}
\ \right \rbrack
\ \left \lbrack
\ \mu_{\ 2} \ \nu_{\ 2}
\ \right \rbrack}
\ =
\ \varepsilon_{\ \mu_{\ 1} \ \mu_{\ 2} \ \nu_{\ 1} \ \nu_{\ 2}}
\end{array}
\end{equation}

\noindent
In eq. (\ref{eq:123}) $K^{\ \pm}$ denote the two Lorentz invariant tensors
with parity $\pm$ respectively and $g_{\ \mu \nu}$ the Lorentz metric
tensor.

\newpage

\noindent
The tensors $K^{\ \pm}$ form projection operations 
on the octet string operators 
\vspace*{-0.3cm} 

\begin{displaymath}
B_{\ \left \lbrack \ \mu_{1} \ \nu_{1} \  \right \rbrack
\ , \ \left \lbrack  \ \mu_{2} \ \nu_{2}  \ \right \rbrack}
\ ( \ x_{\ 1} \ , \ x_{\ 2} \ )
\end{displaymath}

\noindent
introduced in eq. (\ref{eq:101}) ,
described in appendix A.4 .

\noindent
The projections yield
\vspace*{-0.3cm} 

\begin{equation}
\label{eq:124}
\begin{array}{l}
B_{\ \left \lbrack \ \mu_{1} \ \nu_{1} \  \right \rbrack
\ , \ \left \lbrack  \ \mu_{2} \ \nu_{2}  \ \right \rbrack}
\ ( \ x_{\ 1} \ , \ x_{\ 2} \ )
\ =
 \left (
 \hspace*{-0.1cm}
 \begin{array}{r}
 \left (
\ K^{\ +}
\ \right )_{\ 
\ \left \lbrack
\ \mu_{\ 1} \ \nu_{\ 1}
\ \right \rbrack
\ \left \lbrack
\ \mu_{\ 2} \ \nu_{\ 2}
\ \right \rbrack}
\ B^{\ (+)}
\vspace*{0.3cm} \\
+
\ \left (
\ K^{\ -}
\ \right )_{\ 
\ \left \lbrack
\ \mu_{\ 1} \ \nu_{\ 1}
\ \right \rbrack
\ \left \lbrack
\ \mu_{\ 2} \ \nu_{\ 2}
\ \right \rbrack}
\ B^{\ (-)}
\vspace*{0.3cm} \\
 +
\ B^{\ '}_{\ \left \lbrack \ \mu_{1} \ \nu_{1} \  \right \rbrack
\ , \ \ \left \lbrack  \ \mu_{2} \ \nu_{2}  \ \right \rbrack}
\end{array}
 \right )
\vspace*{0.2cm} 
\end{array}
\end{equation}

\noindent
where the quantities $B^{\ (\pm)}$ are derived in appendix A.4 .
They are of the form given in eq. (\ref{eq:1014}) reproduced
below
\vspace*{-0.3cm} 

\begin{equation}
\label{eq:125}
\begin{array}{l}
B^{\ (+)} \ (  \ x_{\ 1} \ , \  x_{\ 2} \ )
\ =
\vspace*{0.3cm} \\
\hspace*{0.6cm} =
\ \frac{1}{12}
\ F_{\ \left \lbrack \ \alpha \ \beta \ \right \rbrack}
\ ( \ x_{\ 1} \ ; \ A \ )
\ U \ ( \ x_{\ 1} \ , \ A \ ; \ x_{\ 2} \ , \ B \ )
\ F^{\ \left \lbrack \ \alpha \ \beta \ \right \rbrack}
\ ( \ x_{\ 2} \ ; \ B \ )
\vspace*{0.3cm} \\
\hline
\vspace*{-0.2cm} \\
B^{\ (-)} \ (  \ x_{\ 1} \ , \  x_{\ 2} \ )
\ =
\vspace*{0.3cm} \\
\hspace*{0.6cm} =
\ - \ \frac{1}{12}
\ F_{\ \left \lbrack \ \alpha \ \beta \ \right \rbrack}
\ ( \ x_{\ 1} \ ; \ A \ )
\ U \ ( \ x_{\ 1} \ , \ A \ ; \ x_{\ 2} \ , \ B \ )
\ \widetilde{F}^{\ \left \lbrack \ \alpha \ \beta \ \right \rbrack}
\ ( \ x_{\ 2} \ ; \ B \ )
\vspace*{0.3cm} \\
\hspace*{0.6cm} =
\ - \ \frac{1}{12}
\ \widetilde{F}_{\ \left \lbrack \ \alpha \ \beta \ \right \rbrack}
\ ( \ x_{\ 1} \ ; \ A \ )
\ U \ ( \ x_{\ 1} \ , \ A \ ; \ x_{\ 2} \ , \ B \ )
\ F^{\ \left \lbrack \ \alpha \ \beta \ \right \rbrack}
\ ( \ x_{\ 2} \ ; \ B \ )
\vspace*{0.3cm} \\
\hline
\vspace*{-0.2cm} \\
\widetilde{F}_{\ \left \lbrack \ \alpha \ \beta \ \right \rbrack}
\ ( \ x_{\ 2} \ ; \ B \ )
\ =
\ \frac{1}{2}
\ \varepsilon_{ \ \alpha \ \beta \ \gamma \ \delta}
\ F^{\ \left \lbrack \ \gamma \ \delta \ \right \rbrack}
\ ( \ x_{\ 2} \ ; \ B \ )
\vspace*{0.3cm} \\
\mbox{and} \hspace*{0.3cm}
( \ x_{\ 2} \ ; \ B \ ) \ \leftrightarrow
\ ( \ x_{\ 1} \ ; \ A \ )
\end{array}
\end{equation}

\noindent
Returning to the (spin-) reduced amplitudes
$\widetilde{t}_{\ I^{\ \pm}} 
\ ( \ z \ , \ p \ , \ J^{\ \pm \ +} \  ; \ . \ )$
introduced in eq. (\ref{eq:123}) we obtain using the
notation of eq. (\ref{eq:104})

\begin{equation}
\label{eq:126}
\begin{array}{l}
\left \langle 
\ \emptyset
\ \right |
\ B^{\ (\pm)}
\ ( \ x_{\ 1} \ , \ x_{\ 2} \ )
\ \left |
\ gb_{\ I^{\ \pm}} \ ( \ J^{\ \pm \ +} \ ) \ ; \ p \ , 
\ \left \lbrace spin \right \rbrace
\ \right \rangle \ =
\vspace*{0.3cm} \\
\hspace*{0.3cm} =
\ \exp^{\ - i p X} 
\ \widetilde{t}_{\ I^{\ \pm}} \ ( \ z \ , \ p \ , \  J^{\ \pm \ +} \ ; \ . \ ) 
\end{array}
\end{equation}

\noindent
with $B^{\ (\pm)}$ given in eq. (\ref{eq:125}) .
In eq. (\ref{eq:126}) the suffix $I^{\ \pm}$ of the states
$gb_{\ I^{\ \pm}}$ indicates, that these are restricted
to the spectral types denoted $I^{\ \pm}$ in eq. (\ref{eq:108}) .

\noindent
In the local limit of $z \ \rightarrow \ 0$ , i.e. shrinking the extension
of the adjoint string to zero length, we recognize in $B^{\ (\pm)}$
two local operators shaping the dynamics of QCD. We ignore
here for clarity all short distance singularities, in this limit.

\begin{equation}
\label{eq:127}
\begin{array}{l}
z \ \rightarrow \ 0
\hspace*{0.3cm} : \hspace*{0.3cm}
\begin{array}{l lr}
B^{\ (+)} & \rightarrow &
\frac{1}{12} 
\ F_{\ \left \lbrack \ \mu \ \nu \ \right \rbrack}
\ ( \ X \ ; \ A \ )
\ F^{\ \left \lbrack \ \mu \ \nu \ \right \rbrack}
\ ( \ X \ ; \ A \ )
\vspace*{0.3cm} \\
B^{\ (-)} & \rightarrow &
- \ \frac{1}{12} 
\ F_{\ \left \lbrack \ \mu \ \nu \ \right \rbrack}
\ ( \ X \ ; \ A \ )
\ \widetilde{F}^{\ \left \lbrack \ \mu \ \nu \ \right \rbrack}
\ ( \ X \ ; \ A \ )
\end{array}
\ \rightarrow
\vspace*{0.4cm} \\
\hline
\vspace*{-0.3cm} \\
\begin{array}{l|l}
\begin{array}{rll}
3 \ \left . B^{\ (+)} \ \right |_{\ 0} 
& = &
{\cal{L}}^{\ (+)} \ ( \ X \ )
\vspace*{0.3cm} \\
- \ 3 \ \left . B^{\ (-)} \ \right |_{\ 0}
& = &
{\cal{L}}^{\ (-)} \ ( \ X \ )
\end{array}
 &
\begin{array}{rll}
{\cal{L}}^{\ (+)} \ ( \ X \ ) 
& = & 
g^{\ 2} \ s \ ( \ X \ )
\vspace*{0.3cm} \\
{\cal{L}}^{\ (-)} \ ( \ X \ ) 
& = & 
8 \ \pi^{\ 2} \ ch_{\ 2} \ ( \ X \ )
\end{array}
\end{array}
\vspace*{0.4cm} \\
\hline
\vspace*{-0.3cm} \\
\begin{array}{rll}
{\cal{L}}^{\ (+)} \ ( \ X \ )
& = &
\ \frac{1}{4} 
\ F_{\ \left \lbrack \ \mu \ \nu \ \right \rbrack}
\ ( \ X \ ; \ A \ )
\ F^{\ \left \lbrack \ \mu \ \nu \ \right \rbrack}
\ ( \ X \ ; \ A \ )
\vspace*{0.3cm} \\
{\cal{L}}^{\ (-)} \ ( \ X \ )
& = &
\frac{1}{4} 
\ F_{\ \left \lbrack \ \mu \ \nu \ \right \rbrack}
\ ( \ X \ ; \ A \ )
\ \widetilde{F}^{\ \left \lbrack \ \mu \ \nu \ \right \rbrack}
\ ( \ X \ ; \ A \ )
\end{array}
\end{array}
\end{equation}

\noindent
In eq. (\ref{eq:127}) $s$ denotes the action density pertaining
to gauge bosons and g the (strong) coupling constant, while 
$ch_{\ 2}$ represents the density of the second Chern character.

\noindent
We return to the wave functions 
$\widetilde{t}_{\ I^{\ \pm}} 
\ ( \ z \ , \ p \ , \ J^{\ \pm \ +} \  ; \ . \ )$ defined in 
eqs. (\ref{eq:123}) and (\ref{eq:126}) pertaining to the gb spectral types
$I^{\ \pm}$ in eq. (\ref{eq:126}) . As a consequence
of eq. (\ref{eq:103}) they satisfy the Bose symmetry relation

\begin{equation}
\label{eq:128}
\begin{array}{l}
\widetilde{t}_{\ I^{\ \pm}} 
\ ( \ z \ , \ p \ , \ J^{\ \pm \ +} \  ; \ . \ )
\ =
\widetilde{t}_{\ I^{\ \pm}} 
\ ( \ - \ z \ , \ p \ , \ J^{\ \pm \ +} \  ; \ . \ )
\end{array}
\end{equation}

\noindent
We meet a problem of interpretation of the bilinear wave functions
$\widetilde{t}_{\ I^{\ \pm}}$ and the symmetry in eq. (\ref{eq:128}) ,
known (also) from the study of
$q \overline{q}$ and 3 q composite systems \cite{PMBar}.

\noindent
This is recognized, decomposing the Lorentz four vector z 
into parallel and transverse components relative to
the c.m. momentum p

\begin{equation}
\label{eq:129}
\begin{array}{l}
z \ = \ z_{\ p} \ + \ \eta \ p \ / \ m^{\ 2}
\hspace*{0.3cm} ; \hspace*{0.3cm}
\eta \ = \ z \ p
\hspace*{0.3cm} \rightarrow \hspace*{0.3cm}
z_{\ p} \ p \ = \ 0
\end{array}
\end{equation}

\noindent
In the c.m. system the scalar product $\eta$ in eq. (\ref{eq:129})
becomes relative time, which is not a genuine degree of freedom
of the dynamical system in question.

\begin{equation}
\label{eq:130}
\begin{array}{l}
\mbox{c.m. :}
\hspace*{0.3cm} 
p \ \rightarrow 
\ p_{\ c.m.} \ ( \ m \ , \ \vec{0} \ )
\vspace*{0.3cm} \\
\eta \ \rightarrow 
\ m \ z_{\ 0} \ = \ m \ t_{\ rel}
\end{array}
\end{equation}

\noindent
Let me illustrate what is addressed here, considering the
decay $\varrho^{\ 0} \ \rightarrow \ 2 \ \pi$ .

\noindent
First we shall assume pions to be absolutely stable.
Then the decay 
\vspace*{-0.3cm} 

\begin{displaymath}
\varrho^{\ 0} \ \rightarrow \ 2 \ \pi^{\ 0}
\end{displaymath}

\noindent
is forbidden by Bose symmetry. Next we take into account, that
$\pi^{\ 0}$ s decay (mainly) into two photons, over the width of 
$\pi^{\ 0}$. The latter is according to the PDG \cite{PDG}  
\vspace*{-0.2cm} 

\begin{equation}
\label{eq:131}
\begin{array}{l}
\tau_{\ \pi^{\ 0}} \ = 
\ \left (
\ 8.4 \ \pm \ 0.6
\ \right ) \ 10^{\ -17} \ \mbox{sec}
\hspace*{0.3cm} \leftrightarrow \hspace*{0.3cm}
\Gamma_{\ \pi^{\ 0}} \ = 
\ \left (
\ 7.84 \ \pm \ 0.53
\ \right ) \ \mbox{eV}
\end{array}
\end{equation}

\noindent
To be specific we consider the reaction
\vspace*{-0.2cm} 

\begin{equation}
\label{eq:132}
\begin{array}{l}
e^+ \ e^- \ \rightarrow \ \varrho^{\ 0} \ \rightarrow \ 4 \ \gamma
\end{array}
\end{equation}

\noindent
and ask the question, whether it can proceed, when the two
pairs of photons, $\gamma_{\ 1} \ \gamma_{\ 2}$ and
$\gamma_{\ 3} \ \gamma_{\ 4}$ say, form each a $\pi^{\ 0}$ , 
with invariant masses $m_{\ 12}$ and $m_{\ 34}$ differing
by a well defined fraction of $\Gamma_{\ \pi^{\ 0}}$  
\vspace*{-0.2cm} 

\begin{equation}
\label{eq:133}
\begin{array}{l}
f_{ +} \ \Gamma_{\ \pi^{\ 0}}
\ \ge 
\left | 
\ m_{\ 12} \ - \  m_{\ 34} 
\ \right | 
\ \ge 
\ f_{ -} \ \Gamma_{\ \pi^{\ 0}}
\vspace*{0.3cm} \\
\mbox{with} 
\hspace*{0.4cm} 
\ f_{+} \ = \ 1 
\hspace*{0.3cm} , \hspace*{0.3cm}
\ f_{-} \ = \ 0.1 
\hspace*{0.4cm} 
 \mbox{say}
\end{array}
\end{equation}

\noindent
The answer relevant here is, that there is
a {\it multitude} of
equivalent irreducible wave functions out of the {\it family} 
defined in  eqs. (\ref{eq:123}) and (\ref{eq:126}) 
\vspace*{-0.2cm} 

\begin{equation}
\label{eq:134}
\begin{array}{l}
\widetilde{t}_{\ I^{\ \pm}} 
\ ( \ z \ , \ p \ , \ J^{\ \pm \ +} \  ; \ . \ )
\ =
\widetilde{t}_{\ I^{\ \pm}} 
\ ( \ z_{\ p} \ , \ p \ , \ \eta \ , \ J^{\ \pm \ +} \  ; \ . \ )
\end{array}
\end{equation}

\noindent
distinguished by the {\it parameter} $\eta$ as defined in
eqs. (\ref{eq:129}) and (\ref{eq:130}). 

\noindent
Thus we choose the representative with $\eta \ = \ 0$
\vspace*{-0.2cm} 

\begin{equation}
\label{eq:135}
\begin{array}{l}
t_{\ I^{\ \pm}} 
\ ( \ z_{\ p} \ , \ p \ , \ J^{\ \pm \ +} \  ; \ . \ )
\ =
\widetilde{t}_{\ I^{\ \pm}} 
\ ( \ z_{\ p} \ , \ p \ , \ \eta \ = \ 0 \ , \ J^{\ \pm \ +} \  ; \ . \ )
\end{array}
\end{equation}

\noindent
The above procedure illustrates the {\it difference}
between decay amplitudes of resonances into two photons
and their selection rules, derived in refs. \cite{Land} and \cite{Yang},
and the wave functions of binary gluonic mesons.

\noindent
The irreducible wave functions $t_{\ I^{\ \pm}}$ can readily
be discussed in the rest system of the momentum p, where
\vspace*{-0.2cm} 

\begin{equation}
\label{eq:136}
\begin{array}{l}
\mbox{c.m. :}
\hspace*{0.3cm} 
\left (
\ \eta \ = \ 0 \ , \ z_{\ p}
\ \right )
\ \rightarrow 
\ z_{\ p} \ = 
\ \left (
\ 0 \ , \ \vec{z}
\ \right )
\ \rightarrow 
\ \vec{z}
\vspace*{0.3cm} \\
t_{\ I^{\ \pm}} 
\ ( \ z_{\ p} \ , \ p \ , \ J^{\ \pm \ +} \  ; \ . \ )
\ \rightarrow 
t_{\ I^{\ \pm}} 
\ ( \ \vec{z} \ , \ J^{\ \pm \ +} \  ; \ . \ )
\end{array}
\end{equation}

\newpage

\noindent
\vspace*{-0.3cm} 
The procedure outlined above implies, that
the octet string bilinear operators
\vspace*{-0.3cm} 

\begin{equation}
\label{eq:137}
\begin{array}{l}
B_{\ \left \lbrack \ \mu_{1} \ \nu_{1} \  \right \rbrack
\ , \ \left \lbrack  \ \mu_{2} \ \nu_{2}  \ \right \rbrack}
\ ( \ x_{\ 1} \ , \ x_{\ 2} \ )
\end{array}
\end{equation}

\noindent
defined in eq. (\ref{eq:101}) are to be evaluated for
spacelike relative positions \\
$z \ = \ x_{\ 1} \ - \ x_{\ 2}$ only.
\vspace*{0.1cm} 

\noindent
Now eq. (\ref{eq:128}) takes the form
\vspace*{-0.3cm} 

\begin{equation}
\label{eq:138}
\begin{array}{l}
t_{\ I^{\ \pm}} 
\ ( \ \vec{z} \ , \ J^{\ \pm \ +} \  ; \ . \ )
\ =
\ t_{\ I^{\ \pm}} 
\ ( \ - \ \vec{z} \ , \ J^{\ \pm \ +} \  ; \ . \ )
\end{array}
\end{equation}

\noindent
The consequence for the angular momentum composition of 
the spectral types $I^{\ \pm}$ is indeed identical to the
situation of decay into two photons
\cite{Land} , \cite{Yang}
\vspace*{-0.3cm} 

\begin{equation}
\label{eq:139}
\begin{array}{l}
t_{\ I^{\ \pm}} 
\ ( \ \vec{z} \ , \ J^{\ \pm \ +} \  ; \ . \ )
\ \rightarrow
\ t_{\ I^{\ \pm}} 
\ ( \ \vec{z} \ , \ J^{\ \pm \ +} \  ; \ M \ )
\vspace*{0.3cm} \\
t_{\ I^{\ \pm}} 
\ ( \ \vec{z} \ , \ J^{\ \pm \ +} \  ; \ M \ )
\ = \ R^{\ J}_{\ I^{\ \pm}} \ ( \ r \ ) \ Y^{\ J}_{\ M} \ ( \ \vec{e} \ )
\hspace*{0.3cm} ; \hspace*{0.3cm}
J \ = \ \mbox{even}
\vspace*{0.3cm} \\
r \ = 
\ \left | \ \vec{z} \ \right |
\hspace*{0.3cm} , \hspace*{0.3cm}
\vec{e} \ = \ \vec{z} \ / \ r
\vspace*{0.4cm} \\
\hline
\vspace*{-0.3cm} \\
\begin{array}{lll}
I^{\ +} & : &
J^{\ P \ C} \ = \ 0^{\ ++} \ , \ 2^{\ ++} \ , \ 4^{\ ++} \ \cdots
\vspace*{0.3cm} \\
I^{\ -} & : &
J^{\ P \ C} \ = \ 0^{\ -+} \ , \ 2^{\ -+} \ , \ 4^{\ -+} \ \cdots
\end{array}
\end{array}
\end{equation}

\noindent
In eq. (\ref{eq:139}) $Y^{\ J}_{\ M}$ denote the orbital spherical harmonics
with angular momentum J , while 
$\left \lbrace \ R^{\ J}_{\ I^{\ \pm}} \ \right \rbrace$
stand for a {\it family} of radial wave functions. 
Neither nature nor extension of this family, nor its ordering in mass can
be deduced from the spectral type.

\noindent
For the sake of absolute clarity let me emphasize that the family
of wave functions (of all types pertinent to binary gluonic mesons)
can be empty, since no first principle proof to the
contrary exists.

\noindent
{\it Quantum numbers of binary gluonic mesons continued}
\vspace*{0.1cm} 

\noindent
{\it The $II^{\ +}$ spectral type}
\vspace*{0.1cm} 

\noindent
We turn to the remaining spectral type denoted $II^{\ +}$
in eq. (\ref{eq:108}) .

\noindent
The properties of this spectral series, represented
by the quantities \\
$B^{\ '}_{\ \left \lbrack \ \mu_{1} \ \nu_{1} \  \right \rbrack
\ , \ \ \left \lbrack  \ \mu_{2} \ \nu_{2}  \ \right \rbrack}$
defined in eq. (\ref{eq:124}) are derived in appendix A.5 .

\noindent
As shown there the wave functions of the gb spectral type
$II^{\ +}$ are uniquely associated with the classical 
energy momentum bilinear pertaining to gauge bosons.

\noindent
We retain here the characteristic composition
of the wave function associated bilinears $B^{\ '}$ in
eqs. (\ref{eq:1072}) and (\ref{eq:1073}) in the summary remarks
of appendix A.5:

\begin{equation}
\label{eq:140}
\begin{array}{l}
\hspace*{-0.2cm}
 B_{\ \left \lbrack \ \mu_{1} \ \nu_{1} \  \right \rbrack
\ , \ \left \lbrack  \ \mu_{2} \ \nu_{2}  \ \right \rbrack}
\ = 
\ \left \lbrace
\begin{array}{c}
\ \frac{1}{2}
 \left (
\begin{array}{l}
 g_{\ \mu_{\ 1} \ \mu_{\ 2}} \ \varrho_{\ \nu_{\ 1} \ \nu_{\ 2}}
- \  g_{\ \nu_{\ 1} \ \mu_{\ 2}} \ \varrho_{\ \mu_{\ 1} \ \nu_{\ 2}}
\vspace*{0.3cm} \\
- \  g_{\ \mu_{\ 1} \ \nu_{\ 2}} \ \varrho_{\ \nu_{\ 1} \ \mu_{\ 2}}
\ + \ g_{\ \nu_{\ 1} \ \nu_{\ 2}} \ \varrho_{\ \mu_{\ 1} \ \mu_{\ 2}}
\end{array}
 \right )
\vspace*{0.3cm} \\
+ 
\ K^{\ +}_{\ \left \lbrack \ \mu_{1} \ \nu_{1} \  \right \rbrack
  \ \left \lbrack  \ \mu_{2} \ \nu_{2}  \ \right \rbrack}
\ B^{\ (+)}
\vspace*{0.3cm} \\
+ 
\ K^{\ -}_{\ \left \lbrack \ \mu_{1} \ \nu_{1} \  \right \rbrack
  \ \left \lbrack  \ \mu_{2} \ \nu_{2}  \ \right \rbrack}
\ B^{\ (-)}
\vspace*{0.3cm} \\
\end{array}
\right \rbrace
\vspace*{0.4cm} \\
\hline
\vspace*{-0.3cm} \\
\mbox{with :}
\ g^{\ \mu_{\ 1} \ \mu_{\ 2}}
\ B_{\ \left \lbrack \ \mu_{1} \ \nu_{1} \  \right \rbrack
\ , \ \left \lbrack  \ \mu_{2} \ \nu_{2}  \ \right \rbrack}
\ =
\ R_{\ \nu_{\ 1} \ \nu_{\ 2}}
\vspace*{0.5cm} \\
- \ \varrho^{\ \mu \ \nu} \ = 
\ - \ R^{\ \mu \ \nu} \ + 
\ \frac{1}{4} 
\ g^{\ \mu \ \nu} \ R \ =
\ \vartheta_{\ cl}^{\ \mu \ \nu} 
\vspace*{0.5cm} \\
R \ =
\ g^{\ \nu_{\ 1} \ \nu_{\ 2}}
\ R_{\ \nu_{\ 1} \ \nu_{\ 2}}
\ =
\ 12 \ B^{\ (+)}
\end{array}
\end{equation}

\noindent
The bilinears $B^{\ '}$ are thus given by
\vspace*{-0.3cm} 

\begin{equation}
\label{eq:141}
\begin{array}{l}
B^{\ '}_{\ \left \lbrack \ \mu_{1} \ \nu_{1} \  \right \rbrack
\ , \ \ \left \lbrack  \ \mu_{2} \ \nu_{2}  \ \right \rbrack}
\ =
\ \frac{1}{2}
 \left (
\begin{array}{l}
 g_{\ \mu_{\ 1} \ \mu_{\ 2}} \ \varrho_{\ \nu_{\ 1} \ \nu_{\ 2}}
- \  g_{\ \nu_{\ 1} \ \mu_{\ 2}} \ \varrho_{\ \mu_{\ 1} \ \nu_{\ 2}}
\vspace*{0.3cm} \\
- \  g_{\ \mu_{\ 1} \ \nu_{\ 2}} \ \varrho_{\ \nu_{\ 1} \ \mu_{\ 2}}
\ + \ g_{\ \nu_{\ 1} \ \nu_{\ 2}} \ \varrho_{\ \mu_{\ 1} \ \mu_{\ 2}}
\end{array}
 \right )
\vspace*{0.3cm} \\
B^{\ '} \ \leftrightarrow \ \left \lbrace \ II^{\ +} \ \right \rbrace
\ \longleftrightarrow \ \vartheta_{\ cl}^{\ \mu \ \nu}
\end{array}
\end{equation}

\newpage

\noindent
The three spectral types are given in eq. (\ref{eq:142}) below,
completing the types $I^{\ \pm}$ in eq. (\ref{eq:139})
\vspace*{-0.3cm} 

\begin{equation}
\label{eq:142}
\begin{array}{l}
t_{\ I^{\ \pm}} 
\ ( \ \vec{z} \ , \ J^{\ \pm \ +} \  ; \ . \ )
\ \rightarrow
\ t_{\ I^{\ \pm}} 
\ ( \ \vec{z} \ , \ J^{\ \pm \ +} \  ; \ M \ )
\vspace*{0.3cm} \\
t_{\ I^{\ \pm}} 
\ ( \ \vec{z} \ , \ J^{\ \pm \ +} \  ; \ M \ )
\ = \ R^{\ J}_{\ I^{\ \pm}} \ ( \ r \ ) \ Y^{\ J}_{\ M} \ ( \ \vec{e} \ )
\hspace*{0.3cm} ; \hspace*{0.3cm}
J \ = \ \mbox{even}
\vspace*{0.3cm} \\
r \ = 
\ \left | \ \vec{z} \ \right |
\hspace*{0.3cm} , \hspace*{0.3cm}
\vec{e} \ = \ \vec{z} \ / \ r
\vspace*{0.4cm} \\
\hline
\vspace*{-0.3cm} \\
\begin{array}{lll}
I^{\ +} & : &
J^{\ P \ C} \ = \ 0^{\ ++} \ , \ 2^{\ ++} \ , \ 4^{\ ++} \ \cdots
\vspace*{0.3cm} \\
I^{\ -} & : &
J^{\ P \ C} \ = \ 0^{\ -+} \ , \ 2^{\ -+} \ , \ 4^{\ -+} \ \cdots
\end{array}
\vspace*{0.4cm} \\
\hline
\hline
\vspace*{-0.3cm} \\
t_{\ II^{\ +}} 
\ ( \ \vec{z} \ , \ J^{\ \pm \ +} \  ; \ . \ )
\ \rightarrow
\ t_{\ II^{\ +}} 
\ ( \ \vec{z} \ , \ J^{\ II \ +} \  ; \ M \ , \ \vec{E}_{\ \pm} \ )
\vspace*{0.4cm} \\
t_{\ II^{\ +}} 
\ ( \ \vec{z} \ , \ J^{\ II \ +} \  ; \ M \ , \ \vec{E}_{\ \pm} \ )
\ =
\vspace*{0.4cm} \\
\hspace*{0.4cm}  = 
\ R^{\ J}_{\ II^{\ +}} \ ( \ r \ , \ \vec{E}_{\ \pm} \ ) 
\ D^{\ J}_{\ M \ \pm 2} \ ( \ \vec{e} \ , \ \vec{E}_{\ \pm} \ )
\vspace*{0.4cm} \\
\begin{array}{lll}
II^{\ +} & : &
J^{\ P \ C} \ = \ 2^{\ ++} \ , \ 3^{\ ++} \ , \ 4^{\ ++} \ , \ 5^{\ ++}
\ \cdots
\end{array} 
\vspace*{0.3cm} \\
\hline
\end{array}
\end{equation}

\noindent
In eq. (\ref{eq:142}) the chromoelectric field strengths
$\vec{E}_{\ \pm}$ are retained in the arguments of the wave functions.

\noindent
The functions 
$D^{\ J}_{\ M \ \sigma} \ ( \ \vec{e} \ , \ \vec{E}_{\ \pm} \ )$ ,
with $\sigma \ = \ \pm 2$ ,
denote the eigenfunctions of a (symmetric) top, 
with the full orientation involving
three Euler angles provided by the correlation between 
the two chromoelectric field strengths $\vec{E}_{\ \pm}$ of
the adjoint string,
discussed in appendix A.5 .

\noindent
Let us end here the theoretical discussion of binary gluonic modes
associated with the octet gauge boson string. 
Theoretical expectations of spectral characteristics of states representing
the spectral types $I^{\ \pm}$ and $II^{\ +}$ shall be addressed in the
next section.

\newpage

\noindent
\section{Spectral patterns of gb - facts and fancy}

\noindent
{\bf a) Lattice QCD calculations}
\vspace*{0.1cm} 

\noindent
The most promising and widely accepted framework to derive
spectral patterns of hadrons, including gluonic
mesons, is lattice gauge theory and therein the restriction
to gauge boson degrees of freedom only. I shall quote 
several papers instead of a review : \cite{Ruefe} , 
\cite{Teper} , \cite{Michael} and \cite{Kuni} .

\noindent
I shall discuss the above papers one by one. In ref. \cite{Ruefe}
a careful and dedicated study is devoted to the determination
of the mass of $gb \ ( \ 0^{\ ++} \ )$ , the lowest lying gluonic meson in pure
Yang-Mills theory based on $SU3_{\ c}$ , and also
$gb \ ( \ 2^{++} \ )$ , with the results
\vspace*{-0.4cm} 

\begin{equation}
\label{eq:143}
\begin{array}{l}
m \ \left ( \ gb \ ( \ 0^{++} \ ) \ \right ) 
\ = \ 1627 \ \pm \ 83 \ \mbox{MeV}
\ \rightarrow 
\ m^{\ 2} \ = \ 2.65 \ \pm \ 0.27 \ \mbox{GeV}^{\ 2}
\vspace*{0.3cm} \\
m \ \left ( \ gb \ ( \ 2^{++} \ ) \ \right )
\ = \ 2354 \ \pm \ 95 \ \mbox{MeV}
\ \rightarrow 
\ m^{\ 2} \ = \ 5.54 \ \pm \ 0.6 \ \mbox{GeV}^{\ 2}
\end{array}
\end{equation}

\noindent
The main result refers to $gb \ ( \ 0^{\ ++} \ )$ and is
in very good agreement with all lattice gauge theory calculations, concerning
the same state. 

\noindent
I compare the above results with the assignment made here in
figure \ref{fig2}
\vspace*{-0.2cm} 

\begin{equation}
\label{eq:144}
\begin{array}{l}
m^{\ 2} \ \left ( \ gb \ ( \ 0^{++} \ ) \ \right ) 
\ = 
\ 1.04 \ \mbox{GeV}^{\ 2}
\vspace*{0.3cm} \\
m^{\ 2} \ \left ( \ gb \ ( \ 2^{++} \ ) \ \right )
\ = 
\ 3.13 \ \mbox{GeV}^{\ 2}
\end{array}
\end{equation}

\noindent
While it is difficult to associate an error with the tentative
pattern represented in figure \ref{fig2} and eq. (\ref{eq:144}) ,
to which I will return below, the essentially smaller 
mass square scale, by factors of $\sim \ 2.5$ and $\sim \ 1.8$ 
for $gb \ ( \ 0^{\ ++} \ )$ and $gb \ ( \ 2^{\ ++} \ )$ respectively,
is indeed a {\it basic} controversy, seemingly disproving
the mass square range considered in eq. (\ref{eq:144}) .

\noindent
In ref. \cite{Teper} an attempt is made to align gb resonances
on the Pomeron trajectory, as done here in figure \ref{fig2},
but with very different assignments : the slope of
the Pomeron trajectory is assumed to be

\begin{equation}
\label{eq:145}
\begin{array}{l}
\cite{Teper} \ :
\ \alpha^{\ '}_{\ P} \ = \ \ \alpha^{\ '}_{\ g \ b}
\ \sim \ 0.22 \ \pm \ 0.4 \ \mbox{GeV}^{\ -2}
\vspace*{0.3cm} \\
\mbox{here} \ :
\ \ \alpha^{\ '}_{\ g \ b} \ = \ \frac{1}{2} \ \alpha^{\ '}
\ = \ 0.5211 \ \pm \ 1.3 \ \% \ \mbox{GeV}^{\ -2}
\end{array}
\end{equation}

\noindent
Again a factor of two opens up, with respect to the
value of $\alpha^{\ '}_{\ gb}$, between ref. \cite{Teper} and
our present discussion, where indeed the relation
$\alpha^{\ '}_{\ gb} \ = \ \frac{1}{2} \ \alpha^{\ '}$ , also discussed below,
can be in doubt.

\newpage

\noindent
As a consequence of the calculations in ref. \cite{Teper} the
deduced mass square value for $gb \ ( \ 2^{\ ++} \ )$ ,
which is supposed to lye on the Pomeron trajectory, becomes

\begin{equation}
\label{eq:146}
\begin{array}{l}
\cite{Teper} \ :
\ m^{\ 2} \ \left ( \ gb \ ( \ 2^{++} \ ) \ \right )
\ = \ 4.4 \ \pm \ 1.2 \ \mbox{GeV}^{\ 2} 
\vspace*{0.3cm} \\
\mbox{here} \ :
\ m^{\ 2} \ \left ( \ gb \ ( \ 2^{++} \ ) \ \right )
\ = 
\ 3.13 \ \mbox{GeV}^{\ 2}
\vspace*{0.3cm} \\
\hspace*{1.0cm} 
\left (
\ m^{\ 2} \ \left ( \ gb \ ( \ 3^{++} \ ) \ \right )
\ = \ 4.17 \  \mbox{GeV}^{\ 2}
\ \right )
\end{array}
\end{equation}

\noindent
Comparing the mass square values of ref. \cite{Teper} in eq. (\ref{eq:146})
with the one of ref. \cite{Ruefe} in eq. (\ref{eq:143}) we see
(marginal) agreement.
\vspace*{0.1cm} 

\noindent
In ref. \cite{Michael} lattice calculations are presented 
to determin the masses of hybrid mesons, composed of
at least one gluon bound with a (nonstrange) $q \overline{q}$ pair,
and exhibiting $q \overline{q}$ exotic quantum numbers, such as

\begin{displaymath}
\begin{array}{l}
J^{\ P C} \ = \ 0^{\ --} \ , \ 0^{\ +-}
\ , \ 1^{\ -+} \ , \ 2^{\ +-} \ \cdots
\end{array}
\end{displaymath}

\noindent
The lightest hybrid states with nonstrange quarks is found with
characteristics

\begin{equation}
\label{eq:147}
\begin{array}{l}
\cite{Michael} \ :
\ J^{\ PC} \ = \ 1^{\ -+}
\hspace*{0.3cm} ; \hspace*{0.3cm}
m_{\ hyb} \ = \ 1.9 \ \pm \ 0.2 \ \mbox{GeV}
\vspace*{0.3cm} \\
\hspace*{1.0cm} \rightarrow  
\ m^{\ 2}_{\ hyb} \ = \ 3.6 \ \pm \ 0.6 \ \mbox{GeV}^{\ 2}
\end{array}
\end{equation}

\noindent
Also in lattice calculations of hybrid meson masses agreement between
different groups is very satisfactory. The above is not directly
related to the discussion of binary gluonic mesons, but
the result in eq. (\ref{eq:147}) is apparently contradicted by
the experimental finding of (at least) two exotic mesons
with $J^{\ PC} \ = \ 1^{\ -+}$ quantum numbers in p wave decay
to $\eta \ \pi$ and $\eta^{\ '} \ \pi$ \cite{Chung} .

\noindent
These resonances carry the name $\pi_{\ 1} \ ( \ 1400 \ )$ and
$\pi_{\ 1} \ ( \ 1600 \ )$ , where the mass in MeV is the argument.

\noindent
The two resonances in question were attributed the following characteristics
\cite{Chung} ( beyond $J^{\ PC} \ = \ 1^{\ -+}$ )

\begin{equation}
\label{eq:148}
\begin{array}{l}
\begin{array}{lll}
\pi_{\ 1} \ ( \ 1400 \ ) 
& : &
m \ = \ 1370 \ \pm \ 16 \ ^{\ + \ 50}_{\ - \ 30} \ \mbox{MeV}
\vspace*{0.3cm} \\
&  &
\Gamma \ = \hspace*{0.4cm} 385 \ \pm \ 40 \ ^{\ + \ 65}_{\ - \ 105} \ \mbox{MeV}
\vspace*{0.3cm} \\
\pi_{\ 1} \ ( \ 1600 \ ) 
& : & m \ = \ 1597 \ \pm \ 10 \  ^{\ + \ 45}_{\ - \ 10} \ \mbox{MeV} 
\vspace*{0.3cm} \\
&  &
\Gamma \ = \hspace*{0.4cm} 340 \ \pm \ 40 \ ^{\ + \ 50}_{\ - \ 50}  \ \mbox{MeV}
\end{array}
\end{array}
\end{equation}

\noindent
In the first paper in ref \cite{Chung} the authors remark, that
the exotic quantum numbers violate $SU3_{\ fl}$ symmetry, in the
decay $\pi_{\ 1} \ ( \ 1400 \ ) \ \rightarrow \ \pi \ \eta_{\ 8}$ ,
assigning pure flavor octet quantum numbers to $\eta$ , unless
it is not a hybrid meson but rather composed of two quarks and two
antiquarks.
There is a $\sim \ 20^{\ \circ}$ singlet octet mixing
between $\eta$ and $\eta^{\ '}$ , which, given the mass of
$\pi_{\ 1} \ ( \ 1400 \ )$ , i.e. below decay threshald for 
$\pi \ \eta^{\ '}$ ( modulo the width ) becomes essential, even though
we would then expect a reduction of the width by $\sim$ a factor of 5.

\noindent
Alternatively, it can not be excluded, that $\pi \ ( \ 1400 \ )$
is in a quark flavor configuration corresponding to 
$q \overline{q} \ q \overline{q}$ and thus is not a hybrid meson in the
first place. This discussion, even if at the side of the issue of
gluonic mesons, gives a taste of the 
{\it interpretation-} difficulties, facing the recognition of gb-s.

\noindent
But even if we assume that precisely $\pi \ ( \ 1600 \ )$ is 
a genuine hybrid meson, and further that the result given in ref. 
\cite{Michael} can be made to agree with a mass value of 1600 MeV,
it is difficult to conceive that $gb \ ( \ 0^{\ ++} \ )$ would
have a mass in excess of 1600 MeV as indicated
in the value given in ref. \cite{Ruefe} . To be fair to all lattice
calculations, let me stress, that the mass values of gluonic mesons refer
to the (unrealistic) case of no quark flavors (or all quark flavors very
heavy), and that a considerable shift in mass of e.g. $gb \ ( \ 0^{\ ++} \ )$
can be the result of the light quark flavors, unaccounted for in \cite{Ruefe}
and all comparable calculations.
\vspace*{0.1cm} 

\noindent
In ref. \cite{Kuni} the calculations focus on the question of low mass
scalar mesons, not gb-s. This issue is a prerequisite
for the successful identification of $gb \ ( \ 0^{\ ++} \ )$,
lowest in mass and thus was examined as to the structure
of the scalar $q \overline{q}$ nonet, lowest in mass,
in ref. \cite{PMWO}, where this nonet was {\it assumed} to be identifyable.

\noindent
In ref. \cite{Kuni} the local, composite field called $\sigma$ was investigated
on the lattice

\begin{equation}
\label{eq:149}
\begin{array}{l}
\sigma \ ( \ x \ ) \ = \ \frac{1}{\sqrt{2}}
\ \left (
\ \overline{u}_{\ c} \ ( \ x \ ) \ u_{\ c} \ ( \ x \ )
\ + \ \overline{d}_{\ c} \ ( \ x \ ) \ d_{\ c} \ ( \ x \ )
\ \right )
\end{array}
\end{equation}

\noindent
where the suffix c denotes triplet color.

\noindent
Irrespective of the pattern of the full nonet it is valid
to consider the two point function of two $\sigma$ fields, on the lattice,
and to deduce the mass of the lowest scalar resonance, coupling
to the $\sigma$ field.  

\noindent
The authors of ref. \cite{Kuni} declare their calculation preliminary,
so it is not yet possible to evaluate the error of their mass determination.
Nevertheless they indicate the following result

\begin{equation}
\label{eq:150}
\begin{array}{l}
\cite{Kuni} \ :
\ m_{\ \pi} \ < \ m_{\ \sigma} \ < \ m_{\ \varrho} 
\ \sim \ 776 \ \mbox{MeV}
\vspace*{0.3cm} \\
\cite{PMWO} \hspace*{0.3cm} : 
\ \sigma \ ( \ q \overline{q} \ ) \ \rightarrow
\ f_{\ 0} \ ( \ 980 \ ) \ ;
\ m_{\ f_{\ 0}} \ \sim \ 980 \ \mbox{MeV} 
\end{array}
\end{equation}

\noindent
We continue the discussion of the scalar $q \overline{q}$ nonet, beyond lattice
calculations only, in the next subsection.

\newpage

\noindent
{\bf b) $\pi \ \pi \ -$ and related ps-ps scattering and scalars}
\vspace*{0.1cm}

\noindent
In this contex let me start with quoting a recent paper
devoted to $\pi \ \pi$ elastic scattering in the framework of chiral
perturbation theory, and the Roy equation for full control of
analyticity, unitarity and crossing relations \cite{CLG}.

\noindent
In ref. \cite{CLG} in a dedicated chapter "Poles on the second sheet" ,
op. cit., the following pole parameters are quoted for the s wave
$I \ = \ 0 \ , \ \pi \pi$ partial wave amplitude

\begin{equation}
\label{eq:151}
\begin{array}{l}
\begin{array}{lll}
\cite{CLG} 
& : &
\sqrt{s} \ = 
\ \left ( 
\ 430 \ \pm \ 30 \ - \ i \ ( \ 295 \ \pm \ 20 \ )
\ \right ) \ \mbox{MeV}
\ \rightarrow
\vspace*{0.3cm} \\
 & &
\hspace*{0.2cm} 
s \ = 
\ \left ( 
\ 0.098 \ \pm 0.037 \ - \ i \ ( \ 0.254 \ \pm \ 0.032 \ )   
\ \right ) \ \mbox{GeV}^{\ 2}
\vspace*{0.3cm} \\
 & &
s_{\ thr} \ = \ 4 \ m_{\ \pi}^{\ 2} \ = \ 0.078 \ \mbox{GeV}^{\ 2}
\end{array}
\end{array}
\end{equation}

\noindent
The result in eq. (\ref{eq:151}) is indeed of highest interest.

\noindent
Within the quoted errors the resonance parameters are compatible
with a {\it threshold} resonance, when considered in the complex
s plane. For the properties of Jost functions in this and in general
cases I refer to Res Jost's original work \cite{Jfunc} .

\noindent
The deeper question related to the existence ( or nonexistence )
of the threshold resonance, as derived in ref. \cite{CLG} is, whether
there exists a symmetry, which would enforce the stability
of the resonance position, in particular in the chiral limit, i.e. of 

\begin{equation}
\label{eq:152}
\begin{array}{l}
s_{\ R} \ = \ \Re \ s \ \sim \ s_{\ thr} \ \rightarrow \ 0
\end{array}
\end{equation}

\noindent
The role of the threshold resonance is then apparently that
of a dilaton zero mode, arising from spontaneous breaking
of dilatation invariance. It is the trace anomaly, which prevents
the dilatation symmetry to be broken {\it exclusively} spontaneously.

\noindent
The relations with respect to the (Lorentz) invariant amplitude are
\vspace*{-0.3cm} 

\begin{equation}
\label{eq:153}
\begin{array}{l}
T_{\ 0} \ ( \ s \ ) \ = 
\ \frac{1}{2} \ {\displaystyle{\int}}_{\ -1}^{\ +1} \ d \ z
\ T \ ( \ s \ , \ z \ )
\hspace*{0.2cm} ; \hspace*{0.2cm} 
f_{\ 0} \ ( \ s \ ) \ = \ \frac{1}{8 \ \pi \ \sqrt{s}} \ T_{\ 0} \ ( \ s \ )
\vspace*{0.3cm} \\
f_{\ 0} \ ( \ s \ ) \ = \ \frac{1}{q} \ t \ ( \ q \ )
\hspace*{0.2cm} ; \hspace*{0.2cm} 
t \ ( \ q \ ) \ = \ \left ( \ S \ ( \ q \ ) \ - \ 1 \ \right ) \ / 
\ ( \ 2 \ i \ )
\vspace*{0.3cm} \\
t \ ( \ q \ ) \ = \ \frac{1}{16 \ \pi}
\ \sqrt{\ 1 \ - \ s_{\ thr} \ / \ s \ } \ T_{\ 0} \ ( \ s \ )
\vspace*{0.3cm} \\
\sigma_{\ el} \ ( \ s \ ) \ = \ 4 \ \pi \ | \ f_{\ 0} \ ( \ s \ ) \ |^{\ 2}
\hspace*{0.2cm} \mbox{for} \hspace*{0.2cm} 
s \ \mbox{real} \  \ge \ s_{\ thr}
\end{array}
\end{equation}

\noindent
In eq. (\ref{eq:153}) q denotes the c.m. momentum.

\noindent
So the function $S \ ( \ q \ )$ is of the form, assuming
indeed a threshold resonance with $s_{\ R} \ =
\ \left . \Re \ s \ \right |_{\ R} \ = \ s_{\ thr}$
and $\Im \ s_{\ R} \ = \ - \ \gamma_{\ R}$
\vspace*{-0.2cm} 

\begin{equation}
\label{eq:154}
\begin{array}{l}
S \ ( \ q \ ) \ = 
\ \begin{array}{c}
s_{\ R} \ + \ i \ \gamma_{\ R} \ - \ s
\vspace*{0.3cm} \\
\hline
\vspace*{-0.4cm} \\
s_{\ R} \ - \ i \ \gamma_{\ R} \ - \ s
\end{array}
\hspace*{0.3cm} S_{\ 1} \ ( \ q \ )
\end{array}
\end{equation}

\noindent
The most interesting situation arises if {\it first} we
assume that $S_{\ 1}$ tends to 1 at threshold

\begin{equation}
\label{eq:155}
\begin{array}{l}
S_{\ 1} \ ( \ q \ \rightarrow \ 0 \ ) \ \rightarrow \ 1
\vspace*{0.3cm} \\
\rightarrow \ \ f_{\ 0} \ \sim \ a \ \rightarrow \ i \ / \ q 
\end{array}
\end{equation}

\noindent
The behaviour displayed in eq. (\ref{eq:155}) is obviously at variance
with the restrictions imposed by (approximate) chiral symmetry,
but this is not the interesting part, due to a threshold
resonance. Rather it is the intrinsic interdependence
of the {\it remaining} contribution $S_{\ 1} \ ( \ q \ \sim \ 0 \ )$
near threshold with the threshold phase of $90^{\ \circ}$ of
the threshold resonance, which is most striking.

\noindent
The latter must be moved either backward or forward by another $90^{\ \circ}$
at threshold in order to achieve a finite scattering length.

\noindent
It is this interdependence, which is unlikely not to move the 
threshold resonance even very far from its initial threshold position.

\noindent
A measure for the width of the deduced threshold resonance is 
the ratio
\vspace*{-0.2cm} 

\begin{equation}
\label{eq:156}
\begin{array}{l}
\gamma_{\ R} \ / \ s_{\ R} \ \sim \ 2.5 
\ \leftrightarrow
\ \Gamma \ = \ 590 \ \pm \ 40 \ \mbox{MeV}
\end{array}
\end{equation}

\noindent
as obtained in ref. \cite{CLG} . While we do not pursue the above discussion
further here, it is necessary to retain, that the value of the mass derived in
ref. \cite{CLG}: $m_{\ R} \ = \ 430 \ \pm \ 30 \ \mbox{MeV}$ ,
especially when the width is just ignored, leads to an increased 
uncertainty concerning the very possibility of
recognizing the mass and mixing pattern of scalar mesons, in the
sense of spectroscopy.

\noindent
{\it Alternative discussions of scalar resonances}

\noindent
Besides the new derivation of the $I \ = \ 0$ s-wave 
$\pi \pi$ scattering amplitude in \cite{CLG}, the phase shifts in this
channel are by now fairly well established from threshold to a c.m. energy
of $\sim \ 1400$ MeV \cite{Lesniak}, but only as far as resolution
of phase ambiguities is concerned. 

\noindent
The  $I \ = \ 0$ s-wave
amplitude from the second reference in \cite{Lesniak} 
is reproduced below

\clearpage

\begin{figure}[htb]
\vskip -2.8cm 
\begin{center}
\vskip -1.5cm 
\mbox{\epsfig{file=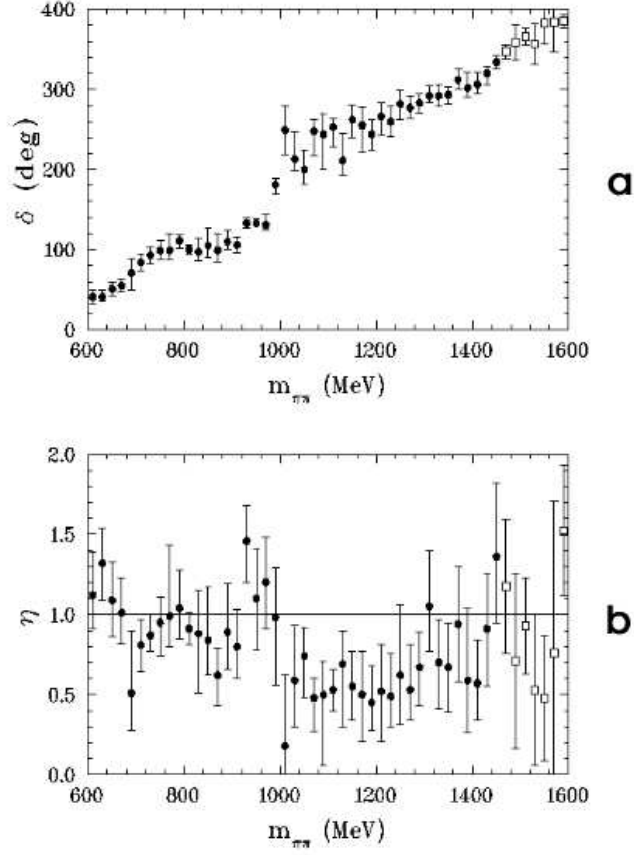,%
width=14cm}}
\end{center}
\vskip -2.9cm 
\caption{a) phase shifts $\delta$ and b) inelasticities $\eta$
for "down-flat" solution (circles). Squares denote data from 
ref. \cite{Les1}. }
\label{fig11}
\end{figure}


\noindent
It becomes clear from the errrors both in the phase shift 
( figure \ref{fig11} a ) as well as
in the inelasticity ( figure \ref{fig11} b ) that the details are, despite
a remarkable effort in analysis, rather uncertain in the
range of c.m. energies $600 \ \mbox{MeV} \ \le \ \sqrt{s} \ \le \ 1600 
\ \mbox{MeV}$.
\vspace*{0.2cm}

\noindent
{\bf The red dragon and "$\sigma$" in $\pi \pi \ ; \ I \ = \ 0$ s wave}
\vspace*{0.2cm}

\noindent
The discussion of the partial wave amplitude, corresponding to the projection
on $I \ = \ 0$ and on the s wave, denoted $t \ ( \ q \ )$ in eq.
\ref{eq:153}

\begin{equation}
\label{eq:157}
\begin{array}{l}
t \ ( \ q \ ) \ = \ \left ( \ S \ ( \ q \ ) \ - \ 1 \ \right ) \ / 
\ ( \ 2 \ i \ )
\end{array}
\end{equation}

\noindent
has been the subject of many recent papers, to which we turn now.
But we first show the result of combining elastic
and quasi elastic pseudoscalar meson scattering,
corresponding to the same quantum numbers, performed in ref. \cite{PMWO}.
For a detailed discussion I refer back to ref. \cite{PMWO}.

\clearpage

\begin{figure}[htb]
\vskip -0.2cm 
\begin{center}
\vskip -0.2cm 
\mbox{\epsfig{file=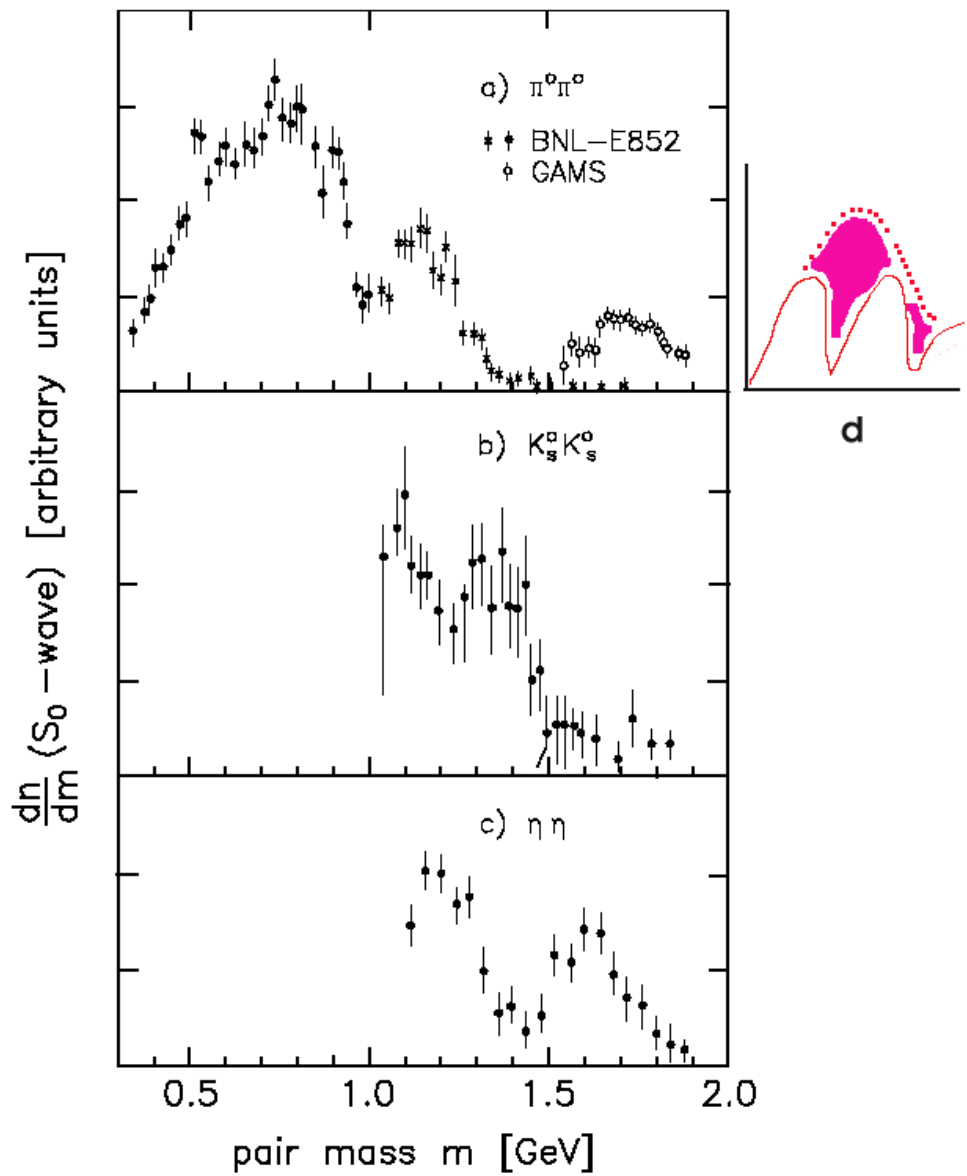,%
width=14cm}}
\end{center}
\vskip -1.0cm 
\caption{a) $\pi^0 \ \pi^0$ , b) $K_{\ s} \ K_{\ s}$ ,
c) $\eta \ \eta$ , d) red dragon in full. }
\label{fig12}
\end{figure}

\clearpage

\newpage

\noindent
The absolute values  $\left | \ t \ ( \ q \ ) \ \right |^{\ 2}$
(with only relative normalization) for 
$\pi \pi \ \rightarrow \ \pi \pi \ , \ K \overline{K} \ , \ \eta \eta$
are shown in figure \ref{fig12} together
with the full shape of the red dragon, amputating the negative interference 
due to $f_{\ 0} \ ( \ 980 \ )$ and  $f_{\ 0} \ ( \ 1500 \ )$.

\noindent
Several comments are necessary here :

\begin{description}
\item i) "data"

The compilation of figures \ref{fig12} a - c makes it appear as if actual
data is displayed. This is by no means the case, rather between the real data
from the reactions

\begin{displaymath}
\begin{array}{l}
\pi \ {\cal{N}} \ \rightarrow
\ \left \lbrace
\ \begin{array}{l}
\pi^{\ 0} \ \pi^{\ 0} \ {\cal{N}} \ ( \ \Delta \ )
\vspace*{0.3cm} \\
K_{\ s} \ K_{\ s} \ {\cal{N}} \ ( \ \Delta \ )
\vspace*{0.3cm} \\
\eta \ \eta \ {\cal{N}} \ ( \ \Delta \ )
\end{array}
\right .
\end{array}
\end{displaymath}

and the displayed absolute values there is a {\it series} of
analysis steps. The latter make it difficult to assess the
overall errors.

\item ii) the second interference minimum due to $f_{\ 0} \ ( \ 1500 \ )$

The pattern showing two interfering narrow states :
$f_{\ 0} \ ( \ 980 \ )$ and $f_{\ 0} \ ( \ 1500 \ )$ by todays notation,
has been inferred from the peripheral $\pi \ {\cal{N}}$ reactions
listed above.

The latter resonance has clearly been observed in $p \overline{p}$
annihilation at rest by the Crystal Barrel collaboration at the Lear facility
of CERN \cite{Amsler} , adding a new element with high statistics 
and precision of analysis.

\item iii) the red dragon proper

The unfolding of the interference due to $f_{\ 0} \ ( \ 980 \ )$ and
$f_{\ 0} \ ( \ 1500 \ )$ reveals a broad structure, the
red dragon proper, as sketched in fig. \ref{fig12} d.

The c.m. energy  over which this structure is extended 
comprises the range 
$400 \ \mbox{MeV} \ \le \ \sqrt{s} \ \le \ 1600 \ \mbox{MeV}$ .
Within all Breit-Wigner like strong interaction resonances, there
does not exist a comparably wide one. This establishes the singular
feature of the $\pi \pi$ s wave scattering amplitude in this range,
and also considerably below 400 MeV, i.e. down to the two pion
threshold, as well as above 1600 MeV.

The combined experimental and theoretical evaluation of data, which led
to the clear picture represented by the red dragon
in figure \ref{fig12} is {\it not} subject to the remaining
large inherent errors of details of the respective
scattering amplitudes. This contrasts with all attempts : \cite{PMWO} ,
\cite{CLG} 
and those discussed below, where further 
{\it interpretation} of details of the red dragon are undertaken.

\end{description}

\noindent
{\bf $\sigma \ ( \ \sim \ 500 \ )$ and/or $\kappa \ ( \ \sim \ 750 \ )$
scalar mesons}
\vspace*{0.1cm} 

\noindent
The claims of the existence of an isospin singlet, nonstrange 
scalar state $\sigma$ in a mass region clearly below 
$f_{\ 0} \ ( \ 980 \ )$ are numerous besides ref. \cite{CLG}. 
Another light scalar state, $\kappa$ with isospin 1/2, well below  
$K^{\ *}_{\ 0} \ ( \ 1430 \ )$ has also received much attention.
These claims have been recently repeated on various grounds. 
We cite two reviews compiled within the PDG \cite{PDG} :
on scalar mesons \cite{spanier} and on non $q \overline{q}$ candidates
\cite{amslerrev} .

\noindent
A new window has opened up in the study of the decay of charmed 
\cite{aitala} , \cite{bediaga} and b flavored mesons \cite{garmash} ,
\cite{babar} .

\noindent
What is emerging from c- and b-flavored meson decays is the
clear fact, that in three pseudoscalar meson ( $\pi$ and $K$ ) decays 
two out of the three pseudoscalars are produced amply 
in their relative s wave. This is quite in line with analogous
decays from $p \overline{p}$ and hence the analysis in terms of two
body amplitudes, the third pseudoscalar beeing treated as 'kinematical
spectator, modulo constraints from Bose statistics',
was performed in all reactions in a similar way. 

\noindent
A few decays are listed below for definiteness

\begin{equation}
\label{eq:158}
\begin{array}{lll ll}
\begin{array}{l}
D^{\ +} 
\vspace*{0.3cm} \\
D^{\ +}_{\ s}
\end{array}
& \rightarrow &
\hspace*{0.3cm} \pi^{\ -} \ \pi^{\ +} \ \pi^{\ +}
\hspace*{0.3cm} & ; & \hspace*{0.3cm}
\cite{aitala} \ , \ \cite{bediaga}
\vspace*{0.3cm} \\
B^{\ +} 
& \rightarrow &
\ \begin{array}{l}
 \pi^{\ -} \ K^{\ +} \ \pi^{\ +}
\vspace*{0.3cm} \\
 K^{\ -} \ K^{\ +} \ K^{\ +}
\end{array}
\hspace*{0.3cm} & ; & \hspace*{0.3cm}
\cite{garmash} \ , \ \cite{babar}
\vspace*{0.3cm} \\
p \ \overline{p}
& \rightarrow &
\hspace*{0.3cm} \pi^{\ -} \ \pi^{\ +} \ \eta
\hspace*{0.3cm} & ; & \hspace*{0.3cm}
\cite{amslerrev2}
\end{array}
\end{equation}

\noindent
The present results from the study of the above decays do favour
the derived existence of a '$\sigma$' isoscalar state, called 
$f_{\ 0} \ ( \ 600 \ )$ by the PDG \cite{PDG} as well as indications
of an isospin 1/2 state called '$\kappa$', with masses 
$\sim \ 500$ MeV for $\sigma$ and $\sim \ 750$ MeV for $\kappa$
respectively. The determination of the widths is rather uncertain,
but follows the widths of peaks in the projected Dalitz plot distributions
of the order of 200-400 MeV. 

\noindent
It is fair to say, that as welcome as these new channels are, 
the present stage of analysis has not led to a clear
picture of scalar meson states. 
\vspace*{0.1cm}

\noindent
{\bf c) Central production experiments}
\vspace*{0.2cm}

\noindent
The first experiment searching for gluonic mesons in central
production was performed at the ISR at CERN \cite{akesson} ,
at $\sqrt{s} \ = \ 63$ GeV.

\noindent
I reproduce here the invariant mass distribution 
of $\pi^{\ +} \ \pi^{\ -}$ pairs as observed in ref. \cite{akesson}

\begin{figure}[htb]
\vskip +0.2cm 
\begin{center}
\vskip -0.2cm 
\mbox{\epsfig{file=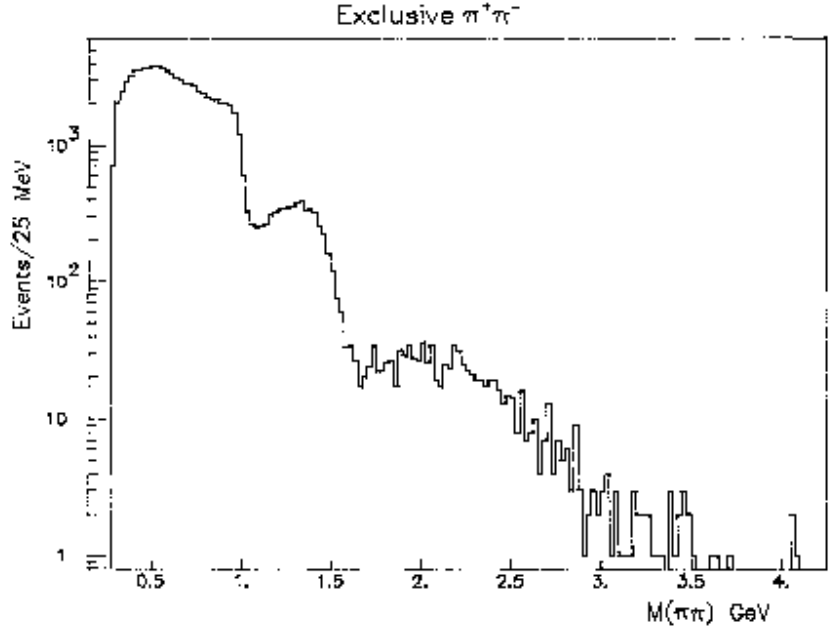,%
width=14cm}}
\end{center}
\vskip -1.0cm 
\caption{Invariant mass distribution of $\pi^{\ \pm}$ pairs in central
$p \ p \ \rightarrow \ p \ p \ X_{\ c}$ production
at $\sqrt{s} \ = \ 63$ GeV \cite{akesson}. }
\label{fig14}
\end{figure}

\noindent
Even though figure \ref{fig14} represents the (absolute) square of an amplitude and figure
\ref{fig12} the square of {\it another} amplitude,
the similarity and shape of the red dragon is clearly visible.
This similarity does not need any further analysis.

\noindent
The more recent experiment WA102 and its predecessor WA76 
are using a fixed target configuration
and thus the c.m. energies studied are lower 
$\sqrt{s} \ \le \ 29$ GeV \cite{WA102} .
\vspace*{0.1cm} 

\noindent
Despite dedicated studies \cite{Close}, no clear understanding
of central production {\it and} spectroscopic information
encoded in ps-ps scattering amplitudes ( section b) of this chapter )
nor any convincing evidence for the mass region from lattice
QCD calculations ( section a) of this chapter ) for 
the gluonic binary $gb ( \ 0^{\ ++} \ )$ is emerging.

\newpage

\noindent
Rather a choice of apparent possibilities is offered, where
in order not to offend any individuals I follow
the PDG \cite{PDG} 

\begin{equation}
\label{eq:159}
\begin{array}{l}
f_{\ 0} \ : \ 600 \ , \ 980 \ , \ 1370 \ , \ 1500 \ , \ 1710
\ \cdots \ \mbox{MeV}
\end{array}
\end{equation}

\noindent
The present {\it controversial} situation does - in my opinion -
reflect human shortcomings more than intrinsic difficulty
to understand the strong interaction dynamics underlying gluonic binaries
as well as $q \ \overline{q}$ scalar mesons.
\vspace*{0.1cm} 

\section{ Conclusion}
\vspace*{0.1cm} 

\noindent
In view of the previous sentence and in summary of the present
outline, I think that a dedicated experiment of
central production, at the highest achievable c.m. energies as well as
with an optimally adapted detector is scientifically worth while.

\newpage
\vspace*{0.1cm} 

\appendix

\section{Appendix}
\subsection{Spinor wave functions, spin states and transformation rules}

\noindent
\vspace*{0.1cm} 

\noindent
We present the spin 1/2 chiral building blocks below, as they determine
the general spin transformation rules defined in 
eq. (\ref{eq:12}) .

\begin{equation}
\label{eq:14}
\begin{array}{l}
S_{\ \underline{\alpha}}^{\hspace*{0.3cm} \underline{\beta}} \ ( \ a \ )
\ = \ \left \lbrace
\ S_{\ \alpha_{1}}^{\hspace*{0.3cm} \beta_{1}} \ ( \ a \ )
\ \times \ \cdots \ \times
\ S_{\ \alpha_{N}}^{\hspace*{0.3cm} \beta_{N}} \ ( \ a \ )
\ \right \rbrace_{\ symm}
\end{array}
\end{equation}

\noindent
The irreducible blocks 
$S_{\ \alpha_{j}}^{\hspace*{0.3cm} \beta_{j}} \ ( \ a \ )$ 
in eq. (\ref{eq:14}) correspond to spin 1/2 

\begin{equation}
\label{eq:15}
\begin{array}{l}
a \ = \ \left ( \ a_{\ 0} \ , \ a_{\ 1} \ , \ a_{\ 2} \ , \ a_{\ 3}
\ \right ) \ = \ a_{\ \mu}
\hspace*{0.3cm} : \hspace*{0.3cm} 
\begin{array}{l}
\mbox{complex four-vector}
\vspace*{0.1cm} \\
\mbox{{\it not} a Lorentz vector}
\end{array}
\vspace*{0.3cm} \\
S_{\ \alpha}^{\hspace*{0.3cm} \beta} \ ( \ a \ ) \ =
\ S_{\ 1}^{\ 1} \ ( \ a \ )
\ = 
\ \left (
\ a_{\ 0} \ \Sigma_{\ 0} \ + \ \frac{1}{i} \ \vec{a}
\ \vec{\Sigma}
\ \right )_{\ \alpha}^{\hspace*{0.3cm} \beta}
\vspace*{0.3cm} \\
 S_{\ 1}^{\ 1} \ ( \ a \ ) \ = 
\ \left (
\begin{array}{rr}
a_{\ 0} \ - \ i \ a_{\ 3} \hspace*{0.2cm} & \hspace*{0.2cm}
- \ a_{\ 2} \ - \ i \ a_{\ 1}
\vspace*{0.3cm} \\
 a_{\ 2} \ - \ i \ a_{\ 1} \hspace*{0.2cm} & \hspace*{0.2cm}
a_{\ 0} \ + \ i \ a_{\ 3}
\end{array}
\ \right )
\vspace*{0.3cm} \\
a^{\ 2} \ = \ a_{\ 0}^{\ 2} \ + \ \vec{a}^{\ 2} \ = \ Det \ S_{\ 1}^{\ 1} 
\ = \ 1
\end{array}
\end{equation}

\noindent
The quadratic constraint restricts $S_{\ 1}^{\ 1}$ as defined in eq.
\ref{eq:15} to be unimodular (i.e. to have Det = 1).
Rotations ( by half angles in bosonic terms )  correspond
to $a_{\ \mu}$ real. This is parametrizing the sphere ( over
the real numbers ) :
$S_{\ 3} \ \equiv \ SU2$. Lorentz boosts ( by hyperbolic
half angles in bosonic terms ) correspond to $a_{\ 0}$ real,
$\vec{a}$ pure imaginary. This is parametrizing
the double hyperboloid ( over the real numbers ) :
$( \ \Re \ a_{\ 0} \ )^{\ 2} \ - \ ( \ \Im \ \vec{a} \ )^{\ 2} \ = \ 1$.

\noindent
The matrices $\Sigma_{\ 0} \ , \ \Sigma_{\ k} \ ; \ k = \ 1,2,3$
are the Pauli matrices, as arising in the right chiral representation
of the full $\gamma \ -$ matrix algebra.

\begin{equation}
\label{eq:16}
\begin{array}{l}
\left ( \ \sigma_{\ \mu \ \nu} \ \right )_{\ \alpha}^{\hspace*{0.3cm} \beta}
\ \leftrightarrow
\ P_{\ R} \ \frac{i}{2}
\ \left \lbrack \ \gamma_{\ \mu} \ , \ \gamma_{\ \nu} \ \right \rbrack
\ P_{\ R}
\hspace*{0.3cm} ; \hspace*{0.3cm} 
P_{\ R} \ = \ \frac{1}{2} \ ( \ 1 \ + \ \gamma_{\ 5 \ R} \ )
\vspace*{0.3cm} \\
\gamma_{\ 5 \ R} \ = \ \frac{1}{i} 
\ \gamma_{\ 0} \ \gamma_{\ 1} \ \gamma_{\ 2} \ \gamma_{\ 3}
\vspace*{0.3cm} \\
\left ( \ \sigma_{\ \mu \ \nu} \ \right )_{\ \alpha}^{\hspace*{0.3cm} \beta}
\ =
\ \begin{array}{ll}
\ \left ( 
\begin{array}{l}
- \ i \ \Sigma_{\ k} 
\vspace*{0.3cm} \\
\varepsilon_{\ m n r} \ \Sigma_{\ r} 
\end{array}
\ \right )_{ \alpha}^{\hspace*{0.3cm} \beta}
\vspace*{1.0cm} 
& 
\vspace*{-0.5cm} 
\begin{array}{l}
\mbox{for} \ \mu \ = \ 0 \ , \ \nu \ = \ k \ = \ 1,2,3
\vspace*{0.3cm} \\
\mbox{for} \ \mu \ = \ m \ , \ \nu \ = \ n \ ; 
\vspace*{-0.2cm} \\
\hspace*{2.7cm} 
 m,n,r \ = \ 1,2,3
\end{array}
\end{array}
\end{array}
\end{equation}

\noindent
The right chiral quantities 
$\left ( \ \sigma_{\ \mu \ \nu} 
\ \right )_{\ \alpha}^{\hspace*{0.3cm} \beta}$ in eq. (\ref{eq:16})
satisfy the duality relation

\begin{equation}
\label{eq:17}
\begin{array}{l}
\left ( \ \sigma_{\ \mu \ \nu}
\ \right )_{\ \alpha}^{\hspace*{0.3cm} \beta}
\ \rightarrow \ \sigma_{\ \mu \ \nu}^{\ R}
\hspace*{0.3cm} ; \hspace*{0.3cm} 
\sigma_{\ \mu \ \nu}^{\ R} \ = \ - \ i 
\ \frac{1}{2} \ \varepsilon_{\ \mu \nu \varrho \tau} 
\ \sigma^{\ \varrho \ \tau \ R}
\end{array}
\end{equation}

\noindent
{\it Half angles (6) , rotational and hyperbolic - a) to the right}

\noindent
An infinitesimal Lorentz transformation is covered by
the spin 1/2 half angles $\omega^{\ \mu \nu}$ , defined below,
multiplying the (right chiral) base transformations 
$\sigma_{\ \mu \ \nu}^{\ R}$
\vspace*{-0.3cm} 

\begin{equation}
\label{eq:18}
\begin{array}{l}
\omega^{\ \mu \nu} \ = \ \frac{1}{2} \ \Omega^{\ \mu \nu}
\vspace*{0.3cm} \\
\omega^{\ \mu \nu} \ =
\ \left (
\begin{array}{ccc c}
0 & \varepsilon_{\ 1} & \varepsilon_{\ 2} & \varepsilon_{\ 3}
\vspace*{0.1cm} \\
- \varepsilon_{\ 1} & 0 & \Theta_{\ 3} & - \Theta_{\ 2}
\vspace*{0.1cm} \\
- \varepsilon_{\ 2} & - \Theta_{\ 3} & 0 & \Theta_{\ 1} 
\vspace*{0.1cm} \\
- \varepsilon_{\ 3} & \Theta_{\ 2} & - \Theta_{\ 1} & 0
\end{array}
\ \right )
\ =
\ \omega^{\ \mu \nu} \ ( \ \vec{\Theta} \ , \ \vec{\varepsilon} \ )
\end{array}
\end{equation}

\noindent
Projecting $\omega$ onto $\sigma^{\ R}$ we obtain

\begin{equation}
\label{eq:19}
\begin{array}{l}
\left ( \ \omega^{\ R} \ \right )_{\ \alpha}^{\hspace*{0.3cm} \beta}
\hspace*{0.3cm} \rightarrow \hspace*{0.3cm} 
\omega_{\ \alpha}^{\hspace*{0.3cm} \beta}
\vspace*{0.3cm} \\
\omega_{\ \alpha}^{\hspace*{0.3cm} \beta}
\ = \ \frac{1}{2}
\ \omega^{\ \mu \nu} 
\ \left ( \ \sigma_{\ \mu \ \nu}
\ \right )_{\ \alpha}^{\hspace*{0.3cm} \beta}
\ =
\ i
\ \left (
\ \left \lbrace
\ \vec{\Theta} \ - i \ \vec{\varepsilon}
\ \right \rbrace
\ \frac{1}{i} \ \vec{\Sigma}
\ \right )_{\ \alpha}^{\hspace*{0.3cm} \beta}
\vspace*{0.3cm} \\
\rightarrow 
\ \vec{\omega} \ \equiv \ \vec{\omega}^{\ R}
\ = \ \vec{\Theta} \ - \ i \ \vec{\varepsilon}
\end{array}
\end{equation}

\noindent
$S_{\ 1}^{\ 1}$ in eq. (\ref{eq:15}) then represents the
exponential of 
$\left ( \ \omega \ \right )_{\ \alpha}^{\hspace*{0.3cm} \beta}$
(multiplied with $\frac{1}{i}$)

\begin{equation}
\label{eq:20}
\begin{array}{l}
S_{\ \alpha}^{\hspace*{0.3cm} \beta} \ ( \ a \ ) \ =
\ \exp 
\ \left (
\ \frac{1}{i} \ \omega
\ \right )_{\ \alpha}^{\hspace*{0.3cm} \beta}
\ =
\ \exp
\ \left (
\ \frac{1}{i} \ \vec{\omega} \ \vec{\Sigma}
\ \right )_{\ \alpha}^{\hspace*{0.3cm} \beta}
\end{array}
\end{equation}

\noindent
Leaving out the (right chiral) spinor indices eq. (\ref{eq:20})
becomes

\begin{equation}
\label{eq:21}
\begin{array}{l}
S \ ( \ a \ ) \ = 
\ \cos \ (  \ \vec{\omega} \ \vec{\Sigma} \ )
\ - \ i \ \sin \ ( \ \vec{\omega} \ \vec{\Sigma} \ )
\end{array}
\end{equation}

\noindent
Thus we introduce the orthogonal complex invariant of $\vec{\omega}$
\vspace*{-0.3cm} 

\begin{equation}
\label{eq:22}
\begin{array}{l}
Z \ ( \ \omega \ ) \ = 
\ z^{\ 2} \ ( \ \omega \ ) \ = 
\ \vec{\omega}^{\ 2} \ = 
\ \left \lbrace
\ \vec{\Theta}^{\ 2} \ - \ \vec{\varepsilon}^{\ 2}
\ \right \rbrace 
\ - \ i 
\ \left \lbrace
\ 2 \ \vec{\Theta} \ \vec{\varepsilon}
\ \right \rbrace 
\vspace*{0.3cm} \\
\rightarrow
\hspace*{0.3cm}
S \ ( \ a \ ) \ =
\ \cos \ (  \ z \ ) \ \Sigma_{\ 0}
\ - \ i \ \left \lbrack 
\ \sin \ ( \ z \ ) \ / \ z 
\ \right \rbrack 
\ \vec{\omega} \ \vec{\Sigma} 
\vspace*{0.3cm} \\
\hspace*{0.3cm}
a \ = \ a \ ( \ \omega \ )
\hspace*{0.3cm} ; \hspace*{0.3cm} 
a_{\ 0} \ = \ \cos \ (  \ z \ )
\hspace*{0.3cm} ; \hspace*{0.3cm} 
\vec{a} \ = \  \left \lbrack
\ \sin \ ( \ z \ ) \ / \ z
\ \right \rbrack \ \vec{\omega}
\end{array}
\end{equation}

\noindent
The square root ambiguity of 
$z \ ( \ \omega \ ) \ = \ \pm \ \sqrt{Z \ ( \ \omega \ )}$
does not affect the functional relation
$a \ = \ a \ ( \ \omega \ )$, as becomes clear from 
eq. (\ref{eq:22}).
\vspace*{0.2cm}

\noindent
{\it From right chiral to left chiral spinors}
\vspace*{0.1cm}

\noindent
The right chiral base representations of $SL2C_{\ R}$
are by construction not parity invariant, nor are the 
matrices $S \ ( \ a \ ) \ \equiv \ {\cal{A}}$ over
the real numbers.

\noindent
Se we shall transform the defining equations (\ref{eq:14}-\ref{eq:16})
to the left chiral side

\begin{equation}
\label{eq:23}
\begin{array}{l}
\hspace*{0.6cm}
S_{\ \underline{\alpha}}^{\hspace*{0.3cm} \underline{\beta}} \ ( \ a \ )
\ = \ \left \lbrace
\ S_{\ \alpha_{1}}^{\hspace*{0.3cm} \beta_{1}} \ ( \ a \ )
\ \times \ \cdots \ \times
\ S_{\ \alpha_{N}}^{\hspace*{0.3cm} \beta_{N}} \ ( \ a \ )
\ \right \rbrace_{\ symm}
\ \rightarrow
\vspace*{0.3cm} \\
\rightarrow
\ \widetilde{S}^{\ \underline{\dot{\gamma}}}_{\hspace*{0.3cm} 
\underline{\dot{\delta}}} \ ( \ b \ )
\ = \ \left \lbrace
\ \widetilde{S}^{\ \dot{\gamma}_{1}}_{\hspace*{0.3cm} \dot{\delta}_{1}}
\ ( \ b \ )
\ \times \ \cdots \ \times
\ \widetilde{S}^{\ \dot{\gamma}_{N}}_{\hspace*{0.3cm} \dot{\delta}_{N}} 
\ ( \ b \ )
\ \right \rbrace_{\ symm}
\end{array}
\end{equation}

\begin{equation}
\label{eq:24}
\begin{array}{l}
b \ = \ \left ( \ b_{\ 0} \ , \ b_{\ 1} \ , \ b_{\ 2} \ , \ b_{\ 3}
\ \right ) \ = \ b_{\ \mu}
\hspace*{0.3cm} : \hspace*{0.3cm} 
\begin{array}{l}
\mbox{complex four-vector}
\vspace*{0.1cm} \\
\mbox{{\it not} a Lorentz vector}
\end{array}
\vspace*{0.3cm} \\
\widetilde{S}^{\ \dot{\gamma}}_{\hspace*{0.3cm} \dot{\delta}} \ ( \ b \ ) \ =
\ \widetilde{S}_{\ 2}^{\ 2} \ ( \ b \ )
\ = 
\ \left (
\ b_{\ 0} \ \widetilde{\Sigma}_{\ 0} \ + \ \frac{1}{i} \ \vec{b}
\ \vec{\widetilde{\Sigma}}
\ \right )^{\ \dot{\gamma}}_{\hspace*{0.3cm} \dot{\delta}}
\vspace*{0.3cm} \\
\widetilde{S}_{\ 2}^{\ 2} \ ( \ b \ ) \ = 
\ \left (
\begin{array}{rr}
b_{\ 0} \ - \ i \ b_{\ 3} \hspace*{0.2cm} & \hspace*{0.2cm}
- \ b_{\ 2} \ - \ i \ b_{\ 1}
\vspace*{0.3cm} \\
 b_{\ 2} \ - \ i \ b_{\ 1} \hspace*{0.2cm} & \hspace*{0.2cm}
b_{\ 0} \ + \ i \ b_{\ 3}
\end{array}
\ \right )
\vspace*{0.3cm} \\
b^{\ 2} \ = \ b_{\ 0}^{\ 2} \ + \ \vec{b}^{\ 2} \ = 
\ Det \ \widetilde{S}_{\ 2}^{\ 2} 
\ = \ 1
\end{array}
\end{equation}

\noindent
The transformation from ${\cal{A}}$ to ${\cal{B}}$
corresponds to the substitution

\begin{equation}
\label{eq:25}
\begin{array}{l}
{\cal{A}} \ \rightarrow 
\ \left ( \ {\cal{A}}^{\ \dagger} \ \right )^{\ -1}
\ = \ {\cal{B}}
\end{array}
\end{equation}

\noindent
The substitution in eq. (\ref{eq:25}) makes use of the
four base representations of SL2C, best represented in the associated
quadrangle 

\begin{equation}
\label{eq:26}
\begin{array}{l}
\begin{array}{ccc}
{\cal{A}} & \longleftrightarrow & 
\left ( \ {\cal{A}}^{\ T} \ \right )^{\ -1}
\vspace*{0.3cm} \\
\updownarrow \ \equiv \ c.c. & & 
\updownarrow \ \equiv \ c.c.
\vspace*{0.3cm} \\
\overline{\cal{A}} &  \longleftrightarrow &
\left ( \ {\cal{A}}^{\ \dagger} \ \right )^{\ -1}
\end{array}
\end{array}
\end{equation}

\noindent
In the quadrangle in eq. (\ref{eq:26})
the up-down operation means complex conjugation of each
matrix element, forming the involutory chains

\begin{displaymath}
\begin{array}{c}
{\cal{A}} \ \rightarrow \ \overline{{\cal{A}}} \ \rightarrow \ {\cal{A}}
\vspace*{0.3cm} \\
\mbox{and}
\vspace*{0.3cm} \\
\left ( \ {\cal{A}}^{\ T} \ \right )^{\ -1} \ \rightarrow
\ \left ( \ {\cal{A}}^{\ \dagger} \ \right )^{\ -1}
\ \rightarrow \ \left ( \ {\cal{A}}^{\ T} \ \right )^{\ -1}
\end{array}
\end{displaymath}

\noindent
whereas the left-right operation associates the symplectic dual,
forming the equally involutary chains

\begin{displaymath}
\begin{array}{c}
{\cal{A}} \ \rightarrow \ \left ( \ {\cal{A}}^{\ T} \ \right )^{\ -1}
\ \rightarrow \ {\cal{A}}
\vspace*{0.3cm} \\
\mbox{and}
\vspace*{0.3cm} \\
 \overline{{\cal{A}}} \ \rightarrow
\ \left ( \ {\cal{A}}^{\ \dagger} \ \right )^{\ -1}
\ \rightarrow \ \overline{{\cal{A}}}
\end{array}
\end{displaymath}

\noindent
Thus both up-down and left-right transformations along the
quadrangle in eq. (\ref{eq:26}) are commutative as well as involutory.

\noindent
Yet the left-right transformation associates equivalent representations,
contrary to the up- down one, which associates inequivalent
representations, of which we have chosen the two residing in the upper left
and lower right corners of the triangle in eq. (\ref{eq:26}.

\noindent
The symplectic equivalence is realized in the right chiral basis
by

\begin{equation}
\label{eq:27}
\begin{array}{l}
\left ( \ {\cal{A}}^{\ T} \ \right )^{\ -1}
\ =
\ s \ {\cal{A}} \ s^{\ -1} 
\hspace*{0.3cm} ; \hspace*{0.3cm} 
s \ = \ ( \ \pm \ ) \ i \ \sigma_{\ 2} \ = 
\ ( \ \pm \ ) \ \left (
\begin{array}{rl}
0 & 1
\vspace*{0.2cm} \\
- 1 & 0
\end{array}
\right )
\end{array}
\end{equation}

\noindent
The base Pauli matrices go into each other under the substitution in eq.
(\ref{eq:25})

\begin{equation}
\label{eq:28}
\begin{array}{l}
\Sigma_{\ \mu} \ = 
\ \Sigma_{\ \mu}^{\ \dagger}
\ =
\ \left ( \ \Sigma_{\ \mu} \ \right )^{\ -1}
\ =
\ \left ( \ \Sigma_{\ \mu}^{\ \dagger} \ \right )^{\ -1}
\vspace*{0.2cm} \\
\rightarrow
\ \left (
\ \widetilde{\Sigma}_{\ 0} \ , 
\ \vec{\widetilde{\Sigma}}
\ \right )
\ =
\ \left (
\ \Sigma_{\ 0} \ , 
\ \vec{\Sigma}
\ \right )
\end{array}
\end{equation}

\noindent
Hence we have

\begin{equation}
\label{eq:29}
\begin{array}{l}
{\cal{A}} \ \rightarrow 
\ \left ( \ {\cal{A}}^{\ \dagger} \ \right )^{\ -1}
\ = \ {\cal{B}} \ ( \ {\cal{A}} \ )
\vspace*{0.3cm} \\
b \ = \ \overline{a} 
\hspace*{0.3cm} ; \hspace*{0.3cm} 
 \forall \ \mbox{components}
\end{array}
\end{equation}

\noindent
Thus the quadrangle in eq. (\ref{eq:26}) leads
to the right- and left-chiral reality restricted form
of $SL2C_{\ R} \ \times \ SL2C_{\ L}$

\begin{equation}
\label{eq:30}
\begin{array}{l}
\left \lbrace spin \right \rbrace \ \rightarrow \ s
\hspace*{0.3cm} , \hspace*{0.3cm} 
\# \ s \ = \ N \ + \ 1
\hspace*{0.3cm} ; \hspace*{0.3cm} 
D_{\ s  \ s^{'} } \ = \ D_{\ s  \ s^{'} }^{\ J} \ ( \ \Lambda \ , \ p \ )
\vspace*{0.5cm} \\
\begin{array}{lll lll l}
R & : & t_{\ \underline{\alpha}} 
\ ( \ \Lambda p \ ; \ s \ ) 
& = &
 S_{\ \underline{\alpha}}^{\hspace*{0.3cm} \underline{\beta}} \ ( \ a \ )
& \hspace*{0.2cm} t_{\ \underline{\beta}} 
\ ( \ p \ ; \ s^{'} \ ) 
& \hspace*{0.2cm} D_{\ s  \ s^{'} }
\vspace*{0.3cm} \\
L & : & \widetilde{t}^{\ \underline{\dot{\gamma}}} 
\ ( \ \Lambda p \ ; \ s \ ) 
& = &
 \widetilde{S}^{\ \underline{\dot{\gamma}}}_{\hspace*{0.4cm} 
\underline{\dot{\delta}}} \ ( \ b \ )
& \hspace*{0.2cm} \widetilde{t}^{\ \underline{\dot{\delta}}} 
\ ( \ p \ ; \ s^{'} \ ) 
& \hspace*{0.2cm} D_{\ s  \ s^{'} }
\end{array}
\end{array}
\end{equation}

\noindent
While we proceed in steps, let me quote
Res Jost \cite{RJ} , illustrating the L-R chiral aspects.

\noindent
The decomposition in eq. (\ref{eq:14}) expands (doubles) into

\begin{equation}
\label{eq:31}
\begin{array}{l}
\begin{array}{lll}
R & : &
S_{\ \underline{\alpha}}^{\hspace*{0.3cm} \underline{\beta}} \ ( \ a \ )
\ = \ \left \lbrace
\ S_{\ \alpha_{1}}^{\hspace*{0.3cm} \beta_{1}} \ ( \ a \ )
\ \times \ \cdots \ \times
\ S_{\ \alpha_{N}}^{\hspace*{0.3cm} \beta_{N}} \ ( \ a \ )
\ \right \rbrace_{\ symm}
\vspace*{0.3cm} \\
L & : & 
\widetilde{S}^{\ \underline{\dot{\gamma}}}_{\hspace*{0.4cm} 
\underline{\dot{\delta}}} \ ( \ b \ )
\ = \ \left \lbrace
\ \widetilde{S}^{\ \dot{\gamma_{1}}}_{\hspace*{0.4cm} \dot{\delta_{1}}} 
\ ( \ b \ )
\ \times \ \cdots \ \times
\ \widetilde{S}^{\ \dot{\gamma_{N}}}_{\hspace*{0.4cm} \dot{\delta_{N}}} 
\ ( \ b \ )
\ \right \rbrace_{\ symm}
\end{array}
\end{array}
\end{equation}

\noindent
and then reduces to the R-L spin 1/2 building blocks

\begin{equation}
\label{eq:32}
\begin{array}{l}
\begin{array}{lll ll ll}
R & : &
S_{\ \alpha}^{\hspace*{0.3cm} \beta} \ ( \ a \ ) 
& = &
S_{\ 1}^{\ 1} \ ( \ a \ ) \ \equiv \ {\cal{A}} \ ( \ a \ )
& = &
\left (
\ a_{\ 0} \ \Sigma_{\ 0} \ + \ \frac{1}{i} \ \vec{a}
\ \vec{\Sigma}
\ \right )_{\ \alpha}^{\hspace*{0.3cm} \beta}
\vspace*{0.3cm} \\
L & : & 
\widetilde{S}^{\ \dot{\gamma}}_{\hspace*{0.3cm} \dot{\delta}} \ ( \ b \ ) 
& = &
\widetilde{S}_{\ 2}^{\ 2} \ ( \ b \ ) \ \equiv \ {\cal{B}} \ ( \ b \ )
& = &
 \left (
\ b_{\ 0} \ \widetilde{\Sigma}_{\ 0} \ + \ \frac{1}{i} \ \vec{b}
\ \vec{\widetilde{\Sigma}}
\ \right )^{\ \dot{\gamma}}_{\hspace*{0.3cm} \dot{\delta}}
\end{array}
\vspace*{0.3cm} \\
\widetilde{\Sigma}_{\ \mu} \ = \ \Sigma_{\ \mu}
\end{array}
\end{equation}

\noindent
The reality condition corresponds to a diagonal in the
quadrangle in eq. (\ref{eq:26})

\begin{equation}
\label{eq:33}
\begin{array}{l}
\begin{array}{rcr c}
{\cal{A}} \ ( \ a \ )  & \longleftrightarrow & 
{\cal{B}} \ ( \ b \ ) &
\vspace*{-0.1cm} \\
\searrow & & \searrow &
\vspace*{-0.0cm} \\
 & {\cal{A}}   & = \hspace*{0.4cm} & 
\left ( \ {\cal{B}}^{\ \dagger} \ \right )^{\ -1}
\vspace*{-0.1cm} \\
& \downarrow & & \downarrow
\vspace*{-0.0cm} \\
 & a & = \hspace*{0.4cm} & \overline{b}
\end{array}
\end{array}
\end{equation}

\noindent
The so constrained pair

\begin{equation}
\label{eq:34}
\begin{array}{l}
\left (
\ {\cal{A}} \ ( \ a \ )
\ , \ {\cal{B}} \ ( \ \overline{a} \ ) 
\ \right )
\ \equiv \ spin \ ( \ 1 \ , \ 3 \ ; \ \Re \ )
\ \simeq \ SL2C
\end{array}
\end{equation}

\noindent
defines the (self covered) group 
$spin \ ( \ 1 \ , \ 3 \ ; \ \Re \ )$ : $\Re$ indicates
that the spin group is over the {\it real} numbers,
whereas $1 \ , \ 3$ denote the signature of the derived metric,
i.e. 1 time and 3 space (real) dimensions.
\vspace*{0.1cm} 

\noindent
{\it Half angles (6) , rotational and hyperbolic - b) to the left}
\vspace*{0.1cm} 

\noindent
The left-chiral representation of $spin \ ( \ 1 \ , \ 3 \ ; \ \Re \ )$
complements the right-chiral one defined in eq. (\ref{eq:16})

\begin{equation}
\label{eq:35}
\begin{array}{l}
\left ( \ \sigma_{\ \mu \ \nu} \ \right )^{\ \dot{\gamma}}_{\hspace*{0.4cm} 
\dot{\delta}}
\ \leftrightarrow
\ P_{\ L} \ \frac{i}{2}
\ \left \lbrack \ \gamma_{\ \mu} \ , \ \gamma_{\ \nu} \ \right \rbrack
\ P_{\ L}
\hspace*{0.3cm} ; \hspace*{0.3cm} 
P_{\ L} \ = \ \frac{1}{2} \ ( \ 1 \ - \ \gamma_{\ 5 \ R} \ )
\vspace*{0.3cm} \\
\gamma_{\ 5 \ R} \ = \ \frac{1}{i} 
\ \gamma_{\ 0} \ \gamma_{\ 1} \ \gamma_{\ 2} \ \gamma_{\ 3}
\vspace*{0.3cm} \\
\left ( \ \sigma_{\ \mu \ \nu} \ \right )^{\ \dot{\gamma}}_{\hspace*{0.4cm} 
\dot{\delta}}
\ =
\ \begin{array}{ll}
\ \left ( 
\begin{array}{l}
 i \ \Sigma_{\ k} 
\vspace*{0.3cm} \\
\varepsilon_{\ m n r} \ \Sigma_{\ r} 
\end{array}
\ \right )^{ \dot{\gamma}}_{\hspace*{0.4cm} \dot{\delta}}
\vspace*{1.0cm} 
& 
\vspace*{-0.5cm} 
\begin{array}{l}
\mbox{for} \ \mu \ = \ 0 \ , \ \nu \ = \ k \ = \ 1,2,3
\vspace*{0.3cm} \\
\mbox{for} \ \mu \ = \ m \ , \ \nu \ = \ n \ ; 
\vspace*{-0.2cm} \\
\hspace*{2.7cm} 
 m,n,r \ = \ 1,2,3
\end{array}
\end{array}
\end{array}
\end{equation}

\noindent
In principle we should have distinguished the left chiral matrices
$\widetilde{\Sigma}_{\ \mu}$ characterizing the left chiral SL2C
basis in eq. (\ref{eq:35}) but we have chosen (without loss of generality)
to identify
$\widetilde{\Sigma}_{\ \mu} \ = \ \Sigma_{\ \mu}$ as specified 
in eq. (\ref{eq:32}).

\noindent
The left chiral variant of eq. (\ref{eq:17}) becomes

\begin{equation}
\label{eq:36}
\begin{array}{l}
\left ( \ \sigma_{\ \mu \ \nu}
\ \right )_{\ \alpha}^{\hspace*{0.3cm} \beta}
\ \rightarrow \ \sigma_{\ \mu \ \nu}^{\ R}
\hspace*{0.3cm} ; \hspace*{0.3cm} 
\sigma_{\ \mu \ \nu}^{\ R} \ = \ - \ i 
\ \frac{1}{2} \ \varepsilon_{\ \mu \nu \varrho \tau} 
\ \sigma^{\ \varrho \ \tau \ R}
\vspace*{0.3cm} \\
\rightarrow 
\ \left ( \ \sigma_{\ \mu \ \nu} \ \right )^{\ \dot{\gamma}}_{\hspace*{0.4cm} 
\dot{\delta}}
\ \rightarrow \ \sigma_{\ \mu \ \nu}^{\ L}
\hspace*{0.3cm} ; \hspace*{0.3cm} 
\sigma_{\ \mu \ \nu}^{\ L} \ = \ + \ i 
\ \frac{1}{2} \ \varepsilon_{\ \mu \nu \varrho \tau} 
\ \sigma^{\ \varrho \ \tau \ L}
\end{array}
\end{equation}

\noindent
Eq. (\ref{eq:19}) when reflected to the left takes on the form

\begin{equation}
\label{eq:37}
\begin{array}{l}
\left ( \ \omega^{\ L} \ \right )^{\ \dot{\gamma}}_{\hspace*{0.4cm} 
\dot{\delta}}
\hspace*{0.3cm} \rightarrow \hspace*{0.3cm} 
\widetilde{\omega}^{\ \dot{\gamma}}_{\hspace*{0.4cm} \dot{\delta}}
\vspace*{0.3cm} \\
\widetilde{\omega}^{\ \dot{\gamma}}_{\hspace*{0.4cm} \dot{\delta}}
\ = \ \frac{1}{2}
\ \omega^{\ \mu \nu} 
\ \left ( \ \sigma_{\ \mu \ \nu}
\ \right )^{\ \dot{\gamma}}_{\hspace*{0.4cm} \dot{\delta}}
\ =
\ i
\ \left (
\ \left \lbrace
\ \vec{\Theta} \ + i \ \vec{\varepsilon}
\ \right \rbrace
\ \frac{1}{i} \ \vec{\Sigma}
\ \right )^{\ \dot{\gamma}}_{\hspace*{0.4cm} \dot{\delta}}
\vspace*{0.3cm} \\
\rightarrow 
\ \vec{\widetilde{\omega}} \ \equiv \ \vec{\omega}^{\ L}
\ = \ \vec{\Theta} \ + \ i \ \vec{\varepsilon}
\hspace*{0.3cm} ; \hspace*{0.3cm}
 \vec{\omega} \ \equiv \ \vec{\omega}^{\ R}
\ = \ \vec{\Theta} \ - \ i \ \vec{\varepsilon}
\end{array}
\end{equation}

\noindent
$\widetilde{S}_{\ 2}^{\ 2}$ in eq. (\ref{eq:32} is
along with the right counterpart in eq. (\ref{eq:20})

\begin{equation}
\label{eq:38}
\begin{array}{l}
S_{\ \alpha}^{\hspace*{0.3cm} \beta} \ ( \ a \ ) \ =
\ \exp 
\ \left (
\ \frac{1}{i} \ \omega
\ \right )_{\ \alpha}^{\hspace*{0.3cm} \beta}
\ =
\ \exp
\ \left (
\ \frac{1}{i} \ \vec{\omega} \ \vec{\Sigma}
\ \right )_{\ \alpha}^{\hspace*{0.3cm} \beta}
\vspace*{0.3cm} \\
\rightarrow 
\ \widetilde{S}^{\ \dot{\gamma}}_{\hspace*{0.3cm} \dot{\delta}} \ ( \ b \ ) 
\ = \ \exp 
\ \left (
\ \frac{1}{i} \ \widetilde{\omega}
\ \right )^{\ \dot{\gamma}}_{\hspace*{0.4cm} \dot{\delta}}
\ =
\ \exp
\ \left (
\ \frac{1}{i} \ \vec{\widetilde{\omega}} \ \vec{\Sigma}
\ \right )^{\ \dot{\gamma}}_{\hspace*{0.4cm} \dot{\delta}}
\end{array}
\end{equation}

\noindent
Eq. (\ref{eq:21}) extends to

\begin{equation}
\label{eq:39}
\begin{array}{l}
S \ ( \ a \ ) \ = 
\ \cos \ (  \ \vec{\omega} \ \vec{\Sigma} \ )
\ - \ i \ \sin \ ( \ \vec{\omega} \ \vec{\Sigma} \ )
\vspace*{0.3cm} \\
\rightarrow 
\ \widetilde{S} \ ( \ b \ ) 
\ = \ S \ ( \ b \ )
\ = \ \exp 
\ \left (
\ \frac{1}{i} \ \widetilde{\omega}
\ \right )
\ = \ \cos \ (  \ \vec{\widetilde{\omega}} \ \vec{\Sigma} \ )
\ - \ i \ \sin \ ( \ \vec{\widetilde{\omega}} \ \vec{\Sigma} \ )
\vspace*{0.3cm} \\
\hspace*{0.5cm}
\vec{\widetilde{\omega}} \ =
\ \overline{\vec{\omega}}
\hspace*{0.3cm} ; \hspace*{0.3cm}
b \ = \ \overline{a}
\vspace*{0.3cm} \\
\hspace*{0.5cm}
S \ ( \ a \ ) \ = \ a_{\ 0} \ \Sigma_{\ 0} \ + \ \frac{1}{i} \ \vec{a}
\ \vec{\Sigma}
\hspace*{0.3cm} ; \hspace*{0.3cm}
S \ ( \ b \ ) \ = \ b_{\ 0} \ \Sigma_{\ 0} \ + \ \frac{1}{i} \ \vec{b}
\ \vec{\Sigma}
\end{array}
\end{equation}

\noindent
Eq. (\ref{eq:22}) completes to

\begin{equation}
\label{eq:40}
\begin{array}{l}
Z \ ( \ \omega \ ) \ = 
\ z^{\ 2} \ ( \ \omega \ ) \ = 
\ \vec{\omega}^{\ 2} \ = 
\ \left \lbrace
\ \vec{\Theta}^{\ 2} \ - \ \vec{\varepsilon}^{\ 2}
\ \right \rbrace 
\ - \ i 
\ \left \lbrace
\ 2 \ \vec{\Theta} \ \vec{\varepsilon}
\ \right \rbrace 
\vspace*{0.3cm} \\
\rightarrow
\hspace*{0.3cm}
S \ ( \ a \ ) \ =
\ \cos \ (  \ z \ ) \ \Sigma_{\ 0}
\ - \ i \ \left \lbrack 
\ \sin \ ( \ z \ ) \ / \ z 
\ \right \rbrack 
\ \vec{\omega} \ \vec{\Sigma} 
\vspace*{0.3cm} \\
\hspace*{0.3cm}
a \ = \ a \ ( \ \omega \ )
\hspace*{0.3cm} ; \hspace*{0.3cm} 
a_{\ 0} \ = \ \cos \ (  \ z \ )
\hspace*{0.3cm} ; \hspace*{0.3cm} 
\vec{a} \ = \  \left \lbrack
\ \sin \ ( \ z \ ) \ / \ z
\ \right \rbrack \ \vec{\omega}
\vspace*{0.5cm} \\
\rightarrow
\hspace*{0.3cm}
Z \ ( \ \widetilde{\omega} \ ) \ = 
\ z^{\ 2} \ ( \ \widetilde{\omega} \ ) \ = 
\ \vec{\widetilde{\omega}}^{\ 2} \ = 
\ \left \lbrace
\ \vec{\Theta}^{\ 2} \ - \ \vec{\varepsilon}^{\ 2}
\ \right \rbrace 
\ + \ i 
\ \left \lbrace
\ 2 \ \vec{\Theta} \ \vec{\varepsilon}
\ \right \rbrace 
\vspace*{0.3cm} \\
Z \ ( \ \widetilde{\omega} \ ) \ = 
\ z^{\ 2} \ ( \ \widetilde{\omega} \ ) \ = 
\ \overline{Z} \ ( \ \omega \ ) 
\ =
\ \overline{z^{\ 2}} \ ( \ \omega \ ) 
\vspace*{0.3cm} \\
S \ ( \ b \ ) \ =
\ \cos \ (  \ \overline{z} \ ) \ \Sigma_{\ 0}
\ - \ i \ \left \lbrack 
\ \sin \ ( \ \overline{z} \ ) \ / \ \overline{z} 
\ \right \rbrack 
\ \overline{\vec{\omega}} \ \vec{\Sigma} 
\vspace*{0.3cm} \\
b \ = \ a \ ( \ \widetilde{\omega} \ ) \ = \ \overline{a \ ( \ \omega \ )}
\end{array}
\end{equation}

\newpage

\noindent
{\bf Realization of (half) angles through an antisymmetric pair of vectors}
\vspace*{0.1cm} 

\noindent
The complex three vectors defining the half angles
$\omega^{\ \mu \nu}$ in eq. (\ref{eq:18})

\begin{equation}
\label{eq:41}
\begin{array}{l}
\vec{\omega}^{\ R}  
\ = \ \vec{\Theta} \ - \ i \ \vec{\varepsilon}
\hspace*{0.3cm} ; \hspace*{0.3cm}
\vec{\omega}^{\ L}
\ = \ \vec{\Theta} \ + \ i \ \vec{\varepsilon}
\end{array}
\end{equation}

\noindent
can be realized 
as antisymmetric combinations of two real Lorentz vectors
$x^{\ \mu} \ , \ y^{\ \nu}$ .
{\it This is however a restricted realization.} \\
Here Lorentz vector does not distinguish
between vector and axial vector. In fact we shall think
of x as a genuine Lorentz four vector and of y
as an axial vector.

\begin{equation}
\label{eq:42}
\begin{array}{l}
\omega_{\ \mu \nu} \ ( \ [ \ x \ , \ y \ ] \ )
\ = 
\ \varepsilon_{ \ \mu \nu \sigma \tau}
\ x^{\ \sigma} \ y^{\ \mu} 
\vspace*{0.3cm} \\
\omega_{\ 0 k} \ = \ \left ( \ \vec{x} \ \wedge \ \vec{y} \ \right )^{\ k}
\hspace*{0.3cm} ; \hspace*{0.3cm}
\omega_{\ m n} \ = 
\ \varepsilon_{ \ m n r }
\ \left ( \ x^{\ 0} \ y^{\ r} \ - \ x^{\ r} \ y^{\ 0} \ \right )
\vspace*{0.3cm} \\
\vec{\varepsilon} \ = \ - \ \vec{x} \ \wedge \ \vec{y}
\hspace*{0.3cm} ; \hspace*{0.3cm}
\vec{\Theta} \ = \ x^{\ 0} \ \vec{y} \ - \ y^{\ 0} \ \vec{x}
\hspace*{0.3cm} : \hspace*{0.3cm}
\vec{\Theta} \ \vec{\varepsilon} \ = \ 0
\vspace*{0.5cm} \\
\begin{array}{ll}
\rightarrow &
\vec{\omega}^{\ R} \ = \ x^{\ 0} \ \vec{y} \ - \ y^{\ 0} \ \vec{x}
\ + \ i \ \left ( \ \vec{x} \ \wedge \ \vec{y} \ \right )
\vspace*{0.3cm} \\
 & \vec{\omega}^{\ L} \ = \ x^{\ 0} \ \vec{y} \ - \ y^{\ 0} \ \vec{x}
\ - \ i \ \left ( \ \vec{x} \ \wedge \ \vec{y} \ \right )
\end{array}
\end{array}
\end{equation}

\noindent
The invariants $Z \ ( \ \omega \ ) \ , \ Z \ ( \ \widetilde{\omega} \ )$
in eqs. (\ref{eq:22}) and (\ref{eq:40}) are purely real

\begin{equation}
\label{eq:43}
\begin{array}{l}
Z \ ( \ \omega \ ) \ = 
\ \left ( \ \vec{\omega}^{\ R} \ \right )^{\ 2}
\ = 
\ \left ( \ x \ y \ \right )^{\ 2}
\ - 
\ x^{\ 2} \ y^{\ 2} 
\ = 
\ \left ( \ \vec{\omega}^{\ L} \ \right )^{\ 2}
\ = 
\ Z \ ( \ \widetilde{\omega} \ )
\end{array}
\end{equation}

\noindent
In eq. (\ref{eq:43}) we used the timelike Lorentz scalar product
$x^{\ 2} \ = \ ( \ x^{\ 0} \ )^{\ 2} \ - \ \vec{x}^{\ 2}$.

\noindent
The realization given in eqs. (\ref{eq:42}) and (\ref{eq:43})
is useful when $x^{\ \mu}$ is proportional
to a four-velocity, i.e.
$x^{\ 0} \ \ge \ \lambda \ > \ 0 \ , \ x^{\ 2} \ = \ \lambda^{\ 2}$ 
and y describes a spin direction, chosen in such a way, that $x y \ = \ 0$,
and $y^{\ 2} \ = \ - \ 1$.

\newpage


\noindent
\subsection{Note on the complex Lorentz group and associated operations}
\vspace*{0.1cm} 

\noindent
We recall the reality constrained covering of the Lorentz group
$spin \ ( \ 1 \ , \ 3 \ ; \ \Re \ )$ defined in eq. (\ref{eq:34})

\begin{equation}
\label{eq:44}
\begin{array}{l}
\left (
\ {\cal{A}} \ ( \ a \ )
\ , \ {\cal{B}} \ ( \ \overline{a} \ ) 
\ \right )
\ \equiv \ spin \ ( \ 1 \ , \ 3 \ ; \ \Re \ )
\ \simeq \ SL2C
\end{array}
\end{equation}

\noindent
I list the operations on amplitudes or fields, which
demand an extension of spin representations
to {\it covering of} the complex Lorentz group. 
This latter extension is denoted by $\stackrel{C}{\rightarrow}$ ,
defined in eq. (\ref{eq:45}) below
\vspace*{-0.3cm}

\begin{equation}
\label{eq:45}
\begin{array}{l}
\begin{array}{ccc cc}
spin \ ( \ 1 \ , \ 3 \ ; \ \Re \ )
& \stackrel{C}{\rightarrow} &
spin \ ( \ 1 \ , \ 3 \ ; \ C \ )
& \simeq & SL2C \ \times \ SL2C
\vspace*{0.3cm} \\
\left (
\ {\cal{A}} \ ( \ a \ )
\ , \ {\cal{B}} \ ( \ \overline{a} \ ) 
\ \right )
& \stackrel{C}{\rightarrow} &
\left (
\ {\cal{A}} \ ( \ a \ )
\ , \ {\cal{B}} \ ( \ b \ ) 
\ \right )
& ; &
a \ , \ b \ \mbox{unrestricted}
\end{array}
\end{array}
\end{equation}

\noindent
In the list below we number and specify the operation in the first
and second columns, the operand in the third, inducing the parallel
operation $\stackrel{C}{\rightarrow}$ 

\begin{equation}
\label{eq:46}
\begin{array}{l}
\begin{array}{|c|c|c|c|}
\hline 
 & \mbox{operation} & \mbox{operand} & \stackrel{C}{\rightarrow}
\\ \hline 
 1 & \mbox{crossing} 
& 
\mbox{scattering amplitudes} 
 & \surd \\
\hline
2 & \mbox{extension to complex momenta} & 
\mbox{scattering amplitudes}   
 & \surd \\
 \hline
3 & \mbox{extension to Euclidean space} & 
\mbox{local fields}   
 & \surd \\
\hline
\end{array}
\end{array}
\end{equation}

\noindent
Operations 1 - 3 in eq. (\ref{eq:46}) are {\it not} independent
of each other. A profound consequence is the
symmetry under the antiunitary CPT transformation
\cite{RJ} for local field theories.

\newpage


\noindent
\subsection{Field strengths, potentials and adjoint string operators}
\vspace*{0.1cm} 

\noindent
Potentials and field strengths have been introduced in eqs.
(\ref{eq:101}) and (\ref{eq:102}). We shall specify their
local gauge transformation properties below. For simplicity
we shall only discuss the octet or adjoint
representation of $SU3_{\ c}$ .

\noindent
The Lie algebra generators of the octet representation
$\left ( \ {\cal{F}}_{\ D} \ \right )_{\ A B} \ = \ i \ f_{\ A \ D \ B}$
in eq. (\ref{eq:102}) lead to the finite (local)
transformations

\begin{equation}
\label{eq:47}
\begin{array}{l}
\Omega_{ \ A \ B} \ ( \ x \ )
\ =
\ \left ( 
\ \exp
\ \frac{1}{i} \ \omega_{\ D} \ ( \ x \ ) \ {\cal{F}}_{\ D}
\ \right )_{\ A \ B}
\vspace*{0.3cm} \\
\left \lbrack
\ {\cal{F}}_{\ A} \ , \ {\cal{F}}_{\ B}
\ \right \rbrack \ = \ i \ f_{\ A \ B \ C} 
\ {\cal{F}}_{\ C}
\end{array}
\end{equation}

\noindent
The $SU3_{\ c}$ {\it angles} $\omega_{\ D} \ ( \ x \ )$
shall not be confused with the Euler half angles $\omega^{\ \mu \nu}$
in eq. (\ref{eq:21}), while the group analogy is obvious.
$\omega_{\ D} \ ( \ x \ )$ shall be chosen varying over
space time $x$ , restricted by differentiability requirements.

\noindent
Let $X \ ( \ x \ , \ A \ )$ be a classical field transforming
under the local octet transformations $\Omega$ 

\begin{equation}
\label{eq:48}
\begin{array}{l}
X^{\ \Omega} \ ( \ x \ , \ A \ )
\ = 
\ \Omega_{ \ A \ B} \ ( \ x \ )
\ X \ ( \ x \ , \ B \ )
\vspace*{0.3cm} \\
\mbox{in short :}
\hspace*{0.3cm}
X^{\ \Omega} \ ( \ x \ )
\ = 
\ \Omega \ ( \ x \ )
\ X \ ( \ x \ )
\ \rightarrow 
\ X^{\ \Omega} \ = \ \Omega \ X
\end{array}
\end{equation}

\noindent
The extension of the local {\it adjoint} transformations in 
eq. (\ref{eq:48}) to other representations of $SU3_{\ c}$
is straightforward. 
$\Omega$ are real, orthogonal $8 \ \times \ 8$ matrices with
determinant 1.

\noindent
Here we treat gauge potentials and field strengths as classical
fields ( test fields in the sence of distributions ). 
The potentials $ V_{\ \mu} \ ( \ x \ , \ D \ )$
are defined through the (octet) covariant derivatives acting on $X$

\begin{equation}
\label{eq:49}
\begin{array}{l}
\left ( \ D_{\ \mu} \ \right )_{\ A \ B}
\ = \ \partial_{\ \mu} \ \delta_{\ A \ B} \ + 
\ \left ( \ {\cal{W}}_{\ \mu} \ \right )_{\ A \ B}
\hspace*{0.3cm} ; \hspace*{0.3cm} \partial_{\ \mu} 
\ = \ \partial \ / \ \partial \ x^{\ \mu} 
\vspace*{0.3cm} \\
\left ( \ {\cal{W}}_{\ \mu} \ \right )_{\ A \ B}
\ = \ i \ V_{\ \mu} \ ( \ x \ , \ D \ )
\ \left ( \ {\cal{F}}_{\ D} \ \right )_{\ A \ B}
\ =
\ V_{\ \mu} \ ( \ x \ , \ D \ ) \ f_{\ D \ A \ B}
\vspace*{0.3cm} \\
\mbox{in short :}
\hspace*{0.3cm}
{\cal{W}}_{\ \mu} \ = \ i \ V_{\ \mu \ D}
\ {\cal{F}}_{\ D}
\hspace*{0.3cm} ; \hspace*{0.3cm}
D_{\ \mu} \ = \ \partial_{\ \mu} \ + \ {\cal{W}}_{\ \mu}
\end{array}
\end{equation}

\noindent
In eq. (\ref{eq:49}) the quantities
$ V_{\ \mu} \ ( \ x \ , \ D \ )$ ,
$\ \left ( \ {\cal{W}}_{\ \mu} \ \right )_{\ A \ B} \ ( \ x \ )$
are real.

\noindent
{\it Parallel transport}
\vspace*{0.1cm}

\noindent
We turn to the parallel transport operators, defined in eq. 
(\ref{eq:102}) repeated below

\begin{equation}
\label{eq:50}
\begin{array}{l}
U \ ( \ x \ , \ A \ ; \ y \ , \ B \ )
\ = 
P \ \exp 
\ \left ( 
\ \left . 
{\displaystyle{\int}}_{\ y}^{\ x} 
\ \right |_{\ {\cal{C}}}
\ d \ z^{\ \mu}
\ \frac{1}{i} \ V_{\ \mu} \ ( \ z \ , \ D \ ) \ {\cal{F}}_{\ D}
\ \right )_{\ A \ B}
\vspace*{0.3cm} \\
\hspace*{0.8cm}  = 
\ P \ \exp 
\ \left ( 
\ \left . 
\ -
\ {\displaystyle{\int}}_{\ y}^{\ x} 
\ \right |_{\ {\cal{C}}}
\ d \ z^{\ \mu}
\  {\cal{W}}_{\ \mu} \ ( \ z \ , \ D \ ) \ {\cal{F}}_{\ D}
\ \right )_{\ A \ B}
\vspace*{0.3cm} \\
\left ( \ {\cal{F}}_{\ D} \ \right )_{\ A B} \ = \ i \ f_{\ A \ D \ B}
\hspace*{0.3cm} ; \hspace*{0.3cm}
{\cal{W}}_{\ \mu} \ ( \ z \ , \ D \ ) \ =
\ i \ V_{\ \mu} \ ( \ z \ , \ D \ ) 
\vspace*{0.5cm} \\
{\cal{W}}_{\ \mu \ ; \ A \ B} \ ( \ z \ ) \ = 
\ {\cal{W}}_{\ \mu} \ ( \ z \ , \ D \ ) 
\ \left ( \ {\cal{F}}_{\ D} \ \right )_{\ A B} 
\vspace*{0.5cm} \\
\mbox{in short :}
\ U \ ( \ x \ ; \ y \ )
\ =
\ P \ \exp 
\ \left ( 
\ \left . 
\ -
\ {\displaystyle{\int}}_{\ y}^{\ x} 
\ \right |_{\ {\cal{C}}}
\ d \ z^{\ \mu}
\  {\cal{W}}_{\ \mu} 
\ \right )
\end{array}
\end{equation}

\noindent
For classical field configurations 
$\left . U \ ( \ x \ ; \ y \ ) \ \right |_{\ {\cal{C}}}$
is the operation of parallel transport of a tangent (octet) vector,
e.g. $X \ ( \ y \ ) \ \left \lbrace 
\ \rightarrow \ X \ ( \ y \ , \ B \ ) \ \right \rbrace$ ,
at the point $y$ along the curve ${\cal{C}}$ to $x$ .

\begin{equation}
\label{eq:51}
\begin{array}{l}
\begin{array}{ccc cc}
X \ ( \ y \ , \ B \ ) 
& \stackrel{{\cal{C}}}{\longrightarrow} &
 X_{\ \parallel} \ ( \ x \ , \ A \ )
& = &
\ U \ ( \ x \ , \ A \ ; \ y \ , \ B \ )
\ X \ ( \ y \ , \ B \ ) 
\vspace*{0.5cm} \\
X \ ( \ y \ ) 
& \stackrel{{\cal{C}}}{\longrightarrow} &
\ X_{\ \parallel} \ ( \ x \ )
& = &
\ U \ ( \ x \ ; \ y \ ) \ X \ ( \ y \ )
\vspace*{0.3cm} \\
y
& \stackrel{{\cal{C}}}{\longrightarrow} &
x & &
\end{array}
\vspace*{0.3cm} \\
U \ ( \ x \ ; \ y \ ) \ = 
\ \left . U \ ( \ x \ ; \ y \ ) \ \right |_{\ {\cal{C}}}
\end{array}
\end{equation}

\noindent
If $X \ ( \ x \ )$ is itself an octet field defined at all $x$ ,
then $\left . X_{\ \parallel} \ ( \ x \ ) 
\ \right |_{\ x \ \stackrel{\leftarrow}{\cal{C}} \ y}$
has to be distinguished from the given value $X \ ( \ x \ )$ .

\noindent
$\left . U \ ( \ x \ ; \ y \ ) \ \right |_{\ {\cal{C}}}$
defined in eqs. (\ref{eq:50}) and (\ref{eq:51})
follows from the parallel transport differential equation,
using a parameter representation of the
curve ${\cal{C}}$
\vspace*{-0.3cm} 

\begin{equation}
\label{eq:52}
\begin{array}{l}
{\cal{C}} \ : 
\ \left \lbrace
\ 1 \ \ge \ \tau \ \ge \ 0
\ \left |
\ \begin{array}{l}
 z \ = \ z \ ( \ \tau \ )
\vspace*{0.3cm} \\
 z \ ( \ 0 \ ) \ = \ y
\vspace*{0.3cm} \\
 z \ ( \ 1 \ ) \ = \ x
\end{array}
\right .
\ \right \rbrace
\vspace*{0.3cm} \\
v \ ( \ \tau \ ) \ = \dot{z} \ ( \ \tau \ )
\ = \ \left ( \ d \ / \ d \ \tau \ \right ) \ z \ ( \ \tau \ )
\end{array}
\end{equation}

\noindent
Lets follow the $\tau$ development of the family
of parallel transports from y along ${\cal{C}}$ to the point
$z \ ( \ \tau \ )$, as the latter moves from $y$ to $x$

\begin{equation}
\label{eq:53}
\begin{array}{l}
U \ ( \ \tau \ ) \ = 
\ \left .
\ U \ ( \ z \ ( \ \tau \ ) \ ; \ y \ ) \ \right |_{\ {\cal{C}}}
\vspace*{0.3cm} \\
{\cal{W}}_{\ \mu} \ ( \ \tau \ )
\ =
\ {\cal{W}}_{\ \mu} \ ( \ z \ ( \ \tau \ ) \ )
\ \rightarrow
\vspace*{0.3cm} \\
\left ( \ d \ / \ d \ \tau \ \right )
\ U \ ( \ \tau \ )  
\ =
\ - \ v^{\ \mu} \ ( \ \tau \ ) 
\ {\cal{W}}_{\ \mu} \ ( \ \tau \ )
\ U \ ( \ \tau \ )  
\vspace*{0.5cm} \\
U \ ( \ 0 \ ) \ = \ \P
\hspace*{0.3cm} \leftrightarrow \hspace*{0.3cm}
\ U \ ( \ y \ , \ A \ ; \ y \ , \ B \ ) \ = \ \delta_{\ A \ B}
\end{array}
\end{equation}

\noindent
The parallel transport equation (\ref{eq:53}) is subjected
to the initial conditions defined in its last line. 

\noindent
It can be integrated by successive iterations 

\begin{equation}
\label{eq:54}
\begin{array}{l}
U \ ( \ \tau \ ) \ = 
\ \sum_{\ n=0}^{\ \infty}
\ {\displaystyle{\int}}_{\ 0}^{\ \tau} \ d \tau_{\ 1}
\ {\displaystyle{\int}}_{\ 0}^{\ \tau_{\ 1}} \ d \tau_{\ 2}
\cdots
\ {\displaystyle{\int}}_{\ 0}^{\ \tau_{\ n-1}} \ d \tau_{\ n}
\ \times
\vspace*{0.3cm} \\
\hspace*{4.0cm} \times
\ w \ ( \ \tau_{\ 1} \ )
\ w \ ( \ \tau_{\ 2} \ )
\cdots
\ w \ ( \ \tau_{\ n} \ )
\vspace*{0.3cm} \\
w \ ( \ \tau \ ) \ =
\ w_{\ A \ B} \ ( \ \tau \ ) \ =
\ - \ v^{\ \mu} \ ( \ \tau \ ) 
\ {\cal{W}}_{\ \mu \ ; \ A \ B} \ ( \ z \ ( \ \tau \ ) \ ) 
\vspace*{0.3cm} \\
\tau 
\ \ge \ \tau_{\ 1} 
\ \ge \ \tau_{\ 2} 
\ \cdots
\hspace*{0.3cm} ; \hspace*{0.3cm}
U \ ( \ 1 \ ) \ = 
\ \left . U \ ( \ x \ ; \ y \ ) \ \right |_{\ {\cal{C}}}
\vspace*{0.5cm} \\
\hspace*{0.5cm} \rightarrow
\left . U \ ( \ x \ A \ ; \ y \ B \ ) \ \right |_{\ {\cal{C}}}
\ =
\vspace*{0.3cm} \\
\hspace*{1.5cm} =
\ P \ \exp 
\ \left ( 
\ \left . 
\ -
\ {\displaystyle{\int}}_{\ y}^{\ x} 
\ \right |_{\ {\cal{C}}}
\ d \ z^{\ \mu}
\  {\cal{W}}_{\ \mu} \ ( \ z \ , \ D \ ) \ {\cal{F}}_{\ D}
\ \right )_{\ A \ B}
\end{array}
\end{equation}

\noindent
The path ordering in eq. (\ref{eq:50}) reflects the path ordered
sequence $\tau \ \ge \ \tau_{\ 1} \ \ge \ \tau_{\ 2} \ \cdots$
in the multiple integrals in eq. (\ref{eq:54}) , thus established.
\vspace*{0.1cm}

\noindent
{\it Parallel transport and gauge transformations}
\vspace*{0.1cm}

\noindent
We go back to eqs. (\ref{eq:48}) and (\ref{eq:49}) ,
implying the action of a local gauge transformation on
the connection ${\cal{W}}_{\ \mu} \ ( \ x \ , \ D \ ) \ {\cal{F}}_{\ D}$

\begin{equation}
\label{eq:55}
\begin{array}{l}
D_{\ \mu} \ = \ \partial_{\ \mu} \ + \ {\cal{W}}_{\ \mu}
\hspace*{0.3cm} ; \hspace*{0.3cm}
D_{\ \mu}^{\ \Omega} \ = \ \partial_{\ \mu} \ + \ {\cal{W}}^{\ \Omega}_{\ \mu}
\vspace*{0.3cm} \\
X^{\ \Omega} \ ( \ x \ )
\ = \ \Omega \ ( \ x \ )
\ X \ ( \ x \ ) \ \rightarrow
\ D_{\ \mu}^{\ \Omega} \ X^{\ \Omega} \ = 
\Omega \ D_{\ \mu} \ X
\end{array}
\end{equation}

\noindent
The local gauge transformation $\Omega$ thus induces the transformation law for
the connection

\begin{equation}
\label{eq:56}
\begin{array}{l}
{\cal{W}}^{\ \Omega}_{\ \mu}
\ = \ \Omega \ \partial_{\ \mu} \ \Omega^{\ -1}
\ +
\ \Omega
\ {\cal{W}}_{\ \mu} \ \Omega^{\ -1}
\end{array}
\end{equation}

\noindent
The parallel transport of tangent vectors $X^{ \Omega} \ ( y )$
along the curve ${\cal{C}}$ with connection
${\cal{W}}^{\ \Omega}_{\ \mu}$ should be equivalent to the same
operation on tangent vectors $X \ ( \ y \ )$ with
${\cal{W}}$ modulo the transformation induced on the tangent vectors.
This implies using the relations in eq. (\ref{eq:51})

\begin{equation}
\label{eq:57}
\begin{array}{l}
X^{\ \Omega} \ ( \ x \ ) \ = \ \Omega \ ( \ x \ ) \ X \ ( \ x \ )
\hspace*{0.3cm} \leftrightarrow \hspace*{0.3cm}
X^{\ \Omega} \ ( \ y \ ) \ = \ \Omega \ ( \ y \ ) \ X \ ( \ y \ )
\ \rightarrow
\vspace*{0.3cm} \\
\begin{array}{ccc cc}
X^{\ \Omega} \ ( \ y \ ) 
& \stackrel{{\cal{C}}}{\longrightarrow} &
\ X_{\ \parallel}^{\ \Omega} \ ( \ x \ )
& = &
\ U^{\ \Omega} \ ( \ x \ ; \ y \ ) \ X^{\ \Omega} \ ( \ y \ )
\vspace*{0.3cm} \\
X \ ( \ y \ ) 
& \stackrel{{\cal{C}}}{\longrightarrow} &
\ X_{\ \parallel} \ ( \ x \ )
& = &
\ U \ ( \ x \ ; \ y \ ) \ X \ ( \ y \ )
\vspace*{0.3cm} \\
y
& \stackrel{{\cal{C}}}{\longrightarrow} &
x & &
\end{array}
\end{array}
\end{equation}

\noindent
Thus we expect the relations

\begin{equation}
\label{eq:58}
\begin{array}{l}
X_{\ \parallel}^{\ \Omega} \ ( \ x \ ) 
\ = \ \Omega \ ( \ x \ ) \ X_{\ \parallel} \ ( \ x \ ) 
\hspace*{0.3cm} \leftrightarrow \hspace*{0.3cm}
X^{\ \Omega} \ ( \ y \ ) 
\ = \ \Omega \ ( \ y \ ) \ X \ ( \ y \ ) 
\ \rightarrow
\vspace*{0.3cm} \\
\Omega \ ( \ x \ ) 
\ U \ ( \ x \ ; \ y \ ) 
\ X \ ( \ y \ )
\ = \ U^{\ \Omega}  \ ( \ x \ ; \ y \ ) 
\ \Omega \ ( \ y \ ) 
\ X \ ( \ y \ )
\ \forall \ X \ ( \ y \ )
\ \rightarrow
\vspace*{0.5cm} \\
\hspace*{2.0cm} 
U^{\ \Omega}  \ ( \ x \ ; \ y \ )
\ =
\ \Omega \ ( \ x \ ) 
\ U \ ( \ x \ ; \ y \ ) 
 \ \Omega^{\ -1} \ ( \ y \ )
\end{array}
\end{equation}

\noindent
We want to verify the relation inferred in eq. (\ref{eq:58}).
To this end we form the two , a priori different, matric
valued functions of $\tau$ along ${\cal{C}}$  

\begin{equation}
\label{eq:59}
\begin{array}{l}
U_{\ 1} \ ( \ \tau \ ) \ = \ U^{\ \Omega} \ ( \ \tau \ )
\hspace*{0.3cm} \leftrightarrow \hspace*{0.3cm}
U_{\ 2} \ ( \ \tau \ ) \ = \ \Omega \ ( \ z_{\ \tau} \ ) 
\ U \ ( \ \tau \ ) \ \Omega^{\ -1} \ ( \ y \ )
\vspace*{0.3cm} \\
z_{\ \tau} \ = \ z \ ( \ \tau \ )
\end{array}
\end{equation}

\noindent
From eq. (\ref{eq:53}) we infer 

\begin{equation}
\label{eq:60}
\begin{array}{l}
\partial_{\ \tau}
\ U_{\ 1} \ ( \ \tau \ )  
\ =
\ - \ v^{\ \mu} \ ( \ \tau \ ) 
\ {\cal{W}}^{\ \Omega}_{\ \mu} \ ( \ \tau \ )
\ U_{\ 1} \ ( \ \tau \ )  
\vspace*{0.3cm} \\
\partial_{\ \tau} 
\ U_{\ 2} \ ( \ \tau \ )  
\ =
 \left \lbrack
 \begin{array}{l}
\left (
\ \partial_{\ \tau} \ \Omega \ ( \ z_{\ \tau} \ ) 
\ \right ) 
\ \Omega^{\ -1} \ ( \ z_{\ \tau} \ ) \ -
\vspace*{0.3cm} \\
- \ v^{\ \mu} \ ( \ \tau \ ) \ \Omega \ ( \ z_{\ \tau} \ ) 
\ {\cal{W}}_{\ \mu} \ ( \ \tau \ )
\ \Omega^{\ -1} \ ( \ z_{\ \tau} \ ) 
\end{array}
 \right \rbrack
 U_{\ 2} \ ( \ \tau \ )
\end{array}
\end{equation}

\noindent
The expression in the first line of the bracket in eq. (\ref{eq:60})
transforms into

\begin{equation}
\label{eq:61}
\begin{array}{l}
\left (
\ \partial_{\ \tau} \ \Omega \ ( \ z_{\ \tau} \ ) 
\ \right ) 
\ \Omega^{\ -1} \ ( \ z_{\ \tau} \ ) \ =
\ -
\ v^{\ \mu} \ ( \ \tau \ )
\ \Omega \ ( \ z_{\ \tau} \ ) \ \partial_{\ z \ \mu}
\ \Omega^{\ -1} \ ( \ z_{\ \tau} \ ) 
\end{array}
\end{equation}

\noindent
Thus the differential equation for $U_{\ 2} \ ( \ \tau \ $
in eq. (\ref{eq:60}) takes the form

\begin{equation}
\label{eq:62}
\begin{array}{l}
\partial_{\ \tau} 
\ U_{\ 2} \ ( \ \tau \ )  
 =
\ - \ v^{\ \mu} \ ( \ \tau \ ) 
 \left \lbrack
 \begin{array}{c}
\ \Omega \ ( \ z_{\ \tau} \ ) 
\ \partial_{\ z \ \mu} 
\ \Omega^{\ -1} \ ( \ z_{\ \tau} \ ) 
\vspace*{0.5cm} \\
+ \ \Omega \ ( \ z_{\ \tau} \ ) 
\ {\cal{W}}_{\ \mu} \ ( \ \tau \ )
\ \Omega^{\ -1} \ ( \ z_{\ \tau} \ ) 
\end{array}
 \right \rbrack
 U_{\ 2} \ ( \ \tau \ )
\vspace*{0.3cm} \\
\hspace*{2.3cm}
 =
\ - \ v^{\ \mu} \ ( \ \tau \ ) 
\ \left \lbrack
\hspace*{2.2cm}
\ {\cal{W}}^{\ \Omega}_{\ \mu} \ ( \ \tau \ )
\hspace*{2.2cm}
\ \right \rbrack
\ U_{\ 2} \ ( \ \tau \ )
\vspace*{0.4cm} \\
\hline
\vspace*{-0.2cm} \\
 {\cal{W}}^{\ \Omega}_{\ \mu} \ ( \ z \ )
\ = \ \Omega \ ( \ z \ ) \ \partial_{\ z \ \mu} \ \Omega^{\ -1} \ ( \ z \ )
\ +
\ \Omega \ ( \ z \ )
\  {\cal{W}}_{\ \mu} \ ( \ z \ )
\ \Omega^{\ -1} \ ( \ z \ )
\end{array}
\end{equation}

\noindent
Comparing eqs. (\ref{eq:60}) and (\ref{eq:62}) we see
that $U_{\ 1}$ and $U_{\ 2}$ fulfill the same differential
equation, as a consequence of the gauge tranformation law
of the connection ${\cal{W}}$ . They also have the same
initial value 

\begin{equation}
\label{eq:63}
\begin{array}{l}
U_{\ 1} \ ( \ 0 \ ) \ = \ U_{\ 2} \ ( 0 \ ) \ = \ \P
\ \rightarrow
\ U_{\ 1} \ ( \ \tau \ ) \ = \ U_{\ 2} \ ( \tau \ ) 
\vspace*{0.3cm} \\
\hspace*{0.3cm} \rightarrow \hspace*{0.3cm}
U^{\ \Omega}  \ ( \ x \ ; \ y \ )
\ =
\ \Omega \ ( \ x \ ) 
\ U \ ( \ x \ ; \ y \ ) 
 \ \Omega^{\ -1} \ ( \ y \ )
\hspace*{0.3cm} \mbox{qed} 
\end{array}
\end{equation}

\newpage

\noindent
{\it On the nonabelian Stokes relation}
\vspace*{0.1cm} 

\noindent
For our purpose here, to describe the degrees of freedom of 
{\it binary} gluonic mesons, the set of parallel transport matrices
( matrix valued bilocal field operators ) as displayed in eq.
(\ref{eq:54})

\begin{equation}
\label{eq:64}
\begin{array}{l}
\left . U \ ( \ x \ A \ ; \ y \ B \ ) \ \right |_{\ {\cal{C}}}
\ =
\ \left (
\left . U \ ( \ x \ ; \ y \ ) \ \right |_{\ {\cal{C}}}
\ \right )_{\ A \ B}
\vspace*{0.3cm} \\
\hspace*{1.5cm} =
 P \ \exp 
\ \left ( 
\ \left . 
\ -
\ {\displaystyle{\int}}_{\ y}^{\ x} 
\ \right |_{\ {\cal{C}}}
\ d \ z^{\ \mu}
\  {\cal{W}}_{\ \mu} \ ( \ z \ , \ D \ ) \ {\cal{F}}_{\ D}
\ \right )_{\ A \ B}
\vspace*{0.5cm} \\
U^{\ \Omega}  \ ( \ x \ ; \ y \ )
\ =
\ \Omega \ ( \ x \ ) 
\ U \ ( \ x \ ; \ y \ ) 
 \ \Omega^{\ -1} \ ( \ y \ )
\end{array}
\end{equation}

\noindent
along {\it straight line} pathes ${\cal{C}}$ restricting
general ones, as defined in eq. (\ref{eq:52}), are sufficient.

\begin{equation}
\label{eq:65}
\begin{array}{l}
\stackrel{{\cal{C}}}{\longleftarrow} \ : 
\ \left \lbrace
\ 1 \ \ge \ \tau \ \ge \ 0
\ \left |
\ \begin{array}{l}
 z \ = \ z \ ( \ \tau \ ) \ = \ y \ + \ \tau \ ( \ x \ - \ y \ )
\vspace*{0.3cm} \\
 z \ ( \ 1 \ ) \ = \ x
\hspace*{0.3cm} \longleftarrow \hspace*{0.3cm}
 z \ ( \ 0 \ ) \ = \ y
\end{array}
\right .
\ \right \rbrace
\vspace*{0.3cm} \\
v \ ( \ \tau \ ) \ = \dot{z} \ ( \ \tau \ )
\ = \ z \ = \ x \ - \ y
\end{array}
\end{equation}

\noindent
Parallel transport beeing generated by the connection
1-form

\begin{equation}
\label{eq:66}
\begin{array}{l}
\left . \mbox{{\Large (}}
\ {\cal{W}}^{\ (1)} 
\ \equiv
d \ z^{\ \mu}
\ {\cal{W}}_{\ \mu} \ ( \ z \ , \ D \ ) \ {\cal{F}}_{\ D}
\ \right . \mbox{{\Large )}}_{\ A \ B}
\ \rightarrow
\vspace*{0.3cm} \\
U \ ( \ x \ ; \ y \ ) 
\ =
 P \ \exp 
\ \left ( 
\ \left . 
\ -
\ {\displaystyle{\int}}_{\ y}^{\ x} 
\ \right |_{\ {\cal{C}}}
\  {\cal{W}}^{\ (1)}
\ \right )
\hspace*{0.3cm} ; \hspace*{0.3cm}
P \ \equiv \ P^{\ (1)}
\end{array}
\end{equation}

\noindent
the matrix valued 1-forms naturally acquire the line ordering,
appropriate for one dimensional integrals.

\noindent
{\it Yet} connection 1-forms and their path $P^{\ (1)}$ ordered integrals 
do not exhaust the range of
r-forms and their r dimensional $P^{\ (r)}$ ordered integrals,
associated with nonabelian degrees of freedom.

\noindent
Next in line are the curvature 2-form and its {\it surface} 
$P^{\ (2)}$ ordered integral.

\noindent
We follow the covariant derivative path with the octet field
$X \ ( \ x \ )$ introduced in eqs. (\ref{eq:47}) - (\ref{eq:49})

\begin{equation}
\label{eq:67}
\begin{array}{l}
D_{\ \mu} \ X \ ( \ x \ ) \ =
\ \left (
\ \partial_{\ \mu} \ +  
\ {\cal{W}}_{\ \mu} 
\ \right )
\ X \ ( \ x \ )
\vspace*{0.3cm} \\
\left ( \ D_{\ \mu} \ D_{\ \nu}
\ - \ D_{\ \nu} \ D_{\ \mu} 
\ \right )
\ X \ ( \ x \ )
\ =
\ {\cal{W}}_{\ \left \lbrack \ \mu \nu \ \right \rbrack}
\ X \ ( \ x \ )
\end{array}
\end{equation}

\noindent
In eq. (\ref{eq:67}) 
${\cal{W}}_{\ \left \lbrack \ \mu \nu \ \right \rbrack}$ 
denotes the (antisymmetric Yang-Mills) curvature tensor, 
i.e. the field strengths

\begin{equation}
\label{eq:68}
\begin{array}{l}
{\cal{W}}_{\ \left \lbrack \ \mu \nu \ \right \rbrack}
\ =
\ \partial_{\ \mu} \ {\cal{W}}_{\ \nu}
\ -
\ \partial_{\ \nu} \ {\cal{W}}_{\ \mu}
\ +
\ \left \lbrack
\ {\cal{W}}_{\ \mu} \ , \ {\cal{W}}_{\ \nu}
\ \right \rbrack
\vspace*{0.3cm} \\
{\cal{W}}_{\ \left \lbrack \ \mu \nu \ \right \rbrack} \ ( \ x \ )
\ =
\ {\cal{W}}_{\ \left \lbrack \ \mu \nu \ \right \rbrack}
\ ( \ x \ , \ D \ ) \ {\cal{F}}_{\ D} \ =
\vspace*{0.3cm} \\
\hspace*{0.8cm} = 
\ \left \lbrack
\begin{array}{l}
\ \left . \mbox{{\Large (}}
\ \partial_{\ \mu} \ {\cal{W}}_{\ \nu} \ ( \ x \ , \ D \ )
\ - \ \partial_{\ \nu} \ {\cal{W}}_{\ \mu} \ ( \ x \ , \ D \ )
\ \mbox{{\Large )}} \right .
\ {\cal{F}}_{\ D}
\vspace*{0.3cm} \\
\ + 
\ {\cal{W}}_{\ \mu} \ ( \ x \ , \ A \ ) 
\ {\cal{W}}_{\ \nu} \ ( \ x \ , \ B \ ) 
\ \left \lbrack
\ {\cal{F}}_{\ A} \ , \ {\cal{F}}_{\ B}
\ \right \rbrack
\end{array}
\right \rbrack
\vspace*{0.3cm} \\
{\cal{W}}_{\ \mu} \ ( \ z \ , \ D \ ) \ =
\ i \ V_{\ \mu} \ ( \ z \ , \ D \ ) 
\hspace*{0.3cm} ; \hspace*{0.3cm}
\left \lbrack
\ {\cal{F}}_{\ A} \ , \ {\cal{F}}_{\ B}
\ \right \rbrack \ = \ i \ f_{\ A \ B \ C} 
\ {\cal{F}}_{\ C}
\end{array}
\end{equation}

\noindent
In eq. (\ref{eq:68}) we have included the relations in eqs. (\ref{eq:47})
and (\ref{eq:50}).

\noindent
The form of the curvature tensor 
${\cal{W}}_{\ \left \lbrack \ \mu \nu \ \right \rbrack}$
in eq. (\ref{eq:68}) becomes

\begin{equation}
\label{eq:69}
\begin{array}{l}
{\cal{W}}_{\ \left \lbrack \ \mu \nu \ \right \rbrack}
\ ( \ x \ , \ D \ ) \ =
\ \frac{1}{i} \ F_{\ \left \lbrack \ \mu \nu \ \right \rbrack}
\ ( \ x \ , \ D \ ) \ =
\vspace*{0.3cm} \\
\hspace*{0.8cm} = 
\ \left \lbrack
\begin{array}{l}
\ \partial_{\ \mu} \ {\cal{W}}_{\ \nu} \ ( \ x \ , \ D \ )
\ - \ \partial_{\ \nu} \ {\cal{W}}_{\ \mu} \ ( \ x \ , \ D \ )
\vspace*{0.3cm} \\
\ + \ i
\ {\cal{W}}_{\ \mu} \ ( \ x \ , \ A \ ) 
\ {\cal{W}}_{\ \nu} \ ( \ x \ , \ B \ ) 
\ f_{\ A \ B \ D}
\end{array}
\right \rbrack
\vspace*{0.5cm} \\
F_{\ \left \lbrack \ \mu \nu \ \right \rbrack}
\ ( \ x \ , \ D \ ) \ =
\vspace*{0.3cm} \\
\hspace*{0.8cm} = 
\ \left \lbrack
\begin{array}{l}
\ \partial_{\ \nu} \ V_{\ \mu} \ ( \ x \ , \ D \ )
\ - \ \partial_{\ \mu} \ V_{\ \nu} \ ( \ x \ , \ D \ )
\vspace*{0.3cm} \\
\ - 
\ V_{\ \nu} \ ( \ x \ , \ A \ ) 
\ V_{\ \mu} \ ( \ x \ , \ B \ ) 
\ f_{\ A \ B \ D}
\end{array}
\right \rbrack
\end{array}
\end{equation}

\noindent
We recast the quantities 
${\cal{W}}_{\ \left \lbrack \ \mu \nu \ \right \rbrack}$
in eqs. (\ref{eq:68}) and (\ref{eq:69}) into their
Lie algebra valued form

\begin{equation}
\label{eq:70}
\begin{array}{l}
\begin{array}{lll}
{\cal{W}}_{\ \left \lbrack \ \mu \nu \ \right \rbrack
\  \left \lbrack \ A B \ \right \rbrack} \ ( \ x \ )
& = &
\ \left (
\ {\cal{W}}_{\ \left \lbrack \ \mu \nu \ \right \rbrack}
\ ( \ x \ , \ D \ ) \ {\cal{F}}_{\ D}
\ \right )_{\ A \ B}
\vspace*{0.3cm} \\
 & = &
\ \left (
\ F_{\ \left \lbrack \ \mu \nu \ \right \rbrack}
\ ( \ x \ , \ D \ ) \ L_{\ D}
\ \right )_{\ A \ B}
\end{array}
\vspace*{0.3cm} \\
\left ( \ L_{\ D} \ \right )_{\ A \ B}
\ = \ \frac{1}{i}
\ \left ( \ {\cal{F}}_{\ D} \ \right )_{\ A \ B}
\ = \ f_{\ A \ D \ B}
\hspace*{0.3cm} ; \hspace*{0.3cm}
\left \lbrack
\ L_{\ R} \ , \ L_{\ S}
\ \right \rbrack 
\ = \ f_{\ R \ S \ T} \ L_{\ T}
\end{array}
\end{equation}

\noindent
We also cast eq. (\ref{eq:66}) into the $L_{\ D}$ form

\begin{equation}
\label{eq:71}
\begin{array}{l}
\left . \mbox{{\Large (}}
\ {\cal{W}}^{\ (1)} 
\ \equiv
d \ z^{\ \mu}
\ {\cal{W}}_{\ \mu} \ ( \ z \ , \ D \ ) \ {\cal{F}}_{\ D}
\ \right . \mbox{{\Large )}}_{\ A \ B}
\ \rightarrow
\vspace*{0.3cm} \\
{\cal{W}}_{\ \mu} \ ( \ z \ , \ D \ ) \ {\cal{F}}_{\ D}
\ = \ i \ {\cal{W}}_{\ \mu} \ ( \ z \ , \ D \ ) \ L_{\ D}
\ = \ - \ V_{\ \mu} \ ( \ z \ , \ D \ ) \ L_{\ D}
\ \rightarrow
\vspace*{0.3cm} \\
{\cal{W}}_{\ \mu \ \left \lbrack \ A B \ \right \rbrack}
\ ( \ x \ ) \ = \ - \  V_{\ \mu} \ ( \ x \ , \ D \ )
\ \left (
\ L_{\ D}
\ \right )_{\ A \ B}
\end{array}
\end{equation}

\noindent
Comparing the connection and curvature representations in 
eqs. (\ref{eq:70}) and (\ref{eq:71}) we learn that
the local quantities 
${\cal{W}}_{\ \mu \ \left \lbrack \ A B \ \right \rbrack} \ ( \ x \ )$
and 
${\cal{W}}_{\ \left \lbrack \ \mu \nu \ \right \rbrack
\  \left \lbrack \ A B \ \right \rbrack} \ ( \ x \ )$,
as well as the components
$- \ V_{\ \mu} \ ( \ x \ , \ D \ )$ and
$F_{\ \left \lbrack \ \mu \nu \ \right \rbrack}
\ ( \ x \ , \ D \ )$ are real. This is usus in the mathematical
literature.

\noindent
Local gauge transformations as defined for the connection
in eqs. (\ref{eq:56}) and (\ref{eq:62}) are naturally
extended to the curvature 

\begin{equation}
\label{eq:72}
\begin{array}{l}
\left \lbrace
\begin{array}{l}
{\cal{W}}_{\ \mu \ \left \lbrack \ A B \ \right \rbrack} 
\vspace*{0.3cm} \\
{\cal{W}}_{\ \left \lbrack \ \mu \nu \ \right \rbrack
\  \left \lbrack \ A B \ \right \rbrack} 
\end{array}
\ \right \rbrace
\ ( \ x \ )
\ \rightarrow 
\ \left \lbrace
\begin{array}{l}
{\cal{W}}_{\ \mu } 
\vspace*{0.3cm} \\
{\cal{W}}_{\ \left \lbrack \ \mu \nu \ \right \rbrack} 
\end{array}
\ \right \rbrace
\ ( \ x \ )
\vspace*{0.3cm} \\
\hline
\vspace*{-0.2cm} \\
{\cal{W}}_{\ \mu }^{\ \Omega} \ ( \ x \ )
\ = \ \Omega \ ( \ x \ ) \ \partial_{\ \mu} \ \Omega^{\ -1} \ ( \ x \ )
\ +
\ \Omega \ ( \ x \ )
\  {\cal{W}}_{\ \mu} \ ( \ x \ )
\ \Omega^{\ -1} \ ( \ x \ )
\vspace*{0.3cm} \\
{\cal{W}}_{\ \left \lbrack \ \mu \nu \ \right \rbrack}^{\ \Omega}
\ ( \ x \ )
\ =
\ \Omega \ ( \ x \ )
\ {\cal{W}}_{\ \left \lbrack \ \mu \nu \ \right \rbrack} \ ( \ x \ )
\ \Omega^{\ -1} \ ( \ x \ )
\end{array}
\end{equation}

\noindent
{\it Lie cohomology and de Rham cohomology}
\vspace*{0.1cm} 

\noindent
With connection and curvature we associate
the Lie algebra valued one and two forms, as defined in eqs.
(\ref{eq:66}) - (\ref{eq:71})

\begin{equation}
\label{eq:73}
\begin{array}{l}
\left (
\ \begin{array}{l}
 {\cal{W}}^{\ (1)} 
\ \equiv
\ d \ x^{\ \mu} \ {\cal{W}}_{\ \mu}
\vspace*{0.3cm} \\
 {\cal{W}}^{\ (2)} 
\ \equiv
\ \frac{1}{2} \ d \ x^{\ \mu} \ \wedge \ d \ x^{\ \nu}
\ {\cal{W}}_{\ \left \lbrack \ \mu \nu \ \right \rbrack} 
\end{array}
\ \right ) 
\ ( \ x \ , \ \left \lbrack \ A B \ \right \rbrack \ )
\vspace*{0.3cm} \\
{\cal{W}}^{\ (2)} \ = \ d \ {\cal{W}}^{\ (1)}
\ + \ {\cal{W}}^{\ (1)} \ \circ \ {\cal{W}}^{\ (1)}
\ \equiv \ D \ {\cal{W}}^{\ (1)}
\end{array}
\end{equation}

\noindent
In eq. (\ref{eq:73}) the symbol $\circ$ denotes normal 
matrix multiplication {\bf to be distinguished} from
the Lie product denoted below by $\odot$.

\noindent
It is the antisymmetric nature of the wedge product 
$d \ x^{\ \mu} \ \wedge \ d \ x^{\ \nu}$ which renders
the $\circ$ product equivalent to a Lie algebra
product $\odot$

\begin{equation}
\label{eq:74}
\begin{array}{l}
{\cal{W}}^{\ (1)} \ \circ \ {\cal{W}}^{\ (1)}
\ = \ \frac{1}{2} 
\ {\cal{W}}^{\ (1)} \ \odot \ {\cal{W}}^{\ (1)}
\end{array}
\end{equation}

\noindent
We shall verify eq. (\ref{eq:74}) by components

\begin{equation}
\label{eq:75}
\begin{array}{l}
{\cal{W}}^{\ (1)} \ \circ \ {\cal{W}}^{\ (1)}
\ ( \ x \ , \ \left \lbrack \ A B \ \right \rbrack \ )
\ = \ d \ x^{\ \mu} \ \wedge \ d \ x^{\ \nu}
\ {\cal{W}}_{\ \mu \ \left \lbrack \ A A' \ \right \rbrack}
\ {\cal{W}}_{\ \nu \ \left \lbrack \ A' B \ \right \rbrack}
\ \rightarrow
\vspace*{0.3cm} \\
\hspace*{0.8cm} =
\ \frac{1}{2} \ d \ x^{\ \mu} \ \wedge \ d \ x^{\ \nu}
\ \left \lbrack
\begin{array}{l}
 {\cal{W}}_{\ \mu \ \left \lbrack \ A A' \ \right \rbrack}
\ {\cal{W}}_{\ \nu \ \left \lbrack \ A' B \ \right \rbrack}
\ -
\vspace*{0.3cm} \\
 - \ {\cal{W}}_{\ \nu \ \left \lbrack \ A \ A' \ \right \rbrack}
\ {\cal{W}}_{\ \mu \ \left \lbrack \ A' B \ \right \rbrack}
\end{array}
 \right \rbrack
\vspace*{0.3cm} \\
\hspace*{0.8cm} =
\ \frac{1}{2} \ d \ x^{\ \mu} \ \wedge \ d \ x^{\ \nu}
\ \left \lbrack
\begin{array}{l}
 {\cal{W}}_{\ \mu} \ \odot
\ {\cal{W}}_{\ \nu}
\end{array}
 \right \rbrack_{\ \left \lbrack \ A \ B \ \right \rbrack}
\vspace*{0.3cm} \\
\hline
\vspace*{-0.2cm} \\
 {\cal{W}}_{\ \mu} \ \odot
\ {\cal{W}}_{\ \nu}
\ = 
\ \left \lbrack
\ {\cal{W}}_{\ \mu} \ , \ {\cal{W}}_{\ \nu}
\ \right \rbrack
\ \equiv 
\ {\cal{W}}_{\ \mu} \ \circ \ {\cal{W}}_{\ \nu}
\ - \ {\cal{W}}_{\ \nu} \ \circ \ {\cal{W}}_{\ \mu}
\end{array}
\end{equation}

\noindent
Eq. (\ref{eq:73}) yields the first relation in the
{\it adaptive} Lie cohomology chain, generated by the the sequence
of operations $D \ \rightarrow \ D^{\ '} \ \neq \ D$

\begin{equation}
\label{eq:76}
\begin{array}{l}
{\cal{W}}^{\ (2)} \ = 
\ D \ {\cal{W}}^{\ (1)}
\hspace*{0.3cm} \rightarrow \hspace*{0.3cm}
{\cal{W}}^{\ (3)} \ = \ D^{\ '} \ {\cal{W}}^{\ (2)} \ = \ 0
\vspace*{0.3cm} \\
\hline
\vspace*{-0.2cm} \\
\begin{array}{lll}
 D & : &
{\cal{W}}^{\ (2)} \ = \ d \ {\cal{W}}^{\ (1)}
\ + \ \frac{1}{2}  \ {\cal{W}}^{\ (1)} \ \odot \ {\cal{W}}^{\ (1)}
\vspace*{0.3cm} \\
 D^{\ '} & : &
{\cal{W}}^{\ (3)} \ =
\ d \ {\cal{W}}^{\ (2)} 
\ + \ {\cal{W}}^{\ (1)} \ \odot \ {\cal{W}}^{\ (2)}
\ = \ 0
\end{array}
\end{array}
\end{equation}

\noindent
The termination of the {\it adaptive} $D \ \rightarrow \ D^{\ '}$
sequence follows from the  
antisymmetry of the wedge product {\it and} the Jacobi
identity of cyclic double commutators

\begin{equation}
\label{eq:77}
\begin{array}{l}
{\cal{W}}^{\ (3)} \ = 
\ \left \lbrace
\begin{array}{l}
 d \ 
\ \left (
\ ( \ {\cal{W}}^{\ (1)} \ \circ \ )^{\ 2}
\ \right )
\ + \ {\cal{W}}^{\ (1)} \ \odot
\ d \ {\cal{W}}^{\ (1)}
\vspace*{0.3cm} \\
 + 
\ {\cal{W}}^{\ (1)} \ \odot \ 
\ ( \ {\cal{W}}^{\ (1)} \ \circ \ )^{\ 2}
\end{array}
\ \right \rbrace
\vspace*{0.3cm} \\
( \ {\cal{W}}^{\ (1)} \ \circ \ )^{\ n}
\ = \ {\cal{W}}^{\ (1)} \ \circ 
\ \left (
\ ( \ {\cal{W}}^{\ (1)} \ \circ \ )^{\ n - 1} 
\ \right )
\ , \ \cdots
\end{array}
\end{equation}

\noindent
Expressing the $\odot$ product in eq. (\ref{eq:77}) in
$\circ$ products it follows

\begin{equation}
\label{eq:78}
\begin{array}{l}
{\cal{W}}^{\ (3)} \ = 
\ \left \lbrace
\begin{array}{l}
 d \ 
\ \left (
\ ( \ {\cal{W}}^{\ (1)} \ \circ \ )^{\ 2}
\ \right )
\vspace*{0.3cm} \\
 + \ {\cal{W}}^{\ (1)} \ \circ 
\ \left ( \ d \ {\cal{W}}^{\ (1)} \ \right )
\ - \ \left ( \  d \ {\cal{W}}^{\ (1)} \ \right )
\ \circ
\ {\cal{W}}^{\ (1)} 
\vspace*{0.3cm} \\
 + 
\ {\cal{W}}^{\ (1)} \ \circ  
\ \left ( \ ( \ {\cal{W}}^{\ (1)} \ \circ \ )^{\ 2} \ \right )
\ -
\ \left ( \ ( \ {\cal{W}}^{\ (1)} \ \circ \ )^{\ 2} \ \right )
\ \circ 
\ {\cal{W}}^{\ (1)}   
\end{array}
\ \right \rbrace
\end{array}
\end{equation}

\noindent
The contribution cubic in ${\cal{W}}^{\ (1)}$ vanishes on the
ground of the associative product $\circ$,
while the first three cancel due to the identity

\begin{equation}
\label{eq:79}
\begin{array}{l}
 d \ 
\ \left (
\ ( \ {\cal{W}}^{\ (1)} \ \circ \ )^{\ 2}
\ \right )
\ = \ \left ( \ d \ {\cal{W}}^{\ (1)} \ \right ) 
\ \circ \  {\cal{W}}^{\ (1)}
\ -
\ {\cal{W}}^{\ (1)}
\ \circ 
\ \left ( \ d \ {\cal{W}}^{\ (1)} \ \right )
\end{array}
\end{equation}

\noindent
{\it Loops of parallel transports, local holonomy groups}
\vspace*{0.1cm} 

\noindent
In the inverse of the differential Lie cohomology chain
the parallel transport matrices

\begin{equation}
\label{eq:80}
\begin{array}{l}
\ U \ ( \ x \ , \ y \ ; \ \stackrel{{\cal{C}}}{\longrightarrow} \ )
\ =
\ P \ \exp 
\ \left ( 
\ \left . 
\ -
\ {\displaystyle{\int}}_{\ y}^{\ x} 
\ \right |_{\ {\cal{C}}}
\ d \ z^{\ \mu}
\  {\cal{W}}_{\ \mu} 
\ \right )
\end{array}
\end{equation}

\noindent
defined in eqs. (\ref{eq:50}) - (\ref{eq:51}) can be combined to form
a closed curve starting and ending at $y$.

\begin{equation}
\label{eq:81}
\begin{array}{l}
U \ \left (
\ y \ , \ y \ ; \ {\cal{CL}}
\hspace*{0.5cm} \begin{array}{c}
\longleftarrow 
\vspace*{-0.42cm} \\
\hspace*{-0.36cm} \swarrow \hspace*{0.5cm} \nearrow
\vspace*{-0.43cm} \\
\hspace*{-0.6cm} \longrightarrow 
\end{array}
\right )_{\ A \ B}
\ \rightarrow 
\ U \ ( \ y \ , \ y \ ; \ {\cal{CL}} \ )
\end{array}
\end{equation}

\noindent
The quantities
$U \ ( \ x \ , \ y \ ; \ \stackrel{{\cal{C}}}{\longrightarrow} \ )$ ,
called adjoint strings here, are rarely
used in lattice discretized Yang-Mills theory.  
The associated fundamental strings, projected on the
fundamental representation of the local gauge group ( the triplet
strings for $SU3_{\ c}$ ) are the dynamical
{\it link} variables therein \cite{latt}.

\noindent
The quantities $U \ ( \ y \ , \ y \ ; \ {\cal{CL}} \ )$ , defined
in eq. (\ref{eq:81}) we shall call closed adjoint (octet) strings.
Their counterparts, projected on the fundamental (triplet)
representation, assigned to a minimal closed lattice loop,
a plaquette, are used to generate the lattice action.

\noindent
Closed loop matrices or operators are widely studied in their own
right. For the fundamental representation they are called Wilson loops
( $W \ ( \ {\cal{C}} \ )$ ) \\
within Yang-Mills theories \cite{alvarez}.

\noindent
We continue to focus on open and closed adjoint strings here.
\marginpar[\hspace*{-0.4cm} universal bundle]
Nevertheless it is tacitly assumed, that the configurations obey the regularity
requirements of extensions of these strings to {\it all}
representations of the gauge group. This framework is called
the universal bundle in the mathematical literature.

\newpage

\noindent
The gauge transformation properties of open and closed (adjoint) strings
in eqs. (\ref{eq:80}) and (\ref{eq:81}) follow from eqs. (\ref{eq:62})
and (\ref{eq:63})

\begin{equation}
\label{eq:82}
\begin{array}{l}
U^{\ \Omega} \ ( \ x \ , \ y \ ; \ \stackrel{{\cal{C}}}{\longrightarrow} \ )
\ =
\ \Omega \ ( \ x \ ) 
\ U \ ( \ x \ , \ y \ ; \ \stackrel{{\cal{C}}}{\longrightarrow} \ )
 \ \Omega^{\ -1} \ ( \ y \ )
\ \rightarrow
\vspace*{0.3cm} \\
U^{\ \Omega} \ ( \ y \ , \ y \ ; \ {\cal{CL}} \ )
\ = 
\ \Omega \ ( \ y \ )
\ U \ ( \ y \ , \ y \ ; \ {\cal{CL}} \ )
\ \Omega^{\ -1} \ ( \ y \ )
\end{array}
\end{equation}

\noindent
The closed curve $ {\cal{CL}}$ is still punctuated at its beginning
and ending. Yet the gauge transformation act {\it locally}
at this point. However the simply connected closed loop
can be repeatedly transcurred, leading to the multiple 
positve as well as negative powers, all transforming the same way under
gauge transformations

\begin{equation}
\label{eq:83}
\begin{array}{l}
U^{\ n} \ ( \ y \ , \ y \ ; \ {\cal{CL}} \ )
\ =
\ U \ ( \ y \ , \ y \ ; \ {\cal{CL}}^{\ (n)} \ )
\hspace*{0.3cm} , \hspace*{0.3cm} \ n \ = \ 0 \ , \ \pm 1 \ , \ \cdots
\vspace*{0.3cm} \\
U^{\ \Omega} \ ( \ y \ , \ y \ ; \ {\cal{CL}}^{\ (n)} \ )
\ = 
\ \Omega \ ( \ y \ )
\ U \ ( \ y \ , \ y \ ; \ {\cal{CL}}^{\ (n)} \ )
\ \Omega^{\ -1} \ ( \ y \ )
\end{array}
\end{equation}

\noindent
The closed curve ${\cal{CL}}^{\ (n)}$ shall represent the n-fold transcurred
simple curev ${\cal{CL}}$, whereby negative powers mean to
reverse the orientation, from clockwise to anticlockwise say.

\noindent
Gauge invariant quantities are thus all (adjoint) traces

\begin{equation}
\label{eq:84}
\begin{array}{l}
W_{\ (n)} \ ( \ {\cal{CL}} \ ) \ = \ tr  
 \left \lbrack 
\ U^{\ n} \ ( \ y \ , \ y \ ; \ {\cal{CL}} \ )
 \right \rbrack
\ =
\ \sum_{ \left \lbrace \lambda \ \right \rbrace} \ \lambda^{\ n} 
\ \left (
\ U \ ( \ y \ , \ y \ ; \ {\cal{CL}} \ )
\ \right )
\end{array}
\end{equation}

\noindent
In eq. (\ref{eq:84}) $\lambda$ runs over all the eigenvalues
of the (real orthogonal) matrix  $U \ ( \ y \ , \ y \ ; \ {\cal{CL}} \ )$ .

\noindent
The quantities $W_{\ (n)} \ ( \ {\cal{CL}} \ )$ in eq. (\ref{eq:84})
do depend on the shape of the simply laced curve $ {\cal{CL}}$ ,
but they are the same for all points along $ {\cal{CL}}$,
when adopted as alternative starting and ending points.

\noindent
They represent the adjoint characters of the (self covering)
Lie group, dependent only on the angles of the Cartan subalgebra.
Thus they depend, for a simple gauge group 
with rank r ( r = 2 for $SU3_{\ c}$ )
on the r Cartan subalgebra angles, characterising
any of the representatives
$U \ ( \ y^{\ '} \ , \ y^{\ '} \ ; \ {\cal{CL}} \ )$ with
$y^{\ '}$ anywhere on the curve ${\cal{CL}}$.
The characteristic coefficients
are determined from the roots of the Lie algebra
and, through its universal extension to all representations,
from its r fundamental weights. 
For $SU3_{\ c}$ these are the weights of the $3$ and
$\overline{3}$ fundamental representations.

\noindent
For $SU3_{\ c}$ let the two Cartan algebra angles be
$\phi \ \leftrightarrow \ I_{\ 3}$ and 
$\psi \ \leftrightarrow \ Y \ / \ ( \ 2 \sqrt{3} \ )$, using standard
weight normalization, where $I_{\ 3}$ and Y 
denote isospin and hypercharge respectively.

\noindent
The $3$ and $\overline{3}$ Cartan matrices shall be  
$u$ and $\overline{u}$ respectively,
$u \ = \ u \ ( \ \phi \ , \ \psi \ )$ . The two fundamental characters
thus become

\begin{equation}
\label{eq:85}
\begin{array}{l}
\chi \ = \ \chi \ ( \ u \ ) \ \rightarrow \ \chi \ ( \ \phi \ , \ \psi \ )
\hspace*{0.3cm} \mbox{and} \hspace*{0.3cm} 
\overline{\chi}
\vspace*{0.3cm} \\
\chi \ = \ \sum_{\ k = 1}^{\ 3} 
\ \exp \ \left ( \ \frac{1}{i} \ \kappa_{\ k} \ \right )
\hspace*{0.3cm} ; \hspace*{0.3cm} 
\kappa_{\ 3} \ = \ - \ \kappa_{\ 1} \ - \ \kappa_{\ 2}
\vspace*{0.3cm} \\
\kappa_{\ 1} \ = \ \frac{1}{2} \ \phi \ + \ \frac{1}{2 \ \sqrt{3}} \ \psi
\hspace*{0.3cm} , \hspace*{0.3cm} 
\kappa_{\ 2} \ = \ - \ \frac{1}{2} \ \phi \ + \ \frac{1}{2 \ \sqrt{3}} \ \psi
\end{array}
\end{equation}

\noindent
Then for a reducible direct product representation
$D_{\ red \ ; \ M \ , \ N}$ of M copies of $u$ 
and N copies of $\overline{u}$, the character is multiplicative

\begin{equation}
\label{eq:86}
\begin{array}{l}
\chi_{\ red \ ; \ M \ , \ N} \ = \ \chi^{\ M} \ \overline{\chi}^{\ N} 
\ =
\ \overline{\chi}_{\ red \ ; \ N \ , \ M} 
\end{array}
\end{equation}

\noindent
From $D_{\ red \ ; \ M \ , \ N}$ the characters of all irreducible
representations of the gauge group can be derived.
We only give the lowest charcters for the i\\
$3$ , $\overline{3}$ ,
$6$ , $\overline{6}$ , $10$ , $\overline{10}$
and adjoint ( $8$ )  representations
of $SU3_{\ c} \ \rightarrow \ SU3_{\ .}$ , with the association

\begin{equation}
\label{eq:87}
\begin{array}{l}
\begin{array}{lll clll}
3 
& = &
 D_{\ ird \ ; \ 1 \ , \ 0} 
 & cc &
\overline{3} 
& = &
D_{\ ird \ ; \ 0 \ , \ 1} 
\vspace*{0.3cm} \\
6 
& = &
 D_{\ ird \ ; \ 2 \ , \ 0} 
 & cc &
\overline{6} 
& = &
D_{\ ird \ ; \ 0 \ , \ 2} 
\vspace*{0.3cm} \\
10 
& = &
 D_{\ ird \ ; \ 3 \ , \ 0} 
 & cc &
\overline{10} 
& = &
D_{\ ird \ ; \ 0 \ , \ 3} 
\vspace*{0.3cm} \\
8 
& = &
 D_{\ ird \ ; \ 1 \ , \ 1} 
 &  \mbox{real} &
&  &
\end{array}
\vspace*{0.0cm} \\
\begin{array}{lll clll}
\chi^{\ ird}_{\ 1 \ , \ 0} 
& = & \chi
 & cc &
\chi^{\ ird}_{\ 0 \ , \ 1} 
& = & 
\overline{\chi}
\vspace*{0.3cm} \\
\chi^{\ ird}_{\ 2 \ , \ 0} 
& = & 
\chi^{\ 2} \ - \ \overline{\chi}
 & cc &
\chi^{\ ird}_{\ 0 \ , \ 2} 
& = & 
\overline{\chi}^{\ 2} \ - \ \chi
\vspace*{0.3cm} \\
\chi^{\ ird}_{\ 3 \ , \ 0} 
& = & 
\left (
\begin{array}{c}
\chi^{\ 3} 
\ - \ 2 \ | \ \chi \ |^{\ 2} 
\vspace*{0.3cm} \\
 + \ 1
\end{array}
\right )
 & cc &
\chi^{\ ird}_{\ 0 \ , \ 3} 
& = & 
\left (
\begin{array}{c}
\overline{\chi}^{\ 3} 
\ - \ 2 \ | \ \chi \ |^{\ 2} 
\vspace*{0.3cm} \\
 + \ 1
\end{array}
\right )
\vspace*{0.3cm} \\
\chi^{\ ird}_{\ 1 \ , \ 1} 
& = & 
| \ \chi \ |^{\ 2} \ - \ 1
 &  \mbox{real} &
&  &
\end{array}
\vspace*{0.5cm} \\
\left (
\ \chi \ , \ \overline{\chi}
\ \right )
\ =
\ \left (
\ \chi \ , \ \overline{\chi}
\ \right ) \ ( \ \phi \ , \ \psi \ )
\end{array}
\end{equation}


\noindent
In our case the quantities 
$U \ ( \ y \ , \ y \ ; \ {\cal{CL}} \ )$ in eq. (\ref{eq:81})
are in the adjoint representation, i.e. in
$D_{\ ird \ ; \ 1 \ , \ 1}$ .
Hence all equivalent representatives 
$U^{\ n} \ ( \ y^{\ '} \ , \ y^{\ '} \ ; \ {\cal{CL}} \ )$ 
with $y^{\ '}$ anywhere on ${\cal{CL}}$ are characterized
by the two Cartan subalgebra angles
\vspace*{-0.3cm} 

\begin{equation}
\label{eq:88}
\begin{array}{l}
U^{\ n} \ ( \ y^{\ '} \ , \ y^{\ '} \ ; \ {\cal{CL}} \ ) \ \rightarrow
\ \left ( \ n \ \phi \ , \ n \ \psi \ \right ) 
\ \left \lbrace 
\ {\cal{CL}}
\ \right \rbrace
\end{array}
\end{equation}

\noindent
The invariant quantities $W_{\ (n)} \ ( \ {\cal{CL}} \ )$   
in eq. (\ref{eq:84}) are thus given by

\begin{equation}
\label{eq:89}
\begin{array}{l}
W_{\ (n)} \ ( \ {\cal{CL}} \ )   
\ = \ | \ \chi_{\ n} 
\ \left \lbrace
\ {\cal{CL}}
\ \right \rbrace
\ |^{\ 2} \ - \ 1
\vspace*{0.3cm} \\
\chi_{\ n} 
\ \left \lbrace 
\ {\cal{CL}}
\ \right \rbrace
\ = \ \chi 
\ \left ( \ n \ \phi \ , \ n \ \psi \ \right ) 
\ \left \lbrace 
\ {\cal{CL}}
\ \right \rbrace
\vspace*{0.3cm} \\
n \ = \ 0 \ , \ \pm \ 1 \ \cdots
\hspace*{0.3cm} ; \hspace*{0.3cm} 
W_{\ (0)} \ ( \ {\cal{CL}} \ ) \ = \ 8  
\end{array}
\end{equation}

\noindent
Because the fundamental $SU3_{\ .}$ matrices $u$ ( and $\overline{u}$ )
are three dimensional, with determinant 1 , only two out of the
infinite $n$ sequence of {\it fundamental} characters 
$\chi_{\ n} \ \left \lbrace\ {\cal{CL}} \ \right \rbrace$ : 
( $n \ = \ 1 \ , \ 2 \ \mod{3}$ ) 
are independent of each other.

\noindent
The fundamental characters of any element u of the
3 representation of $SU3_{\ .}$ obey the {\it elementary} generating
identity, expressing the fundamental polynomial
$P_{\ 3} \ ( \ \mu \ ; \ u \ ) \ = \ Det \ ( \ \P - \ \mu \ u \ )$ 
in terms of fundamental characters 
\vspace*{-0.8cm} 

\begin{equation}
\label{eq:90}
\begin{array}{l}
P_{\ 3} \ ( \ \mu \ ; \ u \ ) \ = \ Det \  ( \ \P - \ \mu \ u \ ) 
\ = \ 1 \ - \ p_{\ 1} \ \mu \ + \ p_{\ 2} \ \mu^{\ 2} \ - \ \mu^{\ 3}
\vspace*{0.3cm} \\
E \ ( \ \mu \ ; \ u \ ) 
\ = \ \exp 
\ \left (
\ - \ \sum_{\ n = 1}^{\ \infty} \ \mu^{\ n} \ \chi_{\ n} \ / \ n
\ \right )
\vspace*{0.3cm} \\
E \ ( \ \mu \ ; \ u \ ) 
\ - \ P_{\ 3} \ ( \ \mu \ ; \ u \ ) \ = \ 0
\hspace*{1.0cm} 
\forall \ \mu
\vspace*{0.3cm} \\
\hline
\vspace*{-0.4cm} \\
p_{\ k} \ = \ p_{\ k} \ ( \ u \ )
\ ; \ k \ = \ 1 \ , \ 2
\hspace*{0.3cm} \leftrightarrow \hspace*{0.3cm} 
\chi_{\ n} \ = \ tr \ u^{\ n} \ = \ \chi_{\ n} \ ( \ u \ )
\vspace*{0.3cm} \\
\Pi_{\ k} \ = \ p_{\ k} \ ( \ u \ = \ \P \ ) 
\vspace*{0.5cm} \\
p_{\ 1} \ = \ \chi_{\ 1}
\hspace*{0.3cm} , \hspace*{0.3cm} 
p_{\ 2} \ = \ \frac{1}{2} 
\ ( \ \chi_{\ 1}^{\ 2} \ - \ \chi_{\ 2} \ )
\vspace*{0.3cm} \\
\chi_{\ - n} \ = \overline{\chi}_{\ n} \ , 
\ \chi_{\ 0} \ = \ 3
\hspace*{0.3cm} ; \hspace*{0.3cm} 
\Xi_{\ n} \ = \ \chi_{\ n} \ ( \ u \ = \ \P \ ) \ = \ 3
\hspace*{0.3cm} \forall \ n
\vspace*{0.3cm} \\
\Xi_{\ n} \ = \ \Xi
\hspace*{0.3cm} ; \hspace*{0.3cm} 
\Pi_{\ k} \ = \ \Xi
\end{array}
\end{equation}

\noindent
The first of the reducing identities is

\begin{equation}
\label{eq:91}
\begin{array}{l}
P_{\ 3} \ ( \ \mu \ = \ u \ ; \ u \ ) \ = \ 0
\ \rightarrow
\ \chi_{\ 3}
\ = 
\ \chi_{\ 0} \ - \ p_{\ 1} \ \chi_{\ 1}
\ + \ p_{\ 2} \ \chi_{\ 2} 
\hspace*{0.3cm} ; \hspace*{0.3cm} 
\ \cdots
\vspace*{0.3cm} \\
\rightarrow
\ \chi_{\ 3}
\ = \ \chi_{\ 0} \ - \ \chi_{\ 1}^{\ 2}
\ + \ \frac{1}{2}
\ ( \ \chi_{\ 1}^{\ 2} \ \chi_{\ 2} \ - \ \chi_{\ 2}^{\ 2} \ )
\end{array}
\end{equation}

\noindent
Substituting 
$\chi^{\ ird}_{\ M \ , \ N} \ ( \ \chi \ , \ \overline{\chi} \ )$
the values for $u \ = \ \P$ we obtain the dimension of the associated
irreducible representation ( eq. (\ref{eq:87}) )

\begin{equation}
\label{eq:92}
\begin{array}{l}
dim \ ( \ D_{\ ird \ ; \ M \ , \ N} \ )
\ =
\ \chi^{\ ird}_{\ M \ , \ N} \ ( \ \Xi \ , \ \Xi \ )
\hspace*{0.3cm} ; \hspace*{0.3cm} 
\Xi \ = \ 3
\end{array}
\end{equation}

\noindent
All this notwithstanding, the adjoint matrices
$U \ ( \ y \ , \ y \ ; \ {\cal{CL}} \ )$ represent the
holonomy group at the point $y$ mapping through ${\cal{CL}}$
parallel transport all adjoint tangent space into itself.

\noindent
The entire range of adjoint matrices forms the group
$SU3_{\ .} \ / \ Z_{\ 3}$. The mapping

\begin{equation}
\label{eq:93}
\begin{array}{l}
\begin{array}[t]{c}
u 
\vspace*{0.3cm} \\
\in \ D_{\ ird \ ; \ 1 , \ 0}
\end{array}
\ \longrightarrow 
\begin{array}[t]{c}
 U
\vspace*{0.3cm} \\
\in \ D_{\ ird \ ; \ 1 , \ 1}
\end{array}
\vspace*{0.3cm} \\
( \ u \ z_{\ 0} \ , \ u \ z_{\ 1} \ , \ u \ z_{\ 2} \ )
\ \rightarrow \ U
\end{array}
\end{equation}

\noindent
is covering the adjoint representation $D_{\ ird \ ; \ 1 , \ 1}$
three times. In eq. (\ref{eq:93}) $z_{\ s} \ , \ s \ = \ 0,1,2$
denote the elements forming the center $Z_{\ 3}$ of $SU3_{\ .}$ .

\noindent
But there is no loss of information in considering
only $U \ ( \ y \ , \ y \ ; \ {\cal{CL}} \ )$ , assuming
the analytic extension of the underlying Lie group to be implementable
in the classical field configurations, which resolves the above threefold 
covering through the analytic extension inherent to the Lie algebra
leading from 
$SU3_{\ .} \ / \ Z_{\ 3} \ \rightarrow \ SL3C \ \rightarrow \ SU3_{\ .}$ .
This is in accordance with the universal fibre bundle structure.

\noindent
At the end of this appendix we shall go back to
$U \ ( \ y \ , \ y \ ; \ {\cal{CL}} \ )$
as defined in eq. (\ref{eq:81}) and state the nonabeliean Stokes
relation \cite{Nachtmann} :

\begin{equation}
\label{eq:94}
\begin{array}{l}
U \ \left (
\ y \ , \ y \ ; \ {\cal{CL}}
\hspace*{0.5cm} \begin{array}{c}
\longleftarrow 
\vspace*{-0.42cm} \\
\hspace*{-0.36cm} \swarrow \hspace*{0.5cm} \nearrow
\vspace*{-0.43cm} \\
\hspace*{-0.6cm} \longrightarrow 
\end{array}
\right )_{\ A \ B}
\ =
\vspace*{0.3cm} \\
\hspace*{0.3cm} =
\ \left (
\ U \ ( \ y \ , \ x \ ; \ \stackrel{{\cal{C}}}{\longrightarrow} \ )
\ \right )_{\ A \ G}
\ \times
\vspace*{0.3cm} \\
\hspace*{1.7cm} \times
\ \left (
\ P_{\ 2}^{\ \Omega} \ \exp  
\ \left ( 
\ -
\ {\displaystyle{\int}}_{\ S_{\ x}} \ {\cal{W}}^{\ (2)} 
\ \right )
\ \right )_{\ G \ H}
\ \times
\vspace*{0.3cm} \\
\hspace*{5.2cm} \times
\ \left (
\ U \ ( \ x \ , \ y \ ; \ \stackrel{{\cal{C}}}{\longrightarrow} \ )
\ \right )_{\ H \ B}
\vspace*{0.3cm} \\
\ \left (
\ U_{\ (2)} \ ( \ x \ ; \ P_{\ 2}^{\ \Omega} 
\ | \ {\cal{CL}} \ = \ \partial \ S \ )
\ \right )_{\ G \ H}
\hspace*{0.4cm} \rightarrow 
\ U_{\ (2)}^{\ \Omega} \ ( \ x \ ; \  {\cal{CL}} \ = \ \partial \ S \ ) 
\vspace*{0.3cm} \\
\hspace*{2.5cm} =
\ \left (
\ P_{\ 2}^{\ \Omega} \ \exp  
\ \left ( 
\ -
\ {\displaystyle{\int}}_{\ S_{\ x}} \ {\cal{W}}^{\ (2)} 
\ \right )
\ \right )_{\ G \ H}
\end{array}
\end{equation}

\noindent
In eq. (\ref{eq:94})
$U_{\ (2)} \ ( \ x \ ; \ P_{\ 2}^{\ \Omega} 
\ | \ {\cal{CL}} \ = \ \partial \ S \ )
\ \rightarrow \ U_{\ (2)}^{\ \Omega} 
\ ( \ x \ ; \  {\cal{CL}} \ = \ \partial \ S \ ) $
denotes the Stokes surface integral proper, punctuated at an internal point
$x$ and oriented in a coil like wiring fashion, denoted
by $P_{\ 2}^{\ \Omega}$ . 

\noindent
The $P_{\ 2}^{\ \Omega}$ ordering for four coils and two wiring layers
is shown in figure \ref{fig3} below.

\begin{figure}[htb]
\vskip 0.5cm 
\begin{center}
\vskip -0.5cm 
\mbox{\epsfig{file=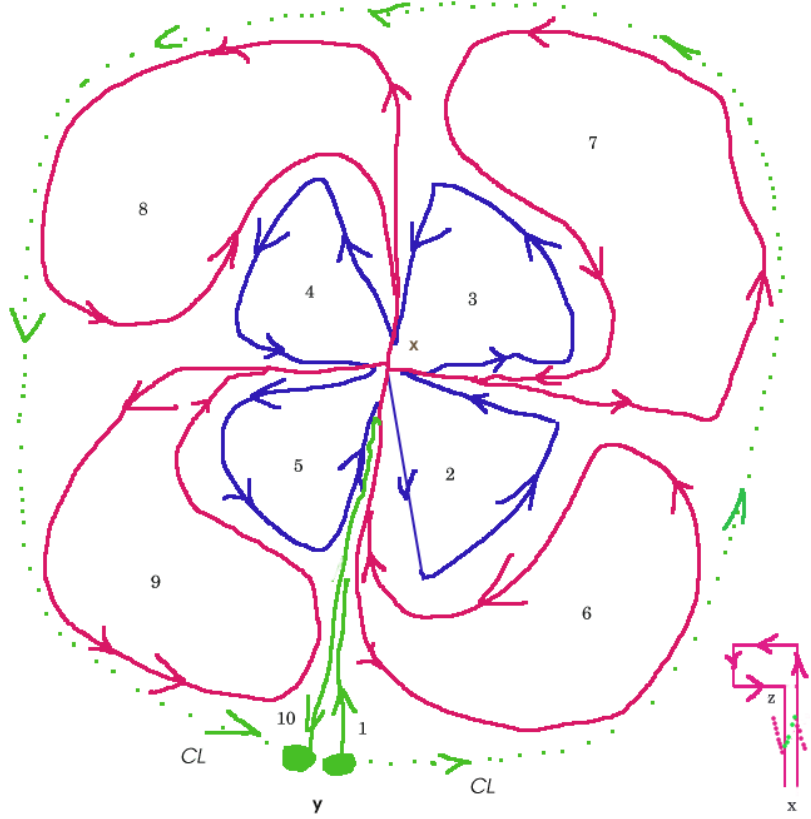,%
width=12cm}}
\end{center}
\vskip -1.5cm 
\caption{ The surface coil-wiring ordering surface integration.
Both number of coils and number of wirings, here 4 and 2,
are to be increased, refining the surface covering. }
\label{fig3}
\end{figure}

\noindent
The ordering $P_{\ 2}^{\ \Omega}$  for the segmnents
at fixed distance from the
base point $x$ converges to a flagpole path, shown in the
lower right corner of figure \ref{fig3} .

\noindent
This path , denoted $\natural$ ,
starts and ends at the base point $x$ and turns around
the plaquette at the point $z$ on the surface ${\cal{S}}$ . 

\noindent
Its contribution inside
the ordering  $P_{\ 2}^{\ \Omega}$ is

\begin{equation}
\label{eq:95}
\begin{array}{l}
\natural \ =
\ U ( \ x \ , \ z \ ; \ \stackrel{{\cal{C}}}{\longrightarrow} \ )
\ P_{\ 2}^{\ \Omega \ | \ z} \ \exp  
\ \left ( 
\ -
\ {\displaystyle{\int}}_{\ S_{\ x}} \ {\cal{W}}^{\ (2)} 
\ \right )
\ U \ ( \ z \ , \ x \ ; \ \stackrel{{\cal{C}}}{\longrightarrow} \ )
\end{array}
\end{equation}

\noindent
The superscript $\Omega$ characterizing the surface ordering
$P_{\ 2}^{\ \Omega}$ is chosen to associate a local gauge transformation
with the surface ${\cal{S}}$ . This follows from
the similarity transformation induced on the
flagpole path $\natural$ as defined in eq. (\ref{eq:95}).

\noindent
To make this explicit we rename the parallel
transport matrix 
$U \ ( \ z \ , \ x \ ; \ \stackrel{{\cal{C}}}{\longrightarrow} \ )$
associated with the fixed base point $x$ and the point $z$
varying over the entire surface ${\cal{S}}$

\begin{equation}
\label{eq:96}
\begin{array}{l}
U \ ( \ z \ , \ x \ ; \ \stackrel{{\cal{C}}}{\longrightarrow} \ )
\ \longrightarrow \ \omega_{\ x} \ ( \ z \ )
\ \rightarrow
\vspace*{0.3cm} \\
\natural \ =
\ \left ( 
\ \omega_{\ x} \ ( \ z \ )
\ \right )^{\ -1}
\ P_{\ 2}^{\ \Omega \ | \ z} \ \exp  
\ \left ( 
\ -
\ {\displaystyle{\int}}_{\ S_{\ x}} \ {\cal{W}}^{\ (2)} 
\ \right )
\ \omega_{\ x} \ ( \ z \ )
\vspace*{0.5cm} \\
\omega_{\ x} \ ( \ z \ )
\ = 
\left .
\omega_{\ x} \ ( \ z \ )
\ \right |_{\ {\cal{C}}}
\end{array}
\end{equation}

\noindent
The last line in eq. (\ref{eq:96}) shall make it explicit,
that the parallel transport matrices $\omega_{\ x} \ ( \ z \ )$
are not local functions of the surface point z.
Rather they depend on the path, one each from $x$ to $z$ .

\noindent
The {\it family} of similarity transformations 
$\left \lbrace \ \omega_{\ x} \ ( \ z \ ) \ \right \rbrace$
induced on the {\it local} field strength differential
${\cal{W}}^{\ (2)} \ = \ {\cal{W}}^{\ (2)} \ ( \ z \ )$ reflects
the nested structure of the weaving pattern defining
$P_{\ 2}^{\ \Omega}$ as a whole \footnote{Nachtmann \cite{Nachtmann}
compares the repetitious return of the weaving pattern
to the base point $x$ with the way
a spider weaves its {\it fan-type anchoring part} of the net.} .

\noindent
Looking at the structure of the similarity transformations
forming the {\it nonlocal} structure $\natural$ in eq. (\ref{eq:96})
the question arises, whether there exists a local gauge transformation
$\widehat{\Omega}$ - {\it adapted} to ${\cal{S}}$ -  which would render the
gauge transformed set 
$\left \lbrace \ \omega_{\ x}^{\ \widehat{\Omega}} 
\ ( \ z \ ) \ \right \rbrace$ trivial

\begin{equation}
\label{eq:97}
\begin{array}{l}
\left .
\omega_{\ x}^{\ \widehat{\Omega}} \ = 
\ \widehat{\Omega} \ ( \ z \ )
\ \omega_{\ x} 
\ \left (
\ \widehat{\Omega} \ ( \ x \ )
\ \right )^{\ -1}
\ \right |_{\ {\cal{C}}}
\ = \ \P
\hspace*{0.3cm} , \hspace*{0.3cm} \forall
\ z \ \in \ {\cal{S}}
\vspace*{0.3cm} \\
\widehat{\Omega} \ \rightarrow  
\ \mbox{Riemann normal gauge}
\end{array}
\end{equation}

\noindent
Indeed the gauge transformation with the requirements in eq.
(\ref{eq:97}) exists and can be found {\it together}
with a coordinate transformation of local coordinates
on ${\cal{S}}$ and the original contour ${\cal{CL}}$
such that ${\cal{S}}$ becomes the inner part of a bounding
circle. The latter forms in the new coordinates the closed contour
${\cal{CL}}$ and the family of curves from the base point
$x$ to $z$ becomes the family of straight {\it radial} lines.
The point $y$ punctuating the contour ${\cal{CL}}$ then can be mapped
on the south pole of the bounding circle ( to be definite ) .

\noindent
The transformed variables are well known in the analogous situation, 
where gauge transformations refer to coordinate transformations, 
i.e. the tangent space (universal) spin bundle.
The respective coordinates are called Riemann normal coordinates.
The gauge equivalent we shall call the Riemann normal gauge
as indicated in eq. (\ref{eq:97}) .

\noindent
The Riemann normal gauge is also known as radial gauge, at least
in the case of an abelian gauge group.

\noindent
It is precisely in the Riemann normal gauge
where the 
$P_{\ 2}^{\ \Omega} \ \rightarrow
\ P_{\ 2}^{\ \widehat{\Omega}}$ ordering becomes 'normal' .
Transforming to the Riemann normal gauge $R.n.g. \ ( \ {\cal{S}} \ )$
we have

\begin{equation}
\label{eq:98}
\begin{array}{l}
R.n.g. \ ( \ {\cal{S}} \ ) \ :
\vspace*{0.3cm} \\
\begin{array}{ccl}
\left .
\omega_{\ x} \ ( \ z \ )
\ \right |_{\ {\cal{C}}} 
& \rightarrow &
\left .
\ \omega_{\ x}^{\ \widehat{\Omega}} \ ( \ z \ ) 
\ \right |_{\ {\cal{C}}} 
\ \equiv \ \P
\vspace*{0.3cm} \\
P_{\ 2}^{\ \Omega} 
& \rightarrow &
 P_{\ 2}^{\ \widehat{\Omega}}
\vspace*{0.3cm} \\
{\cal{W}}^{\ (2)} \ ( \ z \ )
& \rightarrow &
{\cal{W}}^{\ (2) \ \widehat{\Omega}} \ ( \ z \ )
\ = 
\ \widehat{\Omega} \ ( \ z \ )
\ {\cal{W}}^{\ (2)} \ ( \ z \ )
\ \left (
\ \widehat{\Omega} \ ( \ z \ )
\ \right )^{\ -1}
\end{array}
\end{array}
\end{equation}

\noindent
Using the transformad quantities on the right hand side of eq. 
(\ref{eq:98}) we undo first the flagpole sequence $\natural$
in eq. (\ref{eq:96})

\begin{equation}
\label{eq:99}
\begin{array}{l}
\natural \ =
\ \left ( 
\ \omega_{\ x} \ ( \ z \ )
\ \right )^{\ -1}
\ P_{\ 2}^{\ \Omega \ | \ z} \ \exp  
\ \left ( 
\ -
\ {\displaystyle{\int}}_{\ S_{\ x}} \ {\cal{W}}^{\ (2)} 
\ \right )
\ \omega_{\ x} \ ( \ z \ )
\ \rightarrow \ \natural^{\ \widehat{\Omega}}
\vspace*{0.3cm} \\
\natural^{\ \widehat{\Omega}}
\ = 
\ P_{\ 2}^{\ \widehat{\Omega} \ | \ z} \ \exp  
\ \left ( 
\ -
\ {\displaystyle{\int}}_{\ S_{\ x}} \ {\cal{W}}^{\ (2) \ \widehat{\Omega}} 
\ \right )
\end{array}
\end{equation}

\noindent
Next eq. (\ref{eq:94}) becomes

\begin{equation}
\label{eq:100}
\begin{array}{l}
U^{\ \widehat{\Omega}} \ \left (
\ y \ , \ y \ ; \ {\cal{CL}}
\hspace*{0.5cm} \begin{array}{c}
\longleftarrow 
\vspace*{-0.42cm} \\
\hspace*{-0.36cm} \swarrow \hspace*{0.5cm} \nearrow
\vspace*{-0.43cm} \\
\hspace*{-0.6cm} \longrightarrow 
\end{array}
\right )_{\ A \ B}
\ =
\vspace*{0.3cm} \\
\hspace*{0.3cm} =
\ \left (
\ P_{\ 2}^{\ \widehat{\Omega}} \ \exp  
\ \left ( 
\ -
\ {\displaystyle{\int}}_{\ S_{\ x}} \ {\cal{W}}^{\ (2) \ \widehat{\Omega}} 
\ \right )
\ \right )_{\ A \ B}
\vspace*{0.3cm} \\
\hline
\vspace*{-0.2cm} \\
\ \left (
\ U_{\ (2)} \ ( \ x \ ; \ P_{\ 2}^{\ \widehat{\Omega}} 
\ | \ {\cal{CL}} \ = \ \partial \ S \ )
\ \right )_{\ A \ B}
\hspace*{0.4cm} \rightarrow 
\ U_{\ (2)}^{\ \widehat{\Omega}} 
\ ( \ x \ ; \  {\cal{CL}} \ = \ \partial \ S \ ) 
\vspace*{0.3cm} \\
\hspace*{2.5cm} =
\ \left (
\ P_{\ 2}^{\ \widehat{\Omega}} \ \exp  
\ \left ( 
\ -
\ {\displaystyle{\int}}_{\ S_{\ x}} \ {\cal{W}}^{\ (2) \ \widehat{\Omega}} 
\ \right )
\ \right )_{\ A \ B}
\end{array}
\end{equation}

\begin{figure}[htb]
\vskip 0.0cm 
\begin{center}
\vskip -0.2cm 
\mbox{\epsfig{file=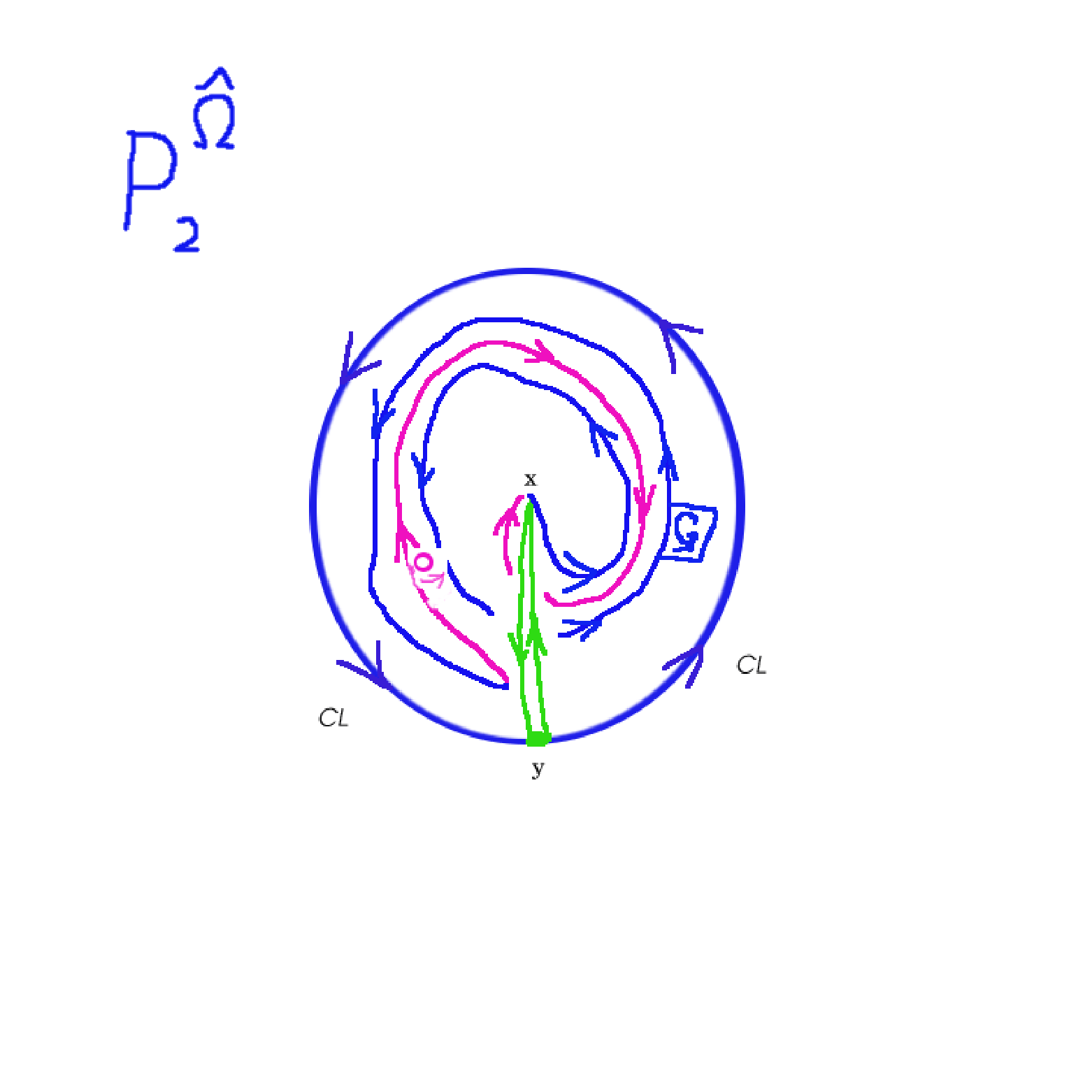,%
width=12cm}}
\end{center}
\vskip -3.6cm 
\caption{The abridged ordering of plaquettes in the Riemann normal gauge. 
It follows a double spiral pattern from the base point $x$ and back. }
\label{fig4}
\end{figure}

\noindent
At this stage, although implicit in the original definition
of the general $P_{\ 2}^{\ \Omega}$ ordering, it remains to indicate
the (or a) simplified ordering in the Riemann normal gauge.
This is shown in figure \ref{fig4} .

\vspace*{-0.0cm}

\noindent
In the Riemann normal gauge the repeated intermediate returns to the base point
$x$ are no more necessary \footnote{This also is the spiders path,
in the second stage : the scaffolding spiral
\cite{encyc} , \cite{spiderfot} .} .

\begin{figure}[htb]
\vskip 0.0cm 
\begin{center}
\vskip -0.2cm 
\mbox{\epsfig{file=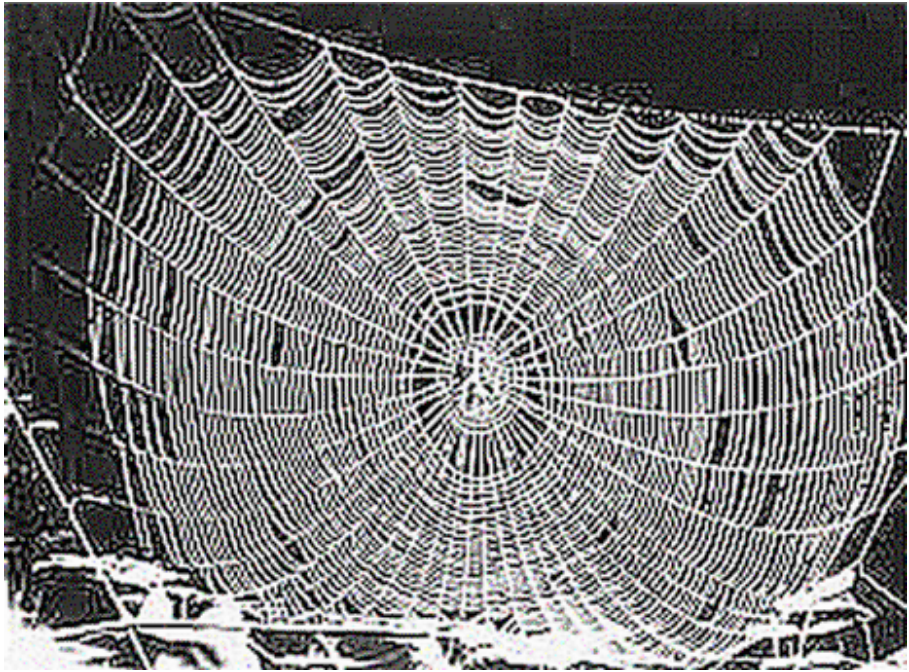,%
width=12cm}}
\end{center}
\caption{The abridged  ordering in an actual spider web.
It follows an abridged double spiral pattern from the base point $x$ and back.
The figure is adapted from a photograph \cite{spiderfot} . }
\label{fig4a}
\end{figure}

\noindent
An interesting shortcut is shown in an actual spiderweb in figure \ref{fig4a}.

\vspace*{0.0cm}

\noindent
Several remarks conclude this discussion :

\begin{description}
\item i) back to the original gauge

In eq. (\ref{eq:100}) we have to transform 
back from Riemann normal gauge on the surface ${\cal{S}}_{\ x \ | \ y}$
to the original gauge
\vspace*{-0.2cm} 

\begin{equation}
\label{eq:1001}
\begin{array}{l}
U \ \left (
\ y \ , \ y \ ; \ {\cal{CL}}
\hspace*{0.5cm} \begin{array}{c}
\longleftarrow 
\vspace*{-0.42cm} \\
\hspace*{-0.36cm} \swarrow \hspace*{0.5cm} \nearrow
\vspace*{-0.43cm} \\
\hspace*{-0.6cm} \longrightarrow 
\end{array}
\right )_{\ A \ B}
\ =
\vspace*{0.3cm} \\
\hspace*{0.3cm} =
\ \left (
\ \left (
\ \widehat{\Omega} \ ( \ y \ )
\ \right )^{\ -1}
\ U^{\ \widehat{\Omega}} \ \left (
\ y \ , \ y \ ; \ {\cal{CL}}
\hspace*{0.5cm} \begin{array}{c}
\longleftarrow 
\vspace*{-0.42cm} \\
\hspace*{-0.36cm} \swarrow \hspace*{0.5cm} \nearrow
\vspace*{-0.43cm} \\
\hspace*{-0.6cm} \longrightarrow 
\end{array}
\ \right )
\ \widehat{\Omega} \ ( \ y \ )
\ \right )_{\ A \ B}
\vspace*{0.1cm} 
\end{array}
\end{equation}

\item ii) cut the edges of the contour ${\cal{CL}}$ 

In order to transform the map of the contour ${\cal{CL}}$ continuously
into a circle, the edges marked in the corresponding symbol in 
eqs. (\ref{eq:81}), (\ref{eq:94}), (\ref{eq:100}) and (\ref{eq:1001})
need to be cut
\vspace*{-0.2cm} 

\begin{equation}
\label{eq:1002}
\begin{array}{l}
\begin{array}{c}
\longleftarrow 
\vspace*{-0.42cm} \\
\hspace*{-0.36cm} \swarrow \hspace*{0.5cm} \nearrow
\vspace*{-0.43cm} \\
\hspace*{-0.6cm} \longrightarrow 
\end{array}
\ \rightarrow 
\ \hspace*{0.05cm} 
\bigcirc \vspace*{0.1cm} \hspace*{-0.14cm} {\scriptstyle{\wedge}}
\end{array}
\end{equation}

\item iii) the full collection of surfaces and Riemann normal gauges

As indicated in point i) the meaning of Stokes relations 
summarized in eqs. (\ref{eq:100}) and (\ref{eq:1001})
is to consider {\it all} surfaces with boundary ${\cal{CL}} \ ( \ y \ )$,
the latter punctuated at the point $y$, the former
with base  point $x$, inheriting the point $y$, {\it and} the
associated Riemann normal gauges $\widehat{\Omega} \ ( \ x \ )$ .

The collection of surfaces and associated Riemann normal
gauges shall be denoted 
\vspace*{-0.2cm} 

\begin{equation}
\label{eq:1003}
\begin{array}{l}
\left \lbrace
\ {\cal{S}}_{\ x \ | \ y} \ ; \ \widehat{\Omega} \ ( \ x \ )
\ \right \rbrace
\end{array}
\end{equation}

\item iv) the Stokes relations proper

Stokes relations in eq. (\ref{eq:100}) in Riemann normal gauges
take the form
\vspace*{-0.3cm} 

\begin{equation}
\label{eq:1004}
\begin{array}{l}
U^{\ \widehat{\Omega}_{\ x}} \ \left (
\ y \ , \ y \ ; \ {\cal{CL}} \ ( \ y \ )
\hspace*{0.2cm} 
\begin{array}{l}
 \hspace*{0.05cm} 
\bigcirc \vspace*{0.1cm} \hspace*{-0.14cm} {\scriptstyle{\wedge}}
\end{array}
\ \right )_{\ A \ B}
\ =
\vspace*{0.3cm} \\
\hspace*{0.3cm} =
\ \left (
\ P_{\ 2}^{\ \widehat{\Omega}_{\ x}} \ \exp  
\ \left ( 
\ -
\ {\displaystyle{\int}}_{\ S_{\ x \ | \ y}} 
\ {\cal{W}}^{\ (2) \ \widehat{\Omega}_{\ x}} 
\ \right )
\ \right )_{\ A \ B}
\vspace*{0.3cm} \\
\hspace*{0.3cm} \forall \hspace*{0.3cm}
\left \lbrace
\ {\cal{S}}_{\ x \ | \ y} \ ; \ \widehat{\Omega}_{\ x} 
\ \right \rbrace
\end{array}
\end{equation}

The true form of Stokes relations returns to a general {\it common} gauge,
combining eqs. (\ref{eq:1001}) and (\ref{eq:1004})
\vspace*{-0.3cm} 

\begin{equation}
\label{eq:1005}
\begin{array}{l}
U \ \left (
\ y \ , \ y \ ; \ {\cal{CL}} \ ( \ y \ )
\hspace*{0.2cm} 
\begin{array}{l}
 \hspace*{0.05cm} 
\bigcirc \vspace*{0.1cm} \hspace*{-0.14cm} {\scriptstyle{\wedge}}
\end{array}
\ \right )_{\ A \ B}
\ =
\vspace*{0.3cm} \\
\hspace*{0.3cm} =
\ \left (
\begin{array}{l}
\ \left (
\ \widehat{\Omega}_{\ x} \ ( \ y \ )
\ \right )^{\ -1}
\ \times
\vspace*{0.3cm} \\
\hspace*{1.5cm} \times 
\ P_{\ 2}^{\ \widehat{\Omega}_{\ x}} \ \exp  
\ \left ( 
\ -
\ {\displaystyle{\int}}_{\ S_{\ x \ | \ y}} 
\ {\cal{W}}^{\ (2) \ \widehat{\Omega}_{\ x}} 
\ \right )
\ \times
\vspace*{0.3cm} \\
\hspace*{3.5cm} \times 
\ \widehat{\Omega}_{\ x} \ ( \ y \ )
\end{array}
\ \right )_{\ A \ B}
\vspace*{0.7cm} \\
\hspace*{0.3cm} \forall \hspace*{0.3cm}
\left \lbrace
\ {\cal{S}}_{\ x \ | \ y} \ ; \ \widehat{\Omega}_{\ x} 
\ \right \rbrace
\end{array}
\end{equation}

The closed contour ${\cal{CL}} \ ( \ y \ )$ integral
$U \ \left (
\ y \ , \ y \ ; \ {\cal{CL}} \ ( \ y \ )
\hspace*{0.2cm} 
\begin{array}{l}
 \hspace*{0.05cm} 
\bigcirc \vspace*{0.1cm} \hspace*{-0.14cm} {\scriptstyle{\wedge}}
\end{array}
\ \right )$
on the left hand side of eq. (\ref{eq:1005}) is dependent
on the point $y$, where the contour begins and ends,
but {\it not} on any surface and associated Riemann normal
gauge forming the collection
$\left \lbrace
\ {\cal{S}}_{\ x \ | \ y} \ ; \ \widehat{\Omega}_{\ x} 
\ \right \rbrace$ .

\item v) the surface integral proper in Riemann normal gauge

The main ingredient in the Stokes relations in eq. (\ref{eq:1005})
is \\
-- selecting a surface and a Riemann normal gauge \\
out of the collection
$\left \lbrace
\ {\cal{S}}_{\ x \ | \ y} \ ; \ \widehat{\Omega}_{\ x} 
\ \right \rbrace$ --
the surface integral proper as summarized in eq. (\ref{eq:1004})

\begin{equation}
\label{eq:1006}
\begin{array}{l}
U^{\ \widehat{\Omega}_{\ x}} \ \left (
\ y \ , \ y \ ; \ {\cal{CL}} \ ( \ y \ )
\hspace*{0.2cm} 
\begin{array}{l}
 \hspace*{0.05cm} 
\bigcirc \vspace*{0.1cm} \hspace*{-0.14cm} {\scriptstyle{\wedge}}
\end{array}
\ \right )_{\ A \ B}
\ =
\vspace*{0.3cm} \\
\hspace*{0.3cm} =
\ \left (
\ P_{\ 2}^{\ \widehat{\Omega}_{\ x}} \ \exp  
\ \left ( 
\ -
\ {\displaystyle{\int}}_{\ S_{\ x \ | \ y}} 
\ {\cal{W}}^{\ (2) \ \widehat{\Omega}_{\ x}} 
\ \right )
\ \right )_{\ A \ B}
\end{array}
\end{equation}

The surface integral on the right hand side of eq. (\ref{eq:1006})
-- {\it in Riemann normal gauge} --
involves the ordering, denoted  
$P_{\ 2}^{\ \widehat{\Omega}_{\ x}}$ , of products of {\it local}
surface differentials 
${\cal{W}}^{\ (2) \ \widehat{\Omega}_{\ x}}$ . 
By this local property the surface 'integral' is indeed an integral.

In any gauge other than a Riemannian normal one, the corresponding
differentials are {\it not} local functions of the plaquette
differentials, rather they depend on the entire set of
flagpole paths, described in figure \ref{fig3} and eqs. (\ref{eq:95})
and (\ref{eq:96}).

\item vi) the ordering of surface elements in Riemann normal gauge

The path ordering
$P_{\ 2}^{\ \widehat{\Omega}_{\ x}}$ of the -- matrix valued --
surface elements is very special to Riemann normal gauges.

It derives  from two steps, starting
in a general (original) gauge. 

They are described in the text following
eq.(\ref{eq:94}) and in figures \ref{fig3} and \ref{fig4}.

An appropriate name for  
$P_{\ 2}^{\ \widehat{\Omega}_{\ x}}$ is 'spider-web ordering' 
illustrated in figure \ref{fig4a} ,
\cite{spiderfot} .

\end{description}

\newpage


\noindent
\subsection{Spin projection operations on adjoint string operators}
\vspace*{0.1cm} 

\noindent
The adjoint string operators forming binary gluonic mesons
are interoduced in eq. (\ref{eq:101}) , repeated below
\vspace*{-0.3cm} 

\begin{equation}
\label{eq:1007}
\begin{array}{l}
B_{\ \left \lbrack \ \mu_{1} \ \nu_{1} \  \right \rbrack
\ , \ \left \lbrack  \ \mu_{2} \ \nu_{2}  \ \right \rbrack}
\ ( \ x_{\ 1} \ , \ x_{\ 2} \ )
\ =
\vspace*{0.3cm} \\
\hspace*{0.5cm}
F_{\ \left \lbrack \ \mu_{1} \ \nu_{1} \ \right \rbrack}
\ ( \ x_{\ 1} \ ; \ A \ )
\ U \ ( \ x_{\ 1} \ , \ A \ ; \ x_{\ 2} \ , \ B \ )
\ F_{\ \left \lbrack \ \mu_{2} \ \nu_{2} \ \right \rbrack}
\ ( \ x_{\ 2} \ ; \ B \ )
\vspace*{0.3cm} \\
A \ , \ B \ , \cdots \ = \ 1, \cdots , 8
\end{array}
\end{equation}

\noindent
The Lorentz invariant tensors $K^{\ \pm}$ are introduced in
eq. (\ref{eq:123}) , repeated below
\vspace*{-0.3cm} 

\begin{equation}
\label{eq:1008}
\begin{array}{l}
\widetilde{t}_{\ \underline{.} \ ; \ I^{\ \pm}} 
\ ( \ z \ , \ p \ , \ J^{\ \pm \ +} \  ; \ . \ )
\ = 
\ \left (
 \begin{array}{l}
\ \left (
\ K^{\ \pm}
\ \right )_{\ 
\ \left \lbrack
\ \mu_{\ 1} \ \nu_{\ 1}
\ \right \rbrack
\ \left \lbrack
\ \mu_{\ 2} \ \nu_{\ 2}
\ \right \rbrack}
\ \times
\vspace*{0.3cm} \\
\hspace*{0.7cm} \times
\ \widetilde{t}_{\ I^{\ \pm}} 
\ ( \ z \ , \ p \ , \ J^{\ \pm \ +} \  ; \ . \ )
\end{array}
\ \right )
\vspace*{0.5cm} \\
\ \left (
\ K^{\ +}
\ \right )_{\ 
\ \left \lbrack
\ \mu_{\ 1} \ \nu_{\ 1}
\ \right \rbrack
\ \left \lbrack
\ \mu_{\ 2} \ \nu_{\ 2}
\ \right \rbrack}
\ =
\ g_{\ \mu_{\ 1} \ \mu_{\ 2}}
\ g_{\ \nu_{\ 1} \ \nu_{\ 2}}
\ -
\ g_{\ \mu_{\ 1} \ \nu_{\ 2}}
\ g_{\ \mu_{\ 2} \ \nu_{\ 1}}
\vspace*{0.5cm} \\
\ \left (
\ K^{\ -}
\ \right )_{\ 
\ \left \lbrack
\ \mu_{\ 1} \ \nu_{\ 1}
\ \right \rbrack
\ \left \lbrack
\ \mu_{\ 2} \ \nu_{\ 2}
\ \right \rbrack}
\ =
\ \varepsilon_{\ \mu_{\ 1} \ \mu_{\ 2} \ \nu_{\ 1} \ \nu_{\ 2}}
\end{array}
\end{equation}

\noindent
We perform the associated projections
\vspace*{-0.3cm} 

\begin{equation}
\label{eq:1009}
\begin{array}{l}
B_{\ \left \lbrack \ \mu_{1} \ \nu_{1} \  \right \rbrack
\ , \ \left \lbrack  \ \mu_{2} \ \nu_{2}  \ \right \rbrack}
\ =
\ \left (
\ \begin{array}{r}
 K^{\ +}_{\ \left \lbrack \ \mu_{1} \ \nu_{1} \  \right \rbrack
  \ \left \lbrack  \ \mu_{2} \ \nu_{2}  \ \right \rbrack}
\ B^{\ (+)} \ + 
\vspace*{0.3cm} \\
 + \ K^{\ -}_{\ \left \lbrack \ \mu_{1} \ \nu_{1} \  \right \rbrack
 \ \left \lbrack  \ \mu_{2} \ \nu_{2}  \ \right \rbrack}
\ B^{\ (-)} \ + 
\vspace*{0.3cm} \\
+ \ B^{\ '}_{\ \left \lbrack \ \mu_{1} \ \nu_{1} \  \right \rbrack
\ , \ \left \lbrack  \ \mu_{2} \ \nu_{2}  \ \right \rbrack}
\end{array}
\ \right )
\end{array}
\end{equation}

\noindent
The decomposition according to eq. (\ref{eq:1009})
is the same for the Riemann curvature tensor, where 
$B^{\ (+)}$ relates to the curvature scalar ,
$B^{\ '}$ to the Ricci and Weyl tensors 
and $B^{\ (-)} \ = \ 0$, unlike here.

\newpage

\noindent
So we form the metric (Ricci-) contraction
\vspace*{-0.2cm} 

\begin{equation}
\label{eq:1010}
\begin{array}{l}
g^{\ \mu_{\ 1} \ \mu_{\ 2}}
\ B_{\ \left \lbrack \ \mu_{1} \ \nu_{1} \  \right \rbrack
\ , \ \left \lbrack  \ \mu_{2} \ \nu_{2}  \ \right \rbrack}
\ = \ R_{\ \nu_{\ 1} \ \nu_{\ 2}}
\vspace*{0.3cm} \\
R_{\ \nu_{\ 1} \ \nu_{\ 2}}
\ = 
\ \left (
\ \begin{array}{l}
3 \ g_{\ \nu_{\ 1} \ \nu_{\ 2}} \ B^{\ (+)} \ +
\vspace*{0.3cm} \\
 + \ g^{\ \mu_{\ 1} \ \mu_{\ 2}} 
\ B^{\ '}_{\ \left \lbrack \ \mu_{1} \ \nu_{1} \  \right \rbrack
\ , \ \ \left \lbrack  \ \mu_{2} \ \nu_{2}  \ \right \rbrack}
\end{array}
\ \right )
\vspace*{0.3cm} \\
R_{\ \nu_{\ 1} \ \nu_{\ 2}} \ = 
\ \varrho_{\ \nu_{\ 1} \ \nu_{\ 2}} \ + \ \frac{1}{4} 
\ g_{\ \nu_{\ 1} \ \nu_{\ 2}} \ R
\hspace*{0.3cm} ; \hspace*{0.3cm}
R \ =
\ g^{\ \nu_{\ 1} \ \nu_{\ 2}} \ R_{\ \nu_{\ 1} \ \nu_{\ 2}}
\vspace*{0.3cm} \\
\varrho_{\ \nu_{\ 1} \ \nu_{\ 2}} \ = 
\ R_{\ \nu_{\ 1} \ \nu_{\ 2}} \ - 
\ \frac{1}{4} 
\ g_{\ \nu_{\ 1} \ \nu_{\ 2}} \ R
\end{array}
\end{equation}

\noindent
It follows from eq. (\ref{eq:1010})
\vspace*{-0.2cm} 

\begin{equation}
\label{eq:1011}
\begin{array}{l}
\varrho_{\ \nu_{\ 1} \ \nu_{\ 2}} \ = 
\ g^{\ \mu_{\ 1} \ \mu_{\ 2}} 
\ B^{\ '}_{\ \left \lbrack \ \mu_{1} \ \nu_{1} \  \right \rbrack
\ , \ \ \left \lbrack  \ \mu_{2} \ \nu_{2}  \ \right \rbrack}
\ \rightarrow
\vspace*{0.3cm} \\
g^{\ \mu_{\ 1} \ \mu_{\ 2}} 
\ g^{\ \nu_{\ 1} \ \nu_{\ 2}} 
\ B^{\ '}_{\ \left \lbrack \ \mu_{1} \ \nu_{1} \  \right \rbrack
\ , \ \ \left \lbrack  \ \mu_{2} \ \nu_{2}  \ \right \rbrack}
\ = \ 0 
\hspace*{0.3cm} \rightarrow \hspace*{0.3cm}
R \ = \ 12 \ B^{\ (+)}
\end{array}
\end{equation}

\noindent
The quantity $B^{\ '}$ with the trace condition in eq. (\ref{eq:1011})
forms the irreducible {\it relativistic} spin two part
$S_{\ 12}^{\ +} \ = \ 2$ as defined in eqs. (\ref{eq:108})
and (\ref{eq:122}) in the main text. Here we concentrate
on the projection on $B^{\ (\pm)}$ .

\noindent
From eq. (\ref{eq:1011}) we obtain
\vspace*{-0.3cm} 

\begin{equation}
\label{eq:1012}
\begin{array}{l}
B^{\ (+)} \ (  \ x_{\ 1} \ , \  x_{\ 2} \ )
\ =
\vspace*{0.3cm} \\
\hspace*{0.6cm} =
\ \frac{1}{12}
\ F_{\ \left \lbrack \ \alpha \ \beta \ \right \rbrack}
\ ( \ x_{\ 1} \ ; \ A \ )
\ U \ ( \ x_{\ 1} \ , \ A \ ; \ x_{\ 2} \ , \ B \ )
\ F^{\ \left \lbrack \ \alpha \ \beta \ \right \rbrack}
\ ( \ x_{\ 2} \ ; \ B \ )
\end{array}
\end{equation}

\noindent
The projection on $B^{\ (-)}$ proceeds in a similar way
\vspace*{-0.3cm} 

\begin{equation}
\label{eq:1013}
\begin{array}{l}
\varepsilon^{\ \mu_{\ 1} \ \mu_{\ 2} \ \nu_{\ 1} \ \nu_{\ 2}}
\ B_{\ \left \lbrack \ \mu_{1} \ \nu_{1} \  \right \rbrack
\ , \ \left \lbrack  \ \mu_{2} \ \nu_{2}  \ \right \rbrack}
\ =
\ - \ 24 \ B^{\ (-)}
\ +
\vspace*{0.3cm} \\
\hspace*{4.5cm} +
\ \varepsilon^{\ \mu_{\ 1} \ \mu_{\ 2} \ \nu_{\ 1} \ \nu_{\ 2}}
\ B^{\ '}_{\ \left \lbrack \ \mu_{1} \ \nu_{1} \  \right \rbrack
\ , \ \ \left \lbrack  \ \mu_{2} \ \nu_{2}  \ \right \rbrack}
\ \rightarrow
\vspace*{0.3cm} \\
\varepsilon^{\ \mu_{\ 1} \ \mu_{\ 2} \ \nu_{\ 1} \ \nu_{\ 2}}
\ B^{\ '}_{\ \left \lbrack \ \mu_{1} \ \nu_{1} \  \right \rbrack
\ , \ \ \left \lbrack  \ \mu_{2} \ \nu_{2}  \ \right \rbrack}
\ = \ 0
\end{array}
\end{equation}

\noindent
The structure pf $B^{\ (-)}$ follows similarly as for $B^{\ (+)}$
in eq. (\ref{eq:1012}) . We thus give both expressions together below

\begin{equation}
\label{eq:1014}
\begin{array}{l}
B^{\ (+)} \ (  \ x_{\ 1} \ , \  x_{\ 2} \ )
\ =
\vspace*{0.3cm} \\
\hspace*{0.6cm} =
\ \frac{1}{12}
\ F_{\ \left \lbrack \ \alpha \ \beta \ \right \rbrack}
\ ( \ x_{\ 1} \ ; \ A \ )
\ U \ ( \ x_{\ 1} \ , \ A \ ; \ x_{\ 2} \ , \ B \ )
\ F^{\ \left \lbrack \ \alpha \ \beta \ \right \rbrack}
\ ( \ x_{\ 2} \ ; \ B \ )
\vspace*{0.3cm} \\
\hline
\vspace*{-0.2cm} \\
B^{\ (-)} \ (  \ x_{\ 1} \ , \  x_{\ 2} \ )
\ =
\vspace*{0.3cm} \\
\hspace*{0.6cm} =
\ - \ \frac{1}{12}
\ F_{\ \left \lbrack \ \alpha \ \beta \ \right \rbrack}
\ ( \ x_{\ 1} \ ; \ A \ )
\ U \ ( \ x_{\ 1} \ , \ A \ ; \ x_{\ 2} \ , \ B \ )
\ \widetilde{F}^{\ \left \lbrack \ \alpha \ \beta \ \right \rbrack}
\ ( \ x_{\ 2} \ ; \ B \ )
\vspace*{0.3cm} \\
\hspace*{0.6cm} =
\ - \ \frac{1}{12}
\ \widetilde{F}_{\ \left \lbrack \ \alpha \ \beta \ \right \rbrack}
\ ( \ x_{\ 1} \ ; \ A \ )
\ U \ ( \ x_{\ 1} \ , \ A \ ; \ x_{\ 2} \ , \ B \ )
\ F^{\ \left \lbrack \ \alpha \ \beta \ \right \rbrack}
\ ( \ x_{\ 2} \ ; \ B \ )
\vspace*{0.3cm} \\
\hline
\vspace*{-0.2cm} \\
\widetilde{F}_{\ \left \lbrack \ \alpha \ \beta \ \right \rbrack}
\ ( \ x_{\ 2} \ ; \ B \ )
\ =
\ \frac{1}{2}
\ \varepsilon_{ \ \alpha \ \beta \ \gamma \ \delta}
\ F^{\ \left \lbrack \ \gamma \ \delta \ \right \rbrack}
\ ( \ x_{\ 2} \ ; \ B \ )
\vspace*{0.3cm} \\
\mbox{and} \hspace*{0.3cm}
( \ x_{\ 2} \ ; \ B \ ) \ \leftrightarrow
\ ( \ x_{\ 1} \ ; \ A \ )
\end{array}
\end{equation}

\noindent

\noindent
\subsection{Spin projection operations on adjoint string operators - extended}
\vspace*{0.1cm} 

\noindent
We continue the projection operations carried out
in appendix A.4 in order to extend them to the remaining gb spectral
series of type $II^{\ +}$ . This is related to the quantities
$B^{\ '}_{\ \left \lbrack \ \mu_{1} \ \nu_{1} \  \right \rbrack
\ , \ \left \lbrack  \ \mu_{2} \ \nu_{2}  \ \right \rbrack}$ defined in 
eq. (\ref{eq:124}) and refined in appendix A.4 ( eq. (\ref{eq:1009}) ) .

\noindent
To that end we perform the {\it full} decomposition of
the tensorial structure of the octet string operators
interoduced in eq. (\ref{eq:101}) and rewritten in eq. (\ref{eq:1007})
\vspace*{-0.3cm} 

\begin{equation}
\label{eq:1015}
\begin{array}{l}
B_{\ \left \lbrack \ \mu_{1} \ \nu_{1} \  \right \rbrack
\ , \ \left \lbrack  \ \mu_{2} \ \nu_{2}  \ \right \rbrack}
\ ( \ x_{\ 1} \ , \ x_{\ 2} \ )
\ \rightarrow
\ B_{\ \left \lbrack \ \mu_{1} \ \nu_{1} \  \right \rbrack
\ , \ \left \lbrack  \ \mu_{2} \ \nu_{2}  \ \right \rbrack}
\end{array}
\end{equation}

\noindent
which is analogous to that of the Riemann curvature tensor,
without the metric constraints of the latter.

\newpage

\noindent
The Ricci contraction introduced in eq. (\ref{eq:1010}) 
yields the follwoing structure 
\vspace*{-0.3cm} 

\begin{equation}
\label{eq:1016}
\begin{array}{l}
B_{\ \left \lbrack \ \mu_{1} \ \nu_{1} \  \right \rbrack
\ , \ \left \lbrack  \ \mu_{2} \ \nu_{2}  \ \right \rbrack}
\ = \ w_{\ \left \lbrack \ \mu_{1} \ \nu_{1} \  \right \rbrack
\ \left \lbrack  \ \mu_{2} \ \nu_{2}  \ \right \rbrack}
+ \Delta 
\ B_{\ \left \lbrack \ \mu_{1} \ \nu_{1} \  \right \rbrack
\ , \ \left \lbrack  \ \mu_{2} \ \nu_{2}  \ \right \rbrack}
\vspace*{0.3cm} \\
\Delta 
\ B_{\ \left \lbrack \ \mu_{1} \ \nu_{1} \  \right \rbrack
\ , \ \left \lbrack  \ \mu_{2} \ \nu_{2}  \ \right \rbrack}
\ = 
\begin{array}{l}
\ \frac{1}{2}
 \left (
\begin{array}{l}
 g_{\ \mu_{\ 1} \ \mu_{\ 2}} \ R_{\ \nu_{\ 1} \ \nu_{\ 2}}
- \  g_{\ \nu_{\ 1} \ \mu_{\ 2}} \ R_{\ \mu_{\ 1} \ \nu_{\ 2}}
\vspace*{0.3cm} \\
- \  g_{\ \mu_{\ 1} \ \nu_{\ 2}} \ R_{\ \nu_{\ 1} \ \mu_{\ 2}}
\ + \ g_{\ \nu_{\ 1} \ \nu_{\ 2}} \ R_{\ \mu_{\ 1} \ \mu_{\ 2}}
\end{array}
 \right )
\vspace*{0.3cm} \\
- 
\ \frac{1}{6}
\ K^{\ +}_{\ \left \lbrack \ \mu_{1} \ \nu_{1} \  \right \rbrack
  \ \left \lbrack  \ \mu_{2} \ \nu_{2}  \ \right \rbrack}
\ R
\end{array}
\vspace*{0.4cm} \\
\hline
\vspace*{-0.3cm} \\
\mbox{with :}
\hspace*{0.3cm}
g^{\ \mu_{\ 1} \ \mu_{\ 2}}
\ w_{\ \left \lbrack \ \mu_{1} \ \nu_{1} \  \right \rbrack
\ \left \lbrack  \ \mu_{2} \ \nu_{2}  \ \right \rbrack}
\ = \ 0
\vspace*{0.5cm} \\
g^{\ \mu_{\ 1} \ \mu_{\ 2}}
\ B_{\ \left \lbrack \ \mu_{1} \ \nu_{1} \  \right \rbrack
\ \ , \ \left \lbrack  \ \mu_{2} \ \nu_{2}  \ \right \rbrack}
\ =
\ g^{\ \mu_{\ 1} \ \mu_{\ 2}}
\ \Delta 
\ B_{\ \left \lbrack \ \mu_{1} \ \nu_{1} \  \right \rbrack
\ , \ \left \lbrack  \ \mu_{2} \ \nu_{2}  \ \right \rbrack}
\ = \ R_{\ \nu_{\ 1} \ \nu_{\ 2}}
\vspace*{0.5cm} \\
R \ =
\ g^{\ \nu_{\ 1} \ \nu_{\ 2}}
\ R_{\ \nu_{\ 1} \ \nu_{\ 2}}
\ =
\ 12 \ B^{\ (+)}
\end{array}
\end{equation}

\noindent
Before proceeding lets express the Ricci bilinear in terms of 
the base octet string operators
$B_{\ \left \lbrack \ \mu_{1} \ \nu_{1} \  \right \rbrack
\ , \ \left \lbrack  \ \mu_{2} \ \nu_{2}  \ \right \rbrack}$

\begin{equation}
\label{eq:1017}
\begin{array}{l}
R_{\ \nu_{\ 1} \ \nu_{\ 2}}
\ =
\vspace*{0.3cm} \\
\hspace*{0.3cm} =
\ - \ F_{\ \nu_{\ 1} \ \alpha}
\ ( \ x_{\ 1} \ ; \ A \ )
\ U \ ( \ x_{\ 1} \ , \ A \ ; \ x_{\ 2} \ , \ B \ )
\ F^{\ \alpha}_{\hspace*{0.3cm} \nu_{\ 2}}
\ ( \ x_{\ 2} \ ; \ B \ )
\end{array}
\end{equation}

\noindent
In order to simplify notation we shall suppress
the position arguments and
use chromoelectric and -magnetic fields for the
field strength tensor.

\begin{equation}
\label{eq:1018}
\begin{array}{l}
- \ R_{\ \nu_{\ 1} \ \nu_{\ 2}}
\ =
\ \left (
\begin{array}{cc}
\vec{E}^{\ A} \ \vec{E}^{\ D}
 & \vec{S}^{\ k \ A \ D} 
 \vspace*{0.5cm} \\
\vec{S}^{\ i \ A \ D} & 
\begin{array}{ll}
- \ \vec{E}^{\ i \ A} \ \vec{E}^{\ k \ D}
- \ \vec{B}^{\ i \ A} \ \vec{B}^{\ k \ D}
\vspace*{0.3cm} \\
\ + \ \delta_{\ i k} \ \vec{B}^{\ A} \ \vec{B}^{\ D}
\end{array}
\end{array}
\ \right )
 \ U_{\ A \ D}
\vspace*{0.5cm} \\
R \ = \ 2 
\ \left (
\ \vec{B}^{\ A} \ \vec{B}^{\ D}
\ - \ \vec{E}^{\ A} \ \vec{E}^{\ D}
\ \right )
\ U_{\ A \ D}
\hspace*{0.3cm} ; \hspace*{0.3cm}
\vec{S}^{\ A \ D} \ = \ \vec{E}^{\ A} \ \wedge \ \vec{B}^{\ D}
\vspace*{0.5cm} \\
\vec{E}^{\ i \ A} 
\ = \ F^{\ 0 \ i \ A} 
\hspace*{0.3cm} , \hspace*{0.3cm}
\vec{B}^{\ i \ A} \ = \ \frac{1}{2} \ \varepsilon_{\ ikl}
\ F^{\ k \ l \ A} 
\end{array}
\end{equation}

\noindent
In eq. (\ref{eq:1018}) we recognize the Maxwell energy momentum like
(bilinear) expression, where 
$\vec{S}^{\ A \ D}$ shall be called the bilinear Poynting vector.

\noindent
Next we substitute the traceless part of the Ricci bilinear
( eq. (\ref{eq:1010}) ) in eq. (\ref{eq:1016})

\begin{equation}
\label{eq:1019}
\begin{array}{l}
- \ \varrho^{\ \mu \ \nu} \ = 
\ - \ R^{\ \mu \ \nu} \ + 
\ \frac{1}{4} 
\ g^{\ \mu \ \nu} \ R \ =
\ \vartheta_{\ cl}^{\ \mu \ \nu} \ =
\vspace*{0.5cm} \\
\hspace*{0.4cm} =
\ \left (
\begin{array}{cc}
\frac{1}{2} 
\ \left ( 
\begin{array}{l}
 \vec{E}^{\ A} \ \vec{E}^{\ D} 
\vspace*{0.3cm} \\
\hspace*{0.3cm} + \ \vec{B}^{\ A} \ \vec{B}^{\ D}
\end{array}
\ \right ) 
 & - \ \vec{S}^{\ k \ A \ D} 
 \vspace*{0.5cm} \\
- \ \vec{S}^{\ i \ A \ D} & 
\begin{array}{ll}
- \ \vec{E}^{\ i \ A} \ \vec{E}^{\ k \ D}
- \ \vec{B}^{\ i \ A} \ \vec{B}^{\ k \ D}
\vspace*{0.3cm} \\
\ + \ \frac{1}{2} \ \delta_{\ i k} 
\ \left (
\begin{array}{l}
\vec{E}^{\ A} \ \vec{E}^{\ D} 
\vspace*{0.3cm} \\
\hspace*{0.3cm} + \ \vec{B}^{\ A} \ \vec{B}^{\ D}
\end{array}
\ \right )
\end{array}
\end{array}
\ \right )
 \ U_{\ A \ D}
\end{array}
\end{equation}

\noindent
In eq. (\ref{eq:1019}) we recognize the bilinear with the structure of the
classical (traceless) Maxwell energy momentum tensor of 
nonabelian gauge field strengths.

\noindent
Eq. (\ref{eq:1016}) becomes decomposed into positive parity irreducible parts

\begin{equation}
\label{eq:1020}
\begin{array}{l}
B_{\ \left \lbrack \ \mu_{1} \ \nu_{1} \  \right \rbrack
\ , \ \left \lbrack  \ \mu_{2} \ \nu_{2}  \ \right \rbrack}
\ = \ w_{\ \left \lbrack \ \mu_{1} \ \nu_{1} \  \right \rbrack
\ \left \lbrack  \ \mu_{2} \ \nu_{2}  \ \right \rbrack}
+ \Delta 
\ B_{\ \left \lbrack \ \mu_{1} \ \nu_{1} \  \right \rbrack
\ , \ \left \lbrack  \ \mu_{2} \ \nu_{2}  \ \right \rbrack}
\vspace*{0.3cm} \\
\Delta 
\ B_{\ \left \lbrack \ \mu_{1} \ \nu_{1} \  \right \rbrack
\ , \ \left \lbrack  \ \mu_{2} \ \nu_{2}  \ \right \rbrack}
\ = 
\begin{array}{l}
\ \frac{1}{2}
 \left (
\begin{array}{l}
 g_{\ \mu_{\ 1} \ \mu_{\ 2}} \ \varrho_{\ \nu_{\ 1} \ \nu_{\ 2}}
- \  g_{\ \nu_{\ 1} \ \mu_{\ 2}} \ \varrho_{\ \mu_{\ 1} \ \nu_{\ 2}}
\vspace*{0.3cm} \\
- \  g_{\ \mu_{\ 1} \ \nu_{\ 2}} \ \varrho_{\ \nu_{\ 1} \ \mu_{\ 2}}
\ + \ g_{\ \nu_{\ 1} \ \nu_{\ 2}} \ \varrho_{\ \mu_{\ 1} \ \mu_{\ 2}}
\end{array}
 \right )
\vspace*{0.3cm} \\
+ 
\ \frac{1}{12}
\ K^{\ +}_{\ \left \lbrack \ \mu_{1} \ \nu_{1} \  \right \rbrack
  \ \left \lbrack  \ \mu_{2} \ \nu_{2}  \ \right \rbrack}
\ R
\end{array}
\vspace*{0.4cm} \\
\hline
\vspace*{-0.3cm} \\
\mbox{with :}
\hspace*{0.3cm}
g^{\ \mu_{\ 1} \ \mu_{\ 2}}
\ w_{\ \left \lbrack \ \mu_{1} \ \nu_{1} \  \right \rbrack
\ \left \lbrack  \ \mu_{2} \ \nu_{2}  \ \right \rbrack}
\ = \ 0
\vspace*{0.5cm} \\
g^{\ \mu_{\ 1} \ \mu_{\ 2}}
\ B_{\ \left \lbrack \ \mu_{1} \ \nu_{1} \  \right \rbrack
\ \ , \ \left \lbrack  \ \mu_{2} \ \nu_{2}  \ \right \rbrack}
\ =
\ g^{\ \mu_{\ 1} \ \mu_{\ 2}}
\ \Delta 
\ B_{\ \left \lbrack \ \mu_{1} \ \nu_{1} \  \right \rbrack
\ , \ \left \lbrack  \ \mu_{2} \ \nu_{2}  \ \right \rbrack}
\ = \ R_{\ \nu_{\ 1} \ \nu_{\ 2}}
\vspace*{0.5cm} \\
R \ =
\ g^{\ \nu_{\ 1} \ \nu_{\ 2}}
\ R_{\ \nu_{\ 1} \ \nu_{\ 2}}
\ =
\ 12 \ B^{\ (+)}
\end{array}
\end{equation}

\noindent
As a side remark to the (Lorentz-) tensorial reduction of 
the bilinear quantities
$ B_{\ \left \lbrack \ \mu_{1} \ \nu_{1} \  \right \rbrack
\ , \ \left \lbrack  \ \mu_{2} \ \nu_{2}  \ \right \rbrack}$
it is necessary to include the spatio-temporal {\it nonlocal}
parallel transport matrices pertaining to a general
connection and metric 
$\Gamma^{\hspace*{0.3cm} \mu}_{\ \sigma \hspace*{0.3cm} \nu}$
and $g_{\ \mu \nu}$ . This is necessary to render 
$ B_{\ \left \lbrack \ \mu_{1} \ \nu_{1} \  \right \rbrack
\ , \ \left \lbrack  \ \mu_{2} \ \nu_{2}  \ \right \rbrack}$
a true {\it nonlocal} Lorentz-tensor. We do not do this
here. In globally flat Minkowski space coordinates this
parallel transport is trivial.

\noindent
The sequence of projections on first
Lorentz spin ($\underline{.}$) and second on rotational spin
($.$) , needs two steps , rearranging the structure
of $- \ \varrho^{\ \mu \nu}$ in eq. (\ref{eq:1019})
\vspace*{-0.5cm}

\begin{equation}
\label{eq:1021}
\begin{array}{l}
- \ \varrho^{\ \mu \ \nu} \ = 
\ \vartheta_{\ cl}^{\ \mu \ \nu} \ =
\vspace*{0.5cm} \\
\hspace*{0.1cm} =
 \left (
\begin{array}{cc}
\frac{1}{2} 
\ \left ( 
\begin{array}{l}
 \vec{E}^{\ A} \ \vec{E}^{\ D} 
\vspace*{0.3cm} \\
\hspace*{0.3cm} + \ \vec{B}^{\ A} \ \vec{B}^{\ D}
\end{array}
\ \right ) 
 & - \ \vec{S}^{\ k \ A \ D} 
 \vspace*{0.5cm} \\
- \ \vec{S}^{\ i \ A \ D} & 
\begin{array}{ll}
\left (
\begin{array}{l}
- \ \vec{E}^{\ i \ A} \ \vec{E}^{\ k \ D}
- \ \vec{B}^{\ i \ A} \ \vec{B}^{\ k \ D}
\vspace*{0.3cm} \\
 + \ \frac{1}{3} \ \delta_{\ i k} 
\ \left (
\begin{array}{l}
\vec{E}^{\ A} \ \vec{E}^{\ D} 
\vspace*{0.3cm} \\
\hspace*{0.3cm} + \ \vec{B}^{\ A} \ \vec{B}^{\ D}
\end{array}
\ \right )
\end{array}
\ \right )
\vspace*{0.6cm} \\
 + \ \frac{1}{6} \ \delta_{\ i k} 
\ \left (
\begin{array}{l}
\vec{E}^{\ A} \ \vec{E}^{\ D} 
\vspace*{0.3cm} \\
\hspace*{0.3cm} + \ \vec{B}^{\ A} \ \vec{B}^{\ D}
\end{array}
\ \right )
\end{array}
\end{array}
 \right )
 \ \times
\vspace*{0.3cm} \\
\hspace*{5.6cm} \times \ U_{\ A \ D}
\end{array}
\end{equation}

\noindent
The tensor structure of $\vartheta_{\ cl}^{\ \mu \ \nu}$
in eq. (\ref{eq:1021}) follows the hydrodynamic nomenclature
\vspace*{-0.3cm}

\begin{equation}
\label{eq:1022}
\begin{array}{l}
- \ \varrho^{\ \mu \ \nu} \ = 
\ \vartheta_{\ cl}^{\ \mu \ \nu} \ =
\ \left (
\begin{array}{cc}
 \varrho_{\ e}  
 & - \ \vec{S}^{\ k} 
 \vspace*{0.5cm} \\
- \ \vec{S}^{\ i} & 
\left (
\ \begin{array}{cc}
\pi_{\ i \ k}
\vspace*{0.3cm} \\
 + \ \delta_{\ i k} \ p
\end{array}
\ \right )
\end{array}
 \right )
\vspace*{0.3cm} \\
\varrho_{\ e} \ = \ 3 \ p \ =  
\ \frac{1}{2} 
\ \left ( 
\ \vec{E}^{\ A} \ \vec{E}^{\ D} 
\ + \ \vec{B}^{\ A} \ \vec{B}^{\ D}
\ \right ) \ U_{\ A \ D}
\vspace*{0.3cm} \\
\pi_{\ i \ k} \ = \ 2 \ p \ \delta_{\ i \ k}
\ - 
\ \left (
\ \vec{E}^{\ i \ A} \ \vec{E}^{\ k \ D} 
\ + \ \vec{B}^{\ i \ A} \ \vec{B}^{\ k \ D} 
\ \right )
\ U_{\ A \ D}
\vspace*{0.3cm} \\
\sum_{\ i} \ \pi_{\ i \ i} \ = \ 0
\end{array}
\end{equation}

\noindent
with the identifications given in eq. (\ref{eq:1022}) .

\noindent
The chain of irreducible components of
$\Delta \ B_{\ \underline{.}}$ is shown in eq. (\ref{eq:1023})
below

\begin{equation}
\label{eq:1023}
\begin{array}{l}
\begin{array}{|c|cc cc|}
\hline 
 &  &  &  & \vspace*{-0.3cm} \\
\mbox{step} & \mbox{name} & \mbox{\# comp.} & \mbox{L.-spin} & \mbox{R.-spin} \\
 &  &  &  & \vspace*{-0.3cm} \\
\hline
 &  &  &  & \vspace*{-0.3cm} \\
1 & \Delta \ B_{\ \underline{.}} & 10 & \mbox{mixed} & \mbox{mixed} \\
2 & B^{\ (+)} & 1 & 1 & 1 \\
2 & \varrho^{\ \underline{.}}   &  9 & D^{\ 1 \ , \ \overline{1}}
 & \mbox{mixed} \\
3 & \varrho_{\ e} & 1 & - & 1 \\
3 & \vec{S} & 3 & - & D^{\ 1} \\
3 & \pi_{\ .} & 5 & - & D^{\ 2} \vspace*{-0.3cm} \\
 &  &  &  &  \\
\hline
\end{array}
\end{array}
\end{equation}
\vspace*{0.2cm} 

\noindent
It is the last term in eq. (\ref{eq:1023}) $\pi_{\ i \ k}$
as displayed in eq. (\ref{eq:1022}) which characterizes the 
$S_{\ 12}^{\ +} \ = \ 2$ spectral series of binary gluonic mesons.
The R-spin 2 tensor $\pi_{\ i \ k}$ is related to
the corresponding components of the Weyl bilinear 
$w_{\ \left \lbrack \ \mu_{1} \ \nu_{1} \  \right \rbrack
\ \left \lbrack  \ \mu_{2} \ \nu_{2}  \ \right \rbrack}$
introduced in eq. (\ref{eq:1016}), which represents the traceless
part of the bilinear Riemann like tensor
$B_{\ \left \lbrack \ \mu_{1} \ \nu_{1} \  \right \rbrack
\ , \ \left \lbrack  \ \mu_{2} \ \nu_{2}  \ \right \rbrack}$ ,
to which we turn next.
\vspace*{0.1cm} 

\noindent
{\it Weyl bilinear and circular polarization basis for gauge field strengths}
\vspace*{0.1cm} 

\noindent
We recall the right and left chiral spin matrices 
defined in appendix A.1 ( eqs. (\ref{eq:16}) and (\ref{eq:17}) )
reproduced below, first for the right circular part.
Here the notion {\it right (left) circular} refers to a
fixed spin axis and {\it not} to the individual momenta of the two
gauge bosons at the end of the octet string, in question.
The spin axis is common to both and an axial vector.

\begin{equation}
\label{eq:1024}
\begin{array}{l}
\left ( \ \sigma_{\ \mu \ \nu} \ \right )_{\ \alpha}^{\hspace*{0.3cm} \beta}
\ \leftrightarrow
\ P_{\ R} \ \frac{i}{2}
\ \left \lbrack \ \gamma_{\ \mu} \ , \ \gamma_{\ \nu} \ \right \rbrack
\ P_{\ R}
\hspace*{0.3cm} ; \hspace*{0.3cm} 
P_{\ R} \ = \ \frac{1}{2} \ ( \ 1 \ + \ \gamma_{\ 5 \ R} \ )
\vspace*{0.3cm} \\
\gamma_{\ 5 \ R} \ = \ \frac{1}{i} 
\ \gamma_{\ 0} \ \gamma_{\ 1} \ \gamma_{\ 2} \ \gamma_{\ 3}
\vspace*{0.3cm} \\
\left ( \ \sigma_{\ \mu \ \nu} \ \right )_{\ \alpha}^{\hspace*{0.3cm} \beta}
\ =
\ \begin{array}{ll}
\ \left ( 
\begin{array}{l}
- \ i \ \Sigma_{\ k} 
\vspace*{0.3cm} \\
\varepsilon_{\ m n r} \ \Sigma_{\ r} 
\end{array}
\ \right )_{ \alpha}^{\hspace*{0.3cm} \beta}
\vspace*{1.0cm} 
& 
\vspace*{-0.5cm} 
\begin{array}{l}
\mbox{for} \ \mu \ = \ 0 \ , \ \nu \ = \ k \ = \ 1,2,3
\vspace*{0.3cm} \\
\mbox{for} \ \mu \ = \ m \ , \ \nu \ = \ n \ ; 
\vspace*{-0.2cm} \\
\hspace*{2.7cm} 
 m,n,r \ = \ 1,2,3
\end{array}
\end{array}
\vspace*{0.3cm} \\
\left ( \ \sigma_{\ \mu \ \nu}
\ \right )_{\ \alpha}^{\hspace*{0.3cm} \beta}
\ \rightarrow \ \sigma_{\ \mu \ \nu}^{\ R}
\hspace*{0.3cm} ; \hspace*{0.3cm} 
\sigma_{\ \mu \ \nu}^{\ R} \ = \ - \ i 
\ \frac{1}{2} \ \varepsilon_{\ \mu \nu \varrho \tau} 
\ \sigma^{\ \varrho \ \tau \ R}
\end{array}
\end{equation}

\noindent
The right circular spinor basis in eq. (\ref{eq:1024})
yields the projection on the gauge field strengths

\begin{equation}
\label{eq:1025}
\begin{array}{l}
\frac{1}{2}
\ \sigma_{\ \mu \ \nu}^{\ R} 
\ F^{\ \left \lbrack \ \mu \ \nu \ \right \rbrack}
\ ( \ x \ ; \ A \ )
\ = \ \Sigma_{\ r} \ \vec{C}^{ \ r \ A} \ ( \ x \ )
\vspace*{0.3cm} \\
\vec{C}^{ \ r \ A} \ ( \ x \ ) 
\ =
\ \left (
\ \vec{B} \ - \ i \ \vec{E}
\ \right )^{\ r \ A} \ ( \ x \ )
\hspace*{0.3cm} \rightarrow \hspace*{0.3cm}
\vec{C}^{ \ r \ A}
\vspace*{0.3cm} \\
r \ = \ 1,2,3
\end{array}
\end{equation}

\noindent
The right circular quantities $\vec{C}^{ \ A}$ in eq. (\ref{eq:1025})
are complex combinations of the hermitian field strengths
$F^{\ \left \lbrack \ \mu \ \nu \ \right \rbrack \ A}$ in the
adjoint representation of $SU3_{\ c}$ .

\noindent
We note the right circular identity, following from eq. (\ref{eq:1024})

\begin{equation}
\label{eq:1026}
\begin{array}{l}
\frac{1}{2}
\ \sigma_{\ \mu \ \nu}^{\ R} 
\ F^{\ \left \lbrack \ \mu \ \nu \ \right \rbrack \ A}
\ \equiv
\ \frac{1}{2}
\ \sigma_{\ \mu \ \nu}^{\ R} 
\ \left ( \ F^{\ R}
\ \right )^{\ \left \lbrack \ \mu \ \nu \ \right \rbrack \ A}
\vspace*{0.3cm} \\
\ \left ( \ F^{\ R}
\ \right )^{\ \left \lbrack \ \mu \ \nu \ \right \rbrack \ A}
\ =
\ \frac{1}{2}
\ \left (
\ F^{\ \left \lbrack \ \mu \ \nu \ \right \rbrack \ A}
\ - \ i
\ \widetilde{F}^{\ \left \lbrack \ \mu \ \nu \ \right \rbrack \ A}
\ \right )
\vspace*{0.3cm} \\
\widetilde{F}_{\ \left \lbrack \ \mu \ \nu \ \right \rbrack}^{\ A}
\ =
\ \frac{1}{2}
\ \varepsilon_{\ \mu \nu \sigma \tau}
\ F^{\ \left \lbrack \ \sigma \ \tau \ \right \rbrack \ A}
\end{array}
\end{equation}

\noindent
When the space-time component $r$ in $\vec{C}^{\ r \ A}$ is
explicitely denoted, the vector symbol of $\vec{C}$ shall be 
omitted for simplicity.

\noindent
Now we recall the left chiral spinor matrices defined in
eqs. (\ref{eq:35}) and (\ref{eq:36}) in appendix A.1

\begin{equation}
\label{eq:1027}
\begin{array}{l}
\left ( \ \sigma_{\ \mu \ \nu} \ \right )^{\ \dot{\gamma}}_{\hspace*{0.4cm} 
\dot{\delta}}
\ \leftrightarrow
\ P_{\ L} \ \frac{i}{2}
\ \left \lbrack \ \gamma_{\ \mu} \ , \ \gamma_{\ \nu} \ \right \rbrack
\ P_{\ L}
\hspace*{0.3cm} ; \hspace*{0.3cm} 
P_{\ L} \ = \ \frac{1}{2} \ ( \ 1 \ - \ \gamma_{\ 5 \ R} \ )
\vspace*{0.3cm} \\
\gamma_{\ 5 \ R} \ = \ \frac{1}{i} 
\ \gamma_{\ 0} \ \gamma_{\ 1} \ \gamma_{\ 2} \ \gamma_{\ 3}
\vspace*{0.3cm} \\
\left ( \ \sigma_{\ \mu \ \nu} \ \right )^{\ \dot{\gamma}}_{\hspace*{0.4cm} 
\dot{\delta}}
\ =
\ \begin{array}{ll}
\ \left ( 
\begin{array}{l}
 i \ \Sigma_{\ k} 
\vspace*{0.3cm} \\
\varepsilon_{\ m n r} \ \Sigma_{\ r} 
\end{array}
\ \right )^{ \dot{\gamma}}_{\hspace*{0.4cm} \dot{\delta}}
\vspace*{1.0cm} 
& 
\vspace*{-0.5cm} 
\begin{array}{l}
\mbox{for} \ \mu \ = \ 0 \ , \ \nu \ = \ k \ = \ 1,2,3
\vspace*{0.3cm} \\
\mbox{for} \ \mu \ = \ m \ , \ \nu \ = \ n \ ; 
\vspace*{-0.2cm} \\
\hspace*{2.7cm} 
 m,n,r \ = \ 1,2,3
\end{array}
\end{array}
\vspace*{0.3cm} \\
 \left ( \ \sigma_{\ \mu \ \nu} \ \right )^{\ \dot{\gamma}}_{\hspace*{0.4cm} 
\dot{\delta}}
\ \rightarrow \ \sigma_{\ \mu \ \nu}^{\ L}
\hspace*{0.3cm} ; \hspace*{0.3cm} 
\sigma_{\ \mu \ \nu}^{\ L} \ = \ + \ i 
\ \frac{1}{2} \ \varepsilon_{\ \mu \nu \varrho \tau} 
\ \sigma^{\ \varrho \ \tau \ L}
\end{array}
\end{equation}

\noindent
Correspondingly to eq. (\ref{eq:1024}) ,
the left circular spinor basis in eq. (\ref{eq:1027})
yields the projection on the left circular gauge field strengths, completing
the right circular one in eq. (\ref{eq:1025})

\begin{equation}
\label{eq:1028}
\begin{array}{l}
\frac{1}{2}
\ \sigma_{\ \mu \ \nu}^{\ L} 
\ F^{\ \left \lbrack \ \mu \ \nu \ \right \rbrack}
\ ( \ x \ ; \ A \ )
\ = \ \Sigma_{\ r} \ \vec{G}^{ \ r \ A} \ ( \ x \ )
\vspace*{0.3cm} \\
\vec{G}^{ \ r \ A} \ ( \ x \ ) 
\ =
\ \left (
\ \vec{B} \ + \ i \ \vec{E}
\ \right )^{\ r \ A} \ ( \ x \ )
\hspace*{0.3cm} \rightarrow \hspace*{0.3cm}
\vec{G}^{ \ r \ A}
\vspace*{0.3cm} \\
r \ = \ 1,2,3
\end{array}
\end{equation}

\noindent
Analogous to the right circular identity in eq. (\ref{eq:1026})
is the left circular one, displayed together in eq. (\ref{eq:1029})
below

\begin{equation}
\label{eq:1029}
\begin{array}{l}
\frac{1}{2}
\ \sigma_{\ \mu \ \nu}^{\ R} 
\ F^{\ \left \lbrack \ \mu \ \nu \ \right \rbrack \ A}
\ \equiv
\ \frac{1}{2}
\ \sigma_{\ \mu \ \nu}^{\ R} 
\ \left ( \ F^{\ R}
\ \right )^{\ \left \lbrack \ \mu \ \nu \ \right \rbrack \ A}
\vspace*{0.3cm} \\
\ \left ( \ F^{\ R}
\ \right )^{\ \left \lbrack \ \mu \ \nu \ \right \rbrack \ A}
\ =
\ \frac{1}{2}
\ \left (
\ F^{\ \left \lbrack \ \mu \ \nu \ \right \rbrack \ A}
\ - \ i
\ \widetilde{F}^{\ \left \lbrack \ \mu \ \nu \ \right \rbrack \ A}
\ \right )
\vspace*{0.3cm} \\
\widetilde{F}_{\ \left \lbrack \ \mu \ \nu \ \right \rbrack}^{\ A}
\ =
\ \frac{1}{2}
\ \varepsilon_{\ \mu \nu \sigma \tau}
\ F^{\ \left \lbrack \ \sigma \ \tau \ \right \rbrack \ A}
\vspace*{0.5cm} \\
\ \sigma_{\ \mu \ \nu}^{\ L} 
\ F^{\ \left \lbrack \ \mu \ \nu \ \right \rbrack \ A}
\ \equiv
\ \frac{1}{2}
\ \sigma_{\ \mu \ \nu}^{\ L} 
\ \left ( \ F^{\ L}
\ \right )^{\ \left \lbrack \ \mu \ \nu \ \right \rbrack \ A}
\vspace*{0.3cm} \\
\ \left ( \ F^{\ L}
\ \right )^{\ \left \lbrack \ \mu \ \nu \ \right \rbrack \ A}
\ =
\ \frac{1}{2}
\ \left (
\ F^{\ \left \lbrack \ \mu \ \nu \ \right \rbrack \ A}
\ + \ i
\ \widetilde{F}^{\ \left \lbrack \ \mu \ \nu \ \right \rbrack \ A}
\ \right )
\end{array}
\end{equation}

\noindent
As long as we remain within the real Lorentz group,
as discussed in appendix A.1, the three vector quantities
$\vec{C}^{\ A}$ and $\vec{G}^{\ A}$ are relative hermitian conjugates
of each other. They transform according to the
$D^{\ 1 \ , \ 0}$ and $D^{\ 0 \ , \ \dot{1}}$ representations of the 
$spin \ ( \ 1 \ , \ 3 \ ; \ \Re \ ) \ \simeq \ SL2C$ group,
as defined in eqs. (\ref{eq:34}) and (\ref{eq:35}) in appendix
A.2 .
\vspace*{-0.3cm} 

\begin{equation}
\label{eq:1030}
\begin{array}{l}
\vec{C}^{\ A} \ = 
\ \left (
\ \vec{G}^{\ A}
\ \right )^{\ *}
\ \rightarrow
\vspace*{0.3cm} \\
D^{\ 1 \ , \ 0} 
\hspace*{0.2cm} : \hspace*{0.2cm}
C^{\ r \ A} \ \rightarrow \ R_{\ rs} \ C^{\ s \ A} 
\hspace*{0.4cm} ; \hspace*{0.4cm}
D^{\ 0 \ , \ \dot{1}}
\hspace*{0.2cm} : \hspace*{0.2cm}
G^{\ r \ A} \ \rightarrow \ L_{\ rs} \ G^{\ s \ A} 
\vspace*{0.3cm} \\
R_{\ rs} \ = \  R_{\ rs} \ ( \ {\cal{A}} \ )
\hspace*{0.3cm} , \hspace*{0.3cm}
L_{\ rs} \ = \  L_{\ rs} \ ( \ {\cal{B}} \ )
\vspace*{0.3cm} \\
R_{\ rs} \ = \ \overline{L}_{\ rs}
\hspace*{0.3cm} \mbox{within} \hspace*{0.3cm}
spin \ ( \ 1 \ , \ 3 \ ; \ \Re \ )
\end{array}
\end{equation}

\noindent
The three by three matrices $R_{\ rs}$ and $L_{\ rs}$
are complex orthogonal with determinant 1, forming
the group $SO3C \ \simeq \ SL2C \ / \ Z_{\ 2}$ ,
where $Z_{\ 2}$ denotes the center of $SL2C$ .

\noindent
We are now ready to decompose the Weyl bilinear 
$w_{\ \left \lbrack \ \mu_{1} \ \nu_{1} \  \right \rbrack
\ \left \lbrack  \ \mu_{2} \ \nu_{2}  \ \right \rbrack}$
in eq. (\ref{eq:1016}), into its irreducible parts.

\begin{equation}
\label{eq:1031}
\begin{array}{l}
w_{\ \left \lbrack \ \mu_{1} \ \nu_{1} \  \right \rbrack
\ \left \lbrack  \ \mu_{2} \ \nu_{2}  \ \right \rbrack}
\ =
\ \left (
\begin{array}{r}
\left (
\ P^{\ RR} \ \left ( \ \vec{C} \ \otimes \ \vec{C} \ \right )
\ \right )_{\ \left \lbrack \ \mu_{1} \ \nu_{1} \  \right \rbrack
\ \left \lbrack  \ \mu_{2} \ \nu_{2}  \ \right \rbrack}
\vspace*{0.3cm} \\
+
\ \left (
\ P^{\ LL} \ \left ( \ \vec{G} \ \otimes \ \vec{G} \ \right )
\ \right )_{\ \left \lbrack \ \mu_{1} \ \nu_{1} \  \right \rbrack
\ \left \lbrack  \ \mu_{2} \ \nu_{2}  \ \right \rbrack}
\vspace*{0.3cm} \\
+
\ B^{\ (-)}
\ \left (
\ K^{\ -}
\ \right )_{\ 
\ \left \lbrack
\ \mu_{ 1} \ \nu_{ 1}
\ \right \rbrack
\ \left \lbrack
\ \mu_{ 2} \ \nu_{ 2}
\ \right \rbrack}
\end{array}
\ \right )
\vspace*{0.4cm} \\
 \left (
\ K^{\ -}
\ \right )_{\ 
\ \left \lbrack
\ \mu_{\ 1} \ \nu_{\ 1}
\ \right \rbrack
\ \left \lbrack
\ \mu_{\ 2} \ \nu_{\ 2}
\ \right \rbrack}
\ =
\ \varepsilon_{\ \mu_{\ 1} \ \mu_{\ 2} \ \nu_{\ 1} \ \nu_{\ 2}}
\end{array}
\end{equation}

\noindent
In eq. (\ref{eq:1031}) the quantities $B^{\ (-)}$ and
$\left ( \ K^{\ -} \ \right )$ are defined in eqs. (\ref{eq:123})
- (\ref{eq:125}) .

\noindent
The projections denoted 
$P^{\ RR} \ \left ( \ \vec{C} \ \otimes \ \vec{C} \ \right )$
and $P^{\ LL} \ \left ( \ \vec{G} \ \otimes \ \vec{G} \ \right )$ ,
operate on the doubly right- and left circular, direct
product combinations indicated as respective arguments
in eq. (\ref{eq:1031}) . They are of the form, omitting the
explicit dependence on the Lorentz indices
$\left \lbrack \ \mu_{\ 1} \ \nu_{\ 1} \ \right \rbrack \ \left \lbrack
\ \mu_{\ 2} \ \nu_{\ 2} \ \right \rbrack$ for simplicity

\begin{equation}
\label{eq:1032}
\begin{array}{l}
P^{\ RR} \ \left ( \ \vec{C} \ \otimes \ \vec{C} \ \right )
\ = \ \pi^{\ (2)}_{\ r \ s} 
\ \left (
\ C^{\ r \ A} \ C^{\ s \ D} \ - 
\ \frac{1}{3} \ \delta^{\ r s} \ \vec{C}^{\ A} \ \vec{C}^{\ D}
\ \right )
\ U_{\ A \ D}
\vspace*{0.3cm} \\
P^{\ LL} \ \left ( \ \vec{G} \ \otimes \ \vec{G} \ \right )
\ = \ \pi^{\ (2)}_{\ r \ s} 
\ \left (
\ G^{\ r \ A} \ G^{\ s \ D} \ - 
\ \frac{1}{3} \ \delta^{\ r s} \ \vec{G}^{\ A} \ \vec{G}^{\ D}
\ \right )
\ U_{\ A \ D}
\end{array}
\end{equation}

\noindent
We thus introduce the abbreviations following eq. (\ref{eq:1032})

\begin{equation}
\label{eq:1033}
\begin{array}{l}
P^{\ RR} \ \left ( \ \vec{C} \ \otimes \ \vec{C} \ \right )
\ \rightarrow 
\ w^{\ RR}_{\ \underline{.}} 
\vspace*{0.3cm} \\
P^{\ LL} \ \left ( \ \vec{G} \ \otimes \ \vec{G} \ \right )
\ \rightarrow 
\ w^{\ LL}_{\ \underline{.}} 
\vspace*{0.3cm} \\
w^{\ RR}_{\ \underline{.}} 
\ =
\ \left (
\ C^{\ r \ A} \ C^{\ s \ D} \ - 
\ \frac{1}{3} \ \delta^{\ r s} \ \vec{C}^{\ A} \ \vec{C}^{\ D}
\ \right )
\ U_{\ A \ D}
\vspace*{0.3cm} \\
w^{\ LL}_{\ \underline{.}} 
\ =
\ \left (
\ G^{\ r \ A} \ G^{\ s \ D} \ - 
\ \frac{1}{3} \ \delta^{\ r s} \ \vec{G}^{\ A} \ \vec{G}^{\ D}
\ \right )
\ U_{\ A \ D}
\vspace*{0.3cm} \\
\hline
\vspace*{-0.2cm} \\
w^{\ LL}_{\ \underline{.}} 
\ =
\ \left (
\ w^{\ RR}_{\ \underline{.}}
\ \right )^{\ *}
\end{array}
\end{equation}

\noindent
The bilinears $w^{\ RR}_{\ \underline{.}}$ and
$w^{\ LL}_{\ \underline{.}}$ defined in eq. (\ref{eq:1033})
transform according to the complex representations
$D^{\ 2 \ , \ 0}$ and $D^{\ 0 \ , \ \dot{2}}$ of  
$spin \ ( \ 1 \ , \ 3 \ ; \ \Re \ ) \ \simeq \ SL2C$ 
respectively. These two representations are complex conjugate
to each other.

\noindent
As in the case of the Lorentz tensor
$\varrho^{\ \mu \nu}$ in eqs. (\ref{eq:1019}) ,  (\ref{eq:1021})
and  (\ref{eq:1022}) the space time indices $r \ s$ for the
quantities $w^{\ LL}_{\ \underline{.}}$ and $w^{\ RR}_{\ \underline{.}}$
in eq. (\ref{eq:1033}) are understood to be symmetrized.

\noindent
This completes the decomposition of the Weyl bilinear.
We compare the structure of the irreducible components
$w_{\ \underline{.}}$ with that of $\Delta \ B_{\ \underline{.}}$ ,
as shown in eq. (\ref{eq:1023}) reproduced below

\begin{equation}
\label{eq:1034}
\begin{array}{l}
\begin{array}{|c|cc cc|}
\hline 
 &  &  &  & \vspace*{-0.3cm} \\
\mbox{step} & \mbox{name} & \mbox{\# comp.} & \mbox{L.-spin} & \mbox{R.-spin} \\
 &  &  &  & \vspace*{-0.3cm} \\
\hline
 &  &  &  & \vspace*{-0.3cm} \\
1 & \Delta \ B_{\ \underline{.}} & 10 & \mbox{mixed} & \mbox{mixed} \\
2 & B^{\ (+)} & 1 & 1 & 1 \\
2 & \varrho^{\ \underline{.}}   &  9 & D^{\ 1 \ , \ \overline{1}}
 & \mbox{mixed} \\
3 & \varrho_{\ e} & 1 & - & 1 \\
3 & \vec{S} & 3 & - & D^{\ 1} \\
3 & \pi_{\ .} & 5 & - & D^{\ 2} \vspace*{-0.3cm} \\
 &  &  &  &  \\
\hline
\end{array}
\end{array}
\end{equation}
\vspace*{0.2cm} 

\noindent
The corresponding structure of the Weyl bilinears is displayed
in eq. (\ref{eq:1035})

\begin{equation}
\label{eq:1035}
\begin{array}{l}
\begin{array}{|c|cc cc|}
\hline 
 &  &  &  & \vspace*{-0.3cm} \\
\mbox{step} & \mbox{name} & \mbox{\# comp.} & \mbox{L.-spin} & \mbox{R.-spin} \\
 &  &  &  & \vspace*{-0.3cm} \\
\hline
 &  &  &  & \vspace*{-0.3cm} \\
1 & w_{\ \underline{.}} & 11 & \mbox{mixed} & \mbox{mixed} \\
2 & B^{\ (-)} & 1 & 1 & 1 \\
2 & w^{\ RR}_{\ \underline{.}}   &  5 & D^{\ 2 \ , \ 0}
 & D^{\ 2} \\
2 & w^{\ LL}_{\ \underline{.}} & \overline{5} &  D^{\ 0 \ , \ \dot{2}} 
& D^{\ 2} \vspace*{-0.3cm} \\
 &  &  &  &  \\
\hline
\end{array}
\end{array}
\end{equation}
\vspace*{0.2cm} 

\noindent
Comparing the counting in the two tables ( eqs. (\ref{eq:1034})
and (\ref{eq:1035}) we should keep in mind that the
colums labeled $\# \ comp.$ are based on counting independent
{\it hermitian operators} among the bilinears $B_{\ \underline{.}}$ .

\noindent
In this respect we verify the correctness of the counting :
the Riemann tensor like bilinears have $6 \ \times \ 7 \ / \ 2 \ = \ 21$
hermitian components, which combine into 10 for the Ricci tensor
like quantities further decomposed according to eq. (\ref{eq:1034})
and 11 for the Weyl tensor like in eq. (\ref{eq:1035}). 

\noindent
For the Ricci tensor the decomposition into $B^{\ (+)}$ corresponding
to the curvature scalar and the traceless part, called
$\varrho^{\ \underline{.}}$ here, is straightforward.

\noindent
For the Weyl tensor the splitting into 10 + 1 hermitian components,
corresponding to the pseudoscalar $B^{\ (-)}$ and the
right- and left circular bilinears 
$w^{\ RR}_{\ \underline{.}}$ and $w^{\ LL}_{\ \underline{.}}$ ,
with together 10 {\it hermitian} components is also
quite clear.

\noindent
What appears impossible, is
to find a common contribution to
the so defined irreducibles : 
$\varrho^{\ \underline{.}}$  with 9 hermitian components
on the one hand and 
$w^{\ RR}_{\ \underline{.}}$ and $w^{\ LL}_{\ \underline{.}}$ 
with 10 on the other. It follows from the discussion below, that
this is indeed impossible.

\noindent
In this connection we have to remember, that we are
considering matrix elements of the form defined in  
eq. (\ref{eq:104}) 
\vspace*{-0.3cm}

\begin{equation}
\label{eq:1036}
\begin{array}{l}
\left \langle 
\ \emptyset
\ \right |
\ B_{\ \left \lbrack \ \mu_{1} \ \nu_{1} \  \right \rbrack
\ , \ \left \lbrack  \ \mu_{2} \ \nu_{2}  \ \right \rbrack}
\ ( \ x_{\ 1} \ , \ x_{\ 2} \ )
\ \left |
\ gb \ ( \ J^{\ P\ C} \ ) \ ; \ p \ , 
\ \left \lbrace spin \right \rbrace
\ \right \rangle \ \rightarrow
\vspace*{0.3cm} \\
\exp^{\ - i p X} 
\ \widetilde{t}_{\ \underline{.}} \ ( \ z \ , \ p \ , \  J^{\ P\ C} \ ; \ . \ ) 
\end{array}
\end{equation}

\noindent
where hermition bilinears induce {\it complex} amplitudes.

\noindent
Next we focus on the continuity equation for the classical energy momentum
tensor pertaining
to the field strengths, 
extended to the nonlocal situation, conditioned by the
c.m. four momentum p. This follows the relations in eqs.
(\ref{eq:1021}) and (\ref{eq:1022})
\vspace*{-0.3cm} 

\begin{equation}
\label{eq:1037}
\begin{array}{l}
- \ \varrho^{\ \mu \ \nu} \ = 
\ \vartheta_{\ cl}^{\ \mu \ \nu} \ ( \ X \ ; \ z \ )
\hspace*{0.3cm} 
 \rightarrow
\hspace*{0.3cm} 
\partial_{\ X \ \mu} 
\ \vartheta_{\ cl}^{\ \mu \ \nu} \ ( \ X \ ; \ z \ )
\ = \ 0
\end{array}
\end{equation}

\noindent
Eq. (\ref{eq:1037}) is valid for classical field configurations and
follows from the analogous classical treatment of the
Stokes relation discussed in appendix A.3. It is not straightforward
for quantized local gauge fields. In the latter case the
identical relation is not sufficiently
established and deserves further study. Nevertheless we use
it here for consistency. Thus it follows that the
quantities $\varrho_{\ e}$ and $\vec{S}$ defined in eq. 
(\ref{eq:1022}) do not contribute to the amplitudes associated
with the classical energy momentum tensor 
$\vartheta_{\ cl}^{\ \mu \ \nu} \ ( \ X \ ; \ z \ )$ .
\vspace*{0.1cm} 

\noindent
Thus we associate each bilinear irreducible to the 
(family of) wave functions, following the notation introduced
in eq. (\ref{eq:104}) and repeated in eq. (\ref{eq:1036}) .
Hereby the complete family of wave functions is accordingly projected

\begin{equation}
\label{eq:1038}
\begin{array}{l}
\begin{array}{lll ll}
\varrho^{\ \underline{.}}  
& \leftrightarrow & 
\ \widetilde{t}_{\ \underline{.}} 
\ ( \ \left \lbrace \ \varrho \ \right \rbrace
\ ; \ z \ , \ p \ , \  J^{\ P\ C} \ ; \ . \ ) 
& \rightarrow & 
 \widetilde{t} 
\ ( \ \left \lbrace \ \varrho \ \right \rbrace \ )
\vspace*{0.3cm} \\
w_{\ \underline{.}} 
& \leftrightarrow & 
\widetilde{t}_{\ \underline{.}} 
\ ( \ \left \lbrace \ w \ \right \rbrace
\ ; \ z \ , \ p \ , \  J^{\ P\ C} \ ; \ . \ ) 
& \rightarrow & 
 \widetilde{t} 
\ ( \ \left \lbrace \ w \ \right \rbrace \ )
\vspace*{0.3cm} \\
w^{\ RR}_{\ \underline{.}} 
& \leftrightarrow & 
\widetilde{t}_{\ \underline{.}} 
\ ( \ \left \lbrace \ w^{\ RR} \ \right \rbrace
\ ; \ z \ , \ p \ , \  J^{\ P\ C} \ ; \ . \ ) 
& \rightarrow & 
 \widetilde{t} 
\ ( \ \left \lbrace \ w^{\ RR} \ \right \rbrace \ )
\vspace*{0.3cm} \\
w^{\ LL}_{\ \underline{.}}
& \leftrightarrow & 
\widetilde{t}_{\ \underline{.}} 
\ ( \ \left \lbrace \ w^{\ LL} \ \right \rbrace
\ ; \ z \ , \ p \ , \  J^{\ P\ C} \ ; \ . \ ) 
& \rightarrow & 
 \widetilde{t} 
\ ( \ \left \lbrace \ w^{\ LL} \ \right \rbrace \ )
\end{array}
\end{array}
\end{equation}

\noindent
The bilinears $B^{\ (\pm)}$ are already fully characterized. They
do not contribute to the wave functions of the
$II^{\ +}$ , i.e. $S_{\ 12}^{\ +} \ = \ 2$ spectral type, and thus
we will not discuss them any further here.

\noindent
The tables in eqs. (\ref{eq:1034}) and (\ref{eq:1035})
are thus reduced and adapted to the wave functions $\widetilde{t}$
defined in eq. (\ref{eq:1038})

\begin{equation}
\label{eq:1039}
\begin{array}{l}
\begin{array}{|c|cc cc|}
\hline 
 &  &  &  & \vspace*{-0.3cm} \\
\mbox{step} & \mbox{name} & \mbox{\# comp.} & \mbox{L.-spin} & \mbox{R.-spin} \\
 &  &  &  & \vspace*{-0.3cm} \\
\hline
 &  &  &  & \vspace*{-0.3cm} \\
2 &  
 \widetilde{t} \ ( \ \left \lbrace \ \varrho \ \right \rbrace \ )
& 9 &  D^{\ 1 \ , \ \overline{1}} & D^{\ 2} \\
3 & \widetilde{t} \ ( \ \left \lbrace \ \varrho_{\ e} \ \right \rbrace \ )
& 0 & - & - \\
3 & \widetilde{t} \ ( \ \left \lbrace \ \vec{S} \ \right \rbrace \ ) 
& 0 & - & - \\
3 & \widetilde{t} \ ( \ \left \lbrace \ \pi \ \right \rbrace \ ) 
& 5 & - & D^{\ 2} \vspace*{-0.3cm} \\
 &  &  &  &  \vspace*{-0.3cm} \\
\hline \vspace*{-0.4cm} \\
2 & \widetilde{t} \ ( \ \left \lbrace \ w^{\ RR} \ \right \rbrace \ )
&  5 & D^{\ 2 \ , \ 0} & D^{\ 2} \\
2 & \widetilde{t} \ ( \ \left \lbrace \ w^{\ LL} \ \right \rbrace \ )
&  5 & D^{\ 0 \ , \ 2} & D^{\ 2}  \vspace*{-0.3cm} \\
 &  &  &  &   \\
\hline 
\end{array}
\end{array}
\end{equation}
\vspace*{0.2cm} 

\noindent
The wave functions denoted 
$\widetilde{t} \ ( \ \left \lbrace \ \varrho_{\ e} \ \right \rbrace \ )$
and $\widetilde{t} \ ( \ \left \lbrace \ \vec{S} \ \right \rbrace \ )$ 
in eq. (\ref{eq:1039}) vanish, as a consequence of the
continuity equation in eq. (\ref{eq:1037}) .

\noindent
Thus the table in eq. (\ref{eq:1039}) reduces to

\begin{equation}
\label{eq:1040}
\begin{array}{l}
\begin{array}{|c|cc cc|}
\hline 
 &  &  &  & \vspace*{-0.3cm} \\
\mbox{step} & \mbox{name} & \mbox{\# comp.} & \mbox{L.-spin} & \mbox{R.-spin} \\
 &  &  &  & \vspace*{-0.3cm} \\
\hline
 &  &  &  & \vspace*{-0.3cm} \\
2 &  
 \widetilde{t} \ ( \ \left \lbrace \ \varrho \ \right \rbrace \ )
& 9 &  D^{\ 1 \ , \ \overline{1}} & D^{\ 2} \\
3 & \widetilde{t} \ ( \ \left \lbrace \ \pi \ \right \rbrace \ ) 
& 5 & - & D^{\ 2} \vspace*{-0.3cm} \\
 &  &  &  &  \vspace*{-0.3cm} \\
\hline \vspace*{-0.4cm} \\
2 & \widetilde{t} \ ( \ \left \lbrace \ w^{\ RR} \ \right \rbrace \ )
&  5 & D^{\ 2 \ , \ 0} & D^{\ 2} \\
2 & \widetilde{t} \ ( \ \left \lbrace \ w^{\ LL} \ \right \rbrace \ )
&  5 & D^{\ 0 \ , \ 2} & D^{\ 2}  \vspace*{-0.3cm} \\
 &  &  &  &   \\
\hline 
\end{array}
\end{array}
\end{equation}
\vspace*{0.2cm} 

\noindent
In the tables ( eqs. (\ref{eq:1039}) and (\ref{eq:1040}) )
the column labelled $\mbox{\# comp.}$ refers to wave function components
over the complex numbers.
\vspace*{0.1cm} 

\noindent
The entries and properties displayed in eq. (\ref{eq:1040})
look more coherent than in the tables 
in eqs. (\ref{eq:1034}) and (\ref{eq:1035}) , but the puzzle
of 5 versus 10 components for
$\widetilde{t} \ ( \ \left \lbrace \ \varrho \ \right \rbrace \ )$
compared to 
$\widetilde{t} \ ( \ \left \lbrace \ w^{\ RR} \ \right \rbrace \ )$
and
$\widetilde{t} \ ( \ \left \lbrace \ w^{\ LL} \ \right \rbrace \ )$
remains.

\noindent
To understand this difference we compare the structure of the
bilinears associated with 
$\widetilde{t} \ ( \ \left \lbrace \ \pi \ \right \rbrace \ )$
( eq. (\ref{eq:1022}) ) with the one pertaining to
$\widetilde{t} \ ( \ \left \lbrace \ w^{\ RR} \ \right \rbrace \ )$
and
$\widetilde{t} \ ( \ \left \lbrace \ w^{\ LL} \ \right \rbrace \ )$
( (\ref{eq:1032}) ) . To this end we use the relations
in eqs. (\ref{eq:1025}) defining the quantities $\vec{C}^{\ A}$
and (\ref{eq:1028}) for $\vec{G}^{\ A}$ respectively.

\begin{equation}
\label{eq:1041}
\begin{array}{l}
- \ \varrho^{\ \mu \ \nu} \ = 
\ \vartheta_{\ cl}^{\ \mu \ \nu} \ \rightarrow
\vspace*{0.3cm} \\
\pi_{\ i \ k} \ = 
\ \left (
\begin{array}{l}
\ \frac{1}{3} \ \delta_{\ i \ k}
\ \left (
\ \vec{E}^{\ A} \ \vec{E}^{\ D} 
\ + \ \vec{B}^{\ A} \ \vec{B}^{\ D} 
\ \right )
\vspace*{0.3cm} \\
\ - 
\ \left (
\ \vec{E}^{\ i \ A} \ \vec{E}^{\ k \ D} 
\ + \ \vec{B}^{\ i \ A} \ \vec{B}^{\ k \ D} 
\ \right )
\end{array}
\ \right )
\ U_{\ A \ D}
\vspace*{0.3cm} \\
\hline
\vspace*{-0.2cm} \\
P^{\ RR} \ \left ( \ \vec{C} \ \otimes \ \vec{C} \ \right )
\ \rightarrow
\vspace*{0.3cm} \\
\left (
\ \pi^{\ R \ R}
\ \right )^{\ r \ s}
\ =
\ \left (
\ C^{\ r \ A} \ C^{\ s \ D} \ - 
\ \frac{1}{3} \ \delta^{\ r s} \ \vec{C}^{\ A} \ \vec{C}^{\ D}
\ \right )
\ U_{\ A \ D}
\vspace*{0.3cm} \\
\hline
\vspace*{-0.2cm} \\
P^{\ LL} \ \left ( \ \vec{G} \ \otimes \ \vec{G} \ \right )
\ \rightarrow
\vspace*{0.3cm} \\
\left (
\ \pi^{\ L \ L}
\ \right )^{\ r \ s}
\ =
\ \left (
\ G^{\ r \ A} \ G^{\ s \ D} \ - 
\ \frac{1}{3} \ \delta^{\ r s} \ \vec{G}^{\ A} \ \vec{G}^{\ D}
\ \right )
\ U_{\ A \ D}
\vspace*{0.3cm} \\
\hline
\vspace*{-0.2cm} \\
\vec{C}^{ \ r \ A} \ ( \ x \ ) 
\ =
\ \left (
\ \vec{B} \ - \ i \ \vec{E}
\ \right )^{\ r \ A} \ ( \ x \ )
\vspace*{0.3cm} \\
\vec{G}^{ \ r \ A} \ ( \ x \ ) 
\ =
\ \left (
\ \vec{B} \ + \ i \ \vec{E}
\ \right )^{\ r \ A} \ ( \ x \ )
\vspace*{0.3cm} \\
\sum_{\ i} \ \pi_{\ i \ i} \ = \ 0
\hspace*{0.3cm} ; \hspace*{0.3cm}
\sum_{\ r} 
\ \left (
\ \pi^{\ R \ R}
\ \right )^{\ r \ r}
\ =
\ \sum_{\ r} 
\left (
\ \pi^{\ L \ L}
\ \right )^{\ r \ r}
\ = \ 0
\end{array}
\end{equation}

\noindent
In eq. (\ref{eq:1041}) the 21 complex components of the octet string bilinear
wave functions
are reduced to 3 complex, traceless and symmetric $3 \ \times \ 3$ matrices,
denoted $\pi$ ,  $\pi^{\ R \ R}$ and $\pi^{\ L \ L}$ respectively.
There is at this stage an essential ingredient missing.
The property distinguishing the above matrices is the 
spatial 'Dreibein' nature of chromoelectric- , chromomagnetic and orientation
axis vectors \cite{MF} .

\noindent
In order to realize the 'Dreibein' property, we reduce the
three quantities $\pi$ ,  $\pi^{\ R \ R}$ and $\pi^{\ L \ L}$ 
in eq. (\ref{eq:41}) to their common (chromo-) electric and magnetic components.
This leaves $\pi$ unchanged

\begin{equation}
\label{eq:1042}
\begin{array}{l}
\pi^{\ r \ s} \ = 
\ \left (
\begin{array}{l}
\ \frac{1}{3} \ \delta^{\ r \ s}
\ \left (
\ \vec{E}^{\ A} \ \vec{E}^{\ D} 
\ + \ \vec{B}^{\ A} \ \vec{B}^{\ D} 
\ \right )
\vspace*{0.3cm} \\
\ - 
\ \left (
\ \vec{E}^{\ r \ A} \ \vec{E}^{\ s \ D} 
\ + \ \vec{B}^{\ r \ A} \ \vec{B}^{\ s \ D} 
\ \right )
\end{array}
\ \right )
\ U_{\ A \ D}
\vspace*{0.3cm} \\
\hline
\vspace*{-0.2cm} \\
\left (
\ \pi^{\ R \ R}
\ \right )^{\ r \ s}
\ = \ a^{\ r \ s} \ - \ i \ b^{\ r \ s}
\vspace*{0.3cm} \\
\hline
\vspace*{-0.2cm} \\
\left (
\ \pi^{\ L \ L}
\ \right )^{\ r \ s}
\ = \ a^{\ r \ s} \ + \ i \ b^{\ r \ s}
\vspace*{0.3cm} \\
\hline
\vspace*{-0.2cm} \\
a^{\ r \ s} \ =
\ \left (
\begin{array}{l}
\ \frac{1}{3} \ \delta^{\ r \ s}
\ \left (
\ - \ \vec{E}^{\ A} \ \vec{E}^{\ D} 
\ + \ \vec{B}^{\ A} \ \vec{B}^{\ D} 
\ \right )
\vspace*{0.3cm} \\
\ - 
\ \left (
\ - \ \vec{E}^{\ r \ A} \ \vec{E}^{\ s \ D} 
\ + \ \vec{B}^{\ r \ A} \ \vec{B}^{\ s \ D} 
\ \right )
\end{array}
\ \right )
\ U_{\ A \ D}
\vspace*{0.3cm} \\
\hline
\vspace*{-0.2cm} \\
b^{\ r \ s} \ =
\ \left (
\begin{array}{l}
\ \frac{1}{3} \ \delta^{\ r \ s}
\ \left (
\ \vec{E}^{\ A} \ \vec{B}^{\ D} 
\ + \ \vec{B}^{\ A} \ \vec{E}^{\ D} 
\ \right )
\vspace*{0.3cm} \\
\ - 
\ \left (
\ \vec{E}^{\ r \ A} \ \vec{B}^{\ s \ D} 
\ + \ \vec{B}^{\ r \ A} \ \vec{E}^{\ s \ D} 
\ \right )
\end{array}
\ \right )
\ U_{\ A \ D}
\end{array}
\end{equation}

\noindent
The adjoint representation indices $A \ , \ D$ and the position
difference $z$ make it necessary to
symmetrize the above expressions with respect to the
indices $r \ , s$ taking into account the
dependence on the relative (Lorentz-) coordinate $z$ , not
explicitely shown in eq. (\ref{eq:1042}) .

\noindent
In order to retain the relevant degrees of freedom we streamline
the displayed indices and kinematic variables to the electromagnetic
case. But in no way is this implying, that
the nonabelian character of the underlying
variables is sacrificed, to the contrary.

\noindent
With this in mind we introduce the abbreviating notation

\begin{equation}
\label{eq:1043}
\begin{array}{l}
\left (
\ \vec{E}^{\ A} \ , \ \vec{B}^{\ A}
\ \right )
\ \rightarrow 
\ \left (
\ \vec{E} \ , \ \vec{B}
\ \right ) \ ( \hspace*{0.3cm} \zeta \hspace*{0.5cm} )
\ \rightarrow 
\ \left (
\ \vec{E} \ , \ \vec{B}
\ \right )_{\ +} 
\vspace*{0.3cm} \\
\left (
\ \vec{E}^{\ D} \ , \ \vec{B}^{\ D}
\ \right )
\ \rightarrow 
\ \left (
\ \vec{E} \ , \ \vec{B}
\ \right ) \ ( \ - \ \zeta \ )
\ \rightarrow 
\ \left (
\ \vec{E} \ , \ \vec{B}
\ \right )_{\ -} 
\vspace*{0.3cm} \\
U_{\ A \ D} \ \rightarrow \ .
\hspace*{0.3cm} ; \hspace*{0.3cm}
\zeta \ = \ z \ / \ 2
\end{array}
\end{equation}

\noindent
The trace parts proportional to $\delta^{\ r \ s}$
of the matrix $\pi$ in eq. (\ref{eq:1042})
can be neglected. This follows from 
eq. (\ref{eq:1037}) . 

\noindent
Thus eq. (\ref{eq:1042}) takes the form using the notation introduced in 
eq.  (\ref{eq:1043})

\begin{equation}
\label{eq:1044}
\begin{array}{l}
\begin{array}{rlr}
- \ \pi^{\ r \ s} 
& \sim &
\vec{E}_{\ +}^{\ r } \ \vec{E}_{\ -}^{\ s} 
\ + \ \vec{B}_{\ +}^{\ r} \ \vec{B}_{\ -}^{\ s} 
\vspace*{0.3cm} \\
a^{\ r \ s} 
& \sim &
\vec{E}_{\ +}^{\ r} \ \vec{E}_{\ -}^{\ s} 
\ - \ \vec{B}_{\ +}^{\ r} \ \vec{B}_{\ -}^{\ s} 
\vspace*{0.3cm} \\
- \ b^{\ r \ s} 
& \sim &
\vec{E}_{\ +}^{\ r} \ \vec{B}_{\ -}^{\ s} 
\ + \ \vec{B}_{\ +}^{\ r} \ \vec{E}_{\ -}^{\ s} 
\end{array}
\ + \ ( \ r \ \leftrightarrow \ s \ )
\vspace*{0.3cm} \\
\mbox{modulo trace parts}
\end{array}
\end{equation}

\noindent
Here we need the variable $\vec{e} \ = \ \vec{z} \ / \ r$
introduced in eq. (\ref{eq:139}) in order to orient
chromoelectric and -magnetic fields, in a radial gauge, where
the parallel transport matrix $U_{\ A \ D} \ \rightarrow \ \delta_{\ A \ D}$ .

\noindent
The chromomagnetic fields then become related to the -electric ones

\begin{equation}
\label{eq:1045}
\begin{array}{l}
\vec{B}_{\ +} \ = \ \vec{e} \ \wedge \ \vec{E}_{\ +}
\hspace*{0.3cm} ; \hspace*{0.3cm}
\vec{B}_{\ -} \ = \ - \ \vec{e} \ \wedge \ \vec{E}_{\ -} 
\vspace*{0.3cm} \\
\vec{e} \ \vec{E}_{\ \pm} 
\ = \ \vec{e} \ \vec{B}_{\ \pm} \ = \ 0
\end{array}
\end{equation}

\noindent
Eq. (\ref{eq:1045}) establishes the nonabelian 'Dreibein' form.

\noindent
As the wave functions associated with the field strengths in eqs.
(\ref{eq:1045}) and (\ref{eq:1045}) are complex, we can choose 
the $\vec{e}$ associated helicity basis. Choosing the $z$ axis along $\vec{e}$
eq. (\ref{eq:1045}) takes the component form

\begin{equation}
\label{eq:1046}
\begin{array}{l}
\vec{E}_{\ \pm} \ = 
\ \left ( \ E_{\ x \ \pm} \ , \  E_{\ y \ \pm} \ , \ 0 \ \right )
\vspace*{0.3cm} \\
\vec{B}_{\ +} \ = 
\ \left ( \ - \ E_{\ y \ +} \ , \ E_{\ x \ +} \ , \ 0 \ \right )
\hspace*{0.3cm} ; \hspace*{0.3cm}
\vec{B}_{\ -} \ = 
\ \left ( \ E_{\ y \ -} \ , \ - \ E_{\ x \ -} \ , \ 0 \ \right )
\ \rightarrow
\vspace*{0.3cm} \\
\left ( \ {\cal{E}} \ , \ {\cal{B}} \ \right )^{\ R}_{\ +} \ =  
\left ( 
\ E \ , \ B
\ \right )_{\ x \ +} \ - \ i 
\left ( 
\  E \ , \ B 
\ \right )_{\ y \ +} 
\vspace*{0.3cm} \\
\left ( \ {\cal{E}} \ , \ {\cal{B}} \ \right )^{\ L}_{\ +} \ =  
\left ( 
\ E \ , \ B
\ \right )_{\ x \ +} \ + \ i 
\left ( 
\  E \ , \ B 
\ \right )_{\ y \ +} 
\vspace*{0.3cm} \\
\left ( \ {\cal{E}} \ , \ {\cal{B}} \ \right )^{\ R}_{\ -} \ =  
\left ( 
\ E \ , \ B
\ \right )_{\ x \ -} \ + \ i 
\left ( 
\  E \ , \ B 
\ \right )_{\ y \ -} 
\vspace*{0.3cm} \\
\left ( \ {\cal{E}} \ , \ {\cal{B}} \ \right )^{\ L}_{\ -} \ =  
\left ( 
\ E \ , \ B
\ \right )_{\ x \ -} \ - \ i 
\left ( 
\  E \ , \ B 
\ \right )_{\ y \ -} 
\end{array}
\end{equation}

\noindent
The (complex) components 
$\left ( \ {\cal{E}} \ , \ {\cal{B}} \ \right )^{\ R \ (L)}_{\ \pm}$  
in eq. (\ref{eq:1046}) \footnote{The helicity basis components
${\cal{B}}$ shall not be confused with the SL2C matrix
${\cal{B}}$ defined in appendix A.1 .}
define the sought helicity basis,
where eq. (\ref{eq:1046}) takes the form

\begin{equation}
\label{eq:1047}
\begin{array}{l}
{\cal{B}}^{\ R}_{\ \pm} \ =  
\ \left ( \  - \ i \ \right )
\ {\cal{E}}^{\ R}_{\ \pm}   
\hspace*{0.3cm} ; \hspace*{0.3cm}
{\cal{B}}^{\ L}_{\ \pm} \ =  
\ \left ( \ i \ \right )
\ {\cal{E}}^{\ L}_{\ \pm}   
\end{array}
\end{equation}

\noindent
The spin component associated with the operation

\begin{equation}
\label{eq:1048}
\begin{array}{l}
\widehat{S}_{\ \vec{e}} \ = \ i \ \vec{e} \ \wedge \ .
\end{array}
\end{equation}

\noindent
which represents an infinitesimal rotation around
the $\vec{e} \ -$ axis (i.e. its derivative with respect to the
rotation angle),
takes the eigenvalues implied by the $R \ , \ L$ components
in eq. (\ref{eq:1047})

\begin{equation}
\label{eq:1049}
\begin{array}{l}
\widehat{S}_{\ \vec{e}} 
\ \left ( 
\ {\cal{E}}^{\ R}_{\ +}   
\ , \ {\cal{B}}^{\ R}_{\ +}   
\ \right ) \ = \ + \ 1
\hspace*{0.3cm} , \hspace*{0.3cm}
\widehat{S}_{\ \vec{e}} 
\ \left ( 
\ {\cal{E}}^{\ L}_{\ -}   
\ , \ {\cal{B}}^{\ L}_{\ -}   
\ \right ) \ = \ + \ 1
\vspace*{0.3cm} \\
\widehat{S}_{\ \vec{e}} 
\ \left ( 
\ {\cal{E}}^{\ L}_{\ +}   
\ , \ {\cal{B}}^{\ L}_{\ +}   
\ \right ) \ = \ - \ 1
\hspace*{0.3cm} , \hspace*{0.3cm}
\widehat{S}_{\ \vec{e}} 
\ \left ( 
\ {\cal{E}}^{\ R}_{\ -}   
\ , \ {\cal{B}}^{\ R}_{\ -}   
\ \right ) \ = \ - \ 1
\end{array}
\end{equation}

\noindent
Hence the direct product components $R_{\ +} \ R_{\ -}$ ,
$R_{\ +} \ L_{\ -}$ , $L_{\ +} \ R_{\ -}$ and  $L_{\ +} \ L_{\ -}$
describe four $S_{\ \vec{e} \ 12}$ spin states

\begin{equation}
\label{eq:1050}
\begin{array}{l}
\begin{array}{clr}
\Delta \ \mbox{helicity components} & & S_{\ \vec{e} \ 12}
\vspace*{0.3cm} \\
\hline
\vspace*{-0.2cm} \\
\ \left ( 
\ {\cal{E}}^{\ R}_{\ +}   
\ , \ {\cal{B}}^{\ R}_{\ +}   
\ \right ) 
\ \otimes
\ \left ( 
\ {\cal{E}}^{\ R}_{\ -}   
\ , \ {\cal{B}}^{\ R}_{\ -}   
\ \right ) 
 &  & 0
\vspace*{0.3cm} \\
\ \left ( 
\ {\cal{E}}^{\ L}_{\ +}   
\ , \ {\cal{B}}^{\ L}_{\ +}   
\ \right ) 
\ \otimes
\ \left ( 
\ {\cal{E}}^{\ L}_{\ -}   
\ , \ {\cal{B}}^{\ L}_{\ -}   
\ \right ) 
 &  & 0 
\vspace*{0.3cm} \\
\ \left ( 
\ {\cal{E}}^{\ R}_{\ +}   
\ , \ {\cal{B}}^{\ R}_{\ +}   
\ \right ) 
\ \otimes
\ \left ( 
\ {\cal{E}}^{\ L}_{\ -}   
\ , \ {\cal{B}}^{\ L}_{\ -}   
\ \right ) 
 &  & 2
\vspace*{0.3cm} \\
\ \left ( 
\ {\cal{E}}^{\ L}_{\ +}   
\ , \ {\cal{B}}^{\ L}_{\ +}   
\ \right ) 
\ \otimes
\ \left ( 
\ {\cal{E}}^{\ R}_{\ -}   
\ , \ {\cal{B}}^{\ R}_{\ -}   
\ \right ) 
 &  & - \ 2
\end{array}
\end{array}
\end{equation}

\noindent
For clarity let us associate a pair of complex, transverse
three vectors $\left ( \vec{v}  ,  \vec{w}  \right )$
\footnote{The auxiliary vector $\vec{w}$ introduced here shall not be
confused with the Weyl bilinears $w_{\ \underline{.}}^{\ RR}$
and  $w_{\ \underline{.}}^{\ RR}$ in eqs. (\ref{eq:1031}) and
(\ref{eq:1033}) .}
with the two sides of the octet string denoted by $+$ and $-$ 

\begin{equation}
\label{eq:1051}
\begin{array}{l}
\begin{array}{lll}
\vec{v}
\hspace*{0.3cm} \leftrightarrow \hspace*{0.3cm}
\ \left ( 
\ \vec{E}   
\ , \ \vec{B}   
\ \right )_{\ +} 
& , &
\vec{w}
\hspace*{0.3cm} \leftrightarrow \hspace*{0.3cm}
\ \left ( 
\ \vec{E}   
\ , \ \vec{B}   
\ \right )_{\ -} 
\vspace*{0.3cm} \\
\vec{e} \ \vec{v} \ = \ 0
& , &
\vec{e} \ \vec{w} \ = \ 0
\vspace*{0.3cm} \\
v^{\ R} \ = \ v_{\ x} \ - \ i \ v_{\ y}
& , & w^{\ R} \ = \ w_{\ x} \ + \ i \ w_{\ y}
\vspace*{0.3cm} \\
v^{\ L} \ = \ v_{\ x} \ + \ i \ v_{\ y}
& , & w^{\ L} \ = \ w_{\ x} \ - \ i \ w_{\ y}
\end{array}
\end{array}
\end{equation}

\noindent
Using the components $R \ , \ L$ as defined
in eqs. (\ref{eq:1046}) and (\ref{eq:1051}) the (complex orthogonal) 
scalar product takes the form

\begin{equation}
\label{eq:1052}
\begin{array}{l}
u \ . \ v \ = \ u_{\ x} \ v_{\ x} \ + \ \ u_{\ y} \ v_{\ y}
\ = \frac{1}{2} 
\ \left (
\ u^{\ R} \ w^{\ R} \ + \  u^{\ L} \ w^{\ L}
\ \right )
\end{array}
\end{equation}

\noindent
We adapt the tensor structure of the symmetric matrices
$\pi \ , \ a \ , b$ in eq. (\ref{eq:1044}) to the $R \ R$ ,
$R \ L$ , $L \ R$ and $L \ L$ basis defined in eq. (\ref{eq:1046})
\vspace*{-0.3cm} 

\begin{equation}
\label{eq:1053}
\begin{array}{l}
\pi^{\ \widetilde{r} \ \widetilde{s}} \ \sim
\ \left (
\begin{array}{ll}
\pi^{\ RR} & \pi^{\ RL}
\vspace*{0.3cm} \\
\pi^{\ LR} & \pi^{\ LL}
\end{array}
\ \right ) \ + 
\ ( \ r \ \leftrightarrow \ s \ )
\hspace*{0.3cm} ; \hspace*{0.3cm}
\hspace*{0.3cm} \mbox{and} \hspace*{0.3cm}
\pi \ \rightarrow \ a \ , \ b
\vspace*{0.3cm} \\
\mbox{modulo trace parts}
\end{array}
\end{equation}

\noindent
It follows from eqs. (\ref{eq:1052}) and (\ref{eq:1053})
that the trace part remains valid in the specific 
$\widetilde{r} \ \widetilde{s}$ assignment chosen. 
However symmetrization with respect to the indices $r \ s$ is
not equivalent to symmetrization with
respect to $\widetilde{r} \ \widetilde{s}$ 
\vspace*{-0.3cm} 

\begin{equation}
\label{eq:1054}
\begin{array}{l}
\left ( 
\ r \ \leftrightarrow \ s 
\ \right )
\ \equiv
\ \left ( 
\begin{array}{c}
 \pi^{\ RR} \ \leftrightarrow \ \pi^{\ LL}
\vspace*{0.3cm} \\
\pi^{\ RL} \ \rightarrow \ \pi^{\ RL}
\hspace*{0.3cm} \mbox{and} \hspace*{0.3cm}
\pi^{\ LR} \ \rightarrow \ \pi^{\ LR}
\end{array}
\ \right )
\vspace*{0.3cm} \\
\mbox{modulo trace parts}
\hspace*{0.3cm} \mbox{and} \hspace*{0.3cm}
\pi \ \rightarrow \ a \ , \ b
\end{array}
\end{equation}

\noindent
Eq. (\ref{eq:1054}) implies for the symmetrized matrices 
$\pi \ , \ a \ , \ b$ 
\vspace*{-0.3cm} 

\begin{equation}
\label{eq:1055}
\begin{array}{l}
\pi^{\ \widetilde{r} \ \widetilde{s}} \ \sim
\ \left (
\begin{array}{cc}
\frac{1}{2} \ \left (
\ \pi^{\ RR} \ + \ \pi^{\ LL} 
\ \right ) & \pi^{\ RL}
\vspace*{0.3cm} \\
\pi^{\ LR} & 
\frac{1}{2} \ \left (
\ \pi^{\ RR} \ + \ \pi^{\ LL} 
\ \right )
\end{array}
\ \right )  
\vspace*{0.3cm} \\
\mbox{modulo trace parts}
\hspace*{0.3cm} \mbox{and} \hspace*{0.3cm}
\pi \ \rightarrow \ a \ , \ b
\end{array}
\end{equation}

\noindent
Now taking the traceless parts simply removes the $R \ R$ and 
$L \ L$ components
\vspace*{-0.2cm} 

\begin{equation}
\label{eq:1056}
\begin{array}{l}
\pi^{\ \widetilde{r} \ \widetilde{s}} \ \sim
\ \left (
\begin{array}{cc}
0 & \pi^{\ RL}
\vspace*{0.3cm} \\
\pi^{\ LR} & 0
\end{array}
\ \right )  
\hspace*{0.3cm} \mbox{and} \hspace*{0.3cm}
\pi \ \rightarrow \ a \ , \ b
\end{array}
\end{equation}

\newpage

\noindent
Now we cast $\pi \ , \ a \ , \ b$ in eq. (\ref{eq:1044})
into the $\widetilde{r} \ \widetilde{s}$ basis
\vspace*{-0.3cm} 

\begin{equation}
\label{eq:1057}
\begin{array}{l}
\begin{array}{rll rrll}
-  \pi^{\ R \ L} 
& \sim &
{\cal{E}}_{\ +}^{\ R } \ {\cal{E}}_{\ -}^{\ L} 
\ + \ {\cal{B}}_{\ +}^{\ R} \ {\cal{B}}_{\ -}^{\ L} 
& , &
-  \pi^{\ L \ R} 
& \sim &
\ {\cal{E}}_{\ +}^{\ L } \ {\cal{E}}_{\ -}^{\ R} 
\ + \ {\cal{B}}_{\ +}^{\ L} \ {\cal{B}}_{\ -}^{\ R} 
\vspace*{0.3cm} \\
a^{\ R \ L} 
& \sim &
\ {\cal{E}}_{\ +}^{\ R} \ {\cal{E}}_{\ -}^{\ L} 
\ - \ {\cal{B}}_{\ +}^{\ R} \ {\cal{B}}_{\ -}^{\ L} 
& , &
a^{\ L \ R} 
& \sim &
 {\cal{E}}_{\ +}^{\ L} \ {\cal{E}}_{\ -}^{\ R} 
\ - \ {\cal{B}}_{\ +}^{\ L} \ {\cal{B}}_{\ -}^{\ R} 
\vspace*{0.3cm} \\
-  b^{\ R \ L} 
& \sim &
 {\cal{E}}_{\ +}^{\ R} \ {\cal{B}}_{\ -}^{\ L} 
\ + \ {\cal{B}}_{\ +}^{\ R} \ {\cal{E}}_{\ -}^{\ L} 
& , &
-  b^{\ L \ R} 
& \sim &
\ {\cal{E}}_{\ +}^{\ L} \ {\cal{B}}_{\ -}^{\ R} 
\ + \ {\cal{B}}_{\ +}^{\ L} \ {\cal{E}}_{\ -}^{\ R} 
\end{array}
\vspace*{0.2cm} 
\end{array}
\end{equation}

\noindent
We substitute the relations in eq. (\ref{eq:1047})
in eq. (\ref{eq:1057})

\begin{equation}
\label{eq:1058}
\begin{array}{l}
{\cal{B}}^{\ R}_{\ \pm} \ =  
\ \left ( \  - \ i \ \right )
\ {\cal{E}}^{\ R}_{\ \pm}   
\hspace*{0.3cm} ; \hspace*{0.3cm}
{\cal{B}}^{\ L}_{\ \pm} \ =  
\ \left ( \ i \ \right )
\ {\cal{E}}^{\ L}_{\ \pm}   
\ \rightarrow
\end{array}
\end{equation}

\noindent
with the result that {\it all contributions from the
spin 2 Weyl bilinear vanish} , as a consequence of the 'Dreibein' conditions

\begin{equation}
\label{eq:1059}
\begin{array}{l}
- \ \pi^{\ R \ L} 
\ \sim 
\ 2 \ {\cal{E}}_{\ +}^{\ R } \ {\cal{E}}_{\ -}^{\ L} 
\hspace*{0.3cm} , \hspace*{0.3cm}
- \ \pi^{\ L \ R} 
\ \sim 
\ 2 \ {\cal{E}}_{\ +}^{\ L } \ {\cal{E}}_{\ -}^{\ R} 
\vspace*{0.3cm} \\
\hline
\vspace*{-0.2cm} \\
\left (
\ a^{\ R \ L} \ , \ a^{\ L \ R} 
\ \right )
\ \sim \ 0
\hspace*{0.3cm} , \hspace*{0.3cm}
\left (
\ b^{\ R \ L} \ , \ b^{\ L \ R} 
\ \right )
\ \sim \ 0
\end{array}
\end{equation}

\noindent
Comparing with the helicity structure in eq. (\ref{eq:1050})
we find 

\begin{equation}
\label{eq:1060}
\begin{array}{l}
\begin{array}{clr}
\Delta \ \mbox{helicity components} & & S_{\ \vec{e} \ 12}
\vspace*{0.3cm} \\
\hline
\vspace*{-0.2cm} \\
- \ \frac{1}{2} \ \pi^{\ R \ L} 
\ =
\ {\cal{E}}_{\ +}^{\ R } \ {\cal{E}}_{\ -}^{\ L} 
 &  & 2
\vspace*{0.3cm} \\
- \ \frac{1}{2} \ \pi^{\ L \ R} 
\ =
\ {\cal{E}}_{\ +}^{\ L } \ {\cal{E}}_{\ -}^{\ R} 
 &  & - \ 2 
\vspace*{0.3cm} \\
a \ = \ b \ = \ 0 & & 0
\end{array}
\end{array}
\end{equation}

\noindent
{\it Summary remarks on the construction of the $II^{\ +}$
spectral gb series}
\vspace*{0.1cm} 

\begin{description}
\item i) Decomposition of adjoint string bilinears and their
associated wave functions

We reproduce here the structure of the adjoint string bilinears
introduced in eq. (\ref{eq:101})

\begin{equation}
\label{eq:1061}
\begin{array}{l}
B_{\ \left \lbrack \ \mu_{1} \ \nu_{1} \  \right \rbrack
\ , \ \left \lbrack  \ \mu_{2} \ \nu_{2}  \ \right \rbrack}
\ ( \ x_{\ 1} \ , \ x_{\ 2} \ )
\ =
\vspace*{0.3cm} \\
\hspace*{0.5cm}
F_{\ \left \lbrack \ \mu_{1} \ \nu_{1} \ \right \rbrack}
\ ( \ x_{\ 1} \ ; \ A \ )
\ U \ ( \ x_{\ 1} \ , \ A \ ; \ x_{\ 2} \ , \ B \ )
\ F_{\ \left \lbrack \ \mu_{2} \ \nu_{2} \ \right \rbrack}
\ ( \ x_{\ 2} \ ; \ B \ )
\vspace*{0.3cm} \\
A \ , \ B \ , \cdots \ = \ 1, \cdots , 8
\end{array}
\end{equation}

The bilinear quantities in eq. (\ref{eq:1061}) yield the
gb wave functions in the three spectral series
$I^{\ +}$ , $I^{\ -}$ and $II^{\ +}$ introduced in
eqs. (\ref{eq:108}) - (\ref{eq:123}) summarized in eq. (\ref{eq:1062})
below

\begin{equation}
\label{eq:1062}
\begin{array}{l}
\widetilde{t}_{\ \underline{.}} 
\ ( \ z \ , \ p \ , \  J^{\ P \ +} \ ; \ . \ )   
\ \rightarrow 
\ \widetilde{t}_{ \ 
\underline{.} \ ; \ S_{\ 12}^{\ \pm}} 
\ ( \ z \ , \ p \ , \ J^{\ \pm \ +} \  ; \ . \ )
\vspace*{0.3cm} \\
\hline
\vspace*{-0.2cm} \\
\widetilde{t}_{\ \underline{.} \ ; \ S_{\ 12}^{\ \pm}} 
\ ( \ z \ , \ p \ , \ J^{\ \pm \ +} \  ; \ . \ )
\begin{array}{ll}
\nearrow & 
\widetilde{t}_{\ \underline{.} \ ; \ II^{\ +}} 
\ ( \ z \ , \ p \ , \ J^{\ + \ +} \  ; \ . \ )
\vspace*{0.3cm} \\
\rightarrow & 
\widetilde{t}_{\ \underline{.} \ , \ I^{\ +}} 
\ ( \ z \ , \ p \ , \ J^{\ + \ +} \  ; \ . \ )
\vspace*{0.3cm} \\
\searrow & 
\widetilde{t}_{\ \underline{.} \ ; \ I^{\ -}} 
\ ( \ z \ , \ p \ , \ J^{\ - \ +} \  ; \ . \ )
\end{array}
\vspace*{0.3cm} \\
\underline{.} \ ; S_{\ 12}^{\ \pm} \ \rightarrow \ \underline{.}
\ \rightarrow 
\ \left \lbrack
\ \mu_{\ 1} \ \nu_{\ 1}
\ \right \rbrack
\ , 
\ \left \lbrack
\ \mu_{\ 2} \ \nu_{\ 2}
\ \right \rbrack
\end{array}
\end{equation}

The decomposition of the adjoint string bilinears is introduced in
eq. (\ref{eq:124}) repeated below
\vspace*{-0.3cm} 

\begin{equation}
\label{eq:1063}
\begin{array}{l}
\hspace*{-1.0cm}
B_{\ \left \lbrack \ \mu_{1} \ \nu_{1} \  \right \rbrack
\ , \ \left \lbrack  \ \mu_{2} \ \nu_{2}  \ \right \rbrack}
\ ( \ x_{\ 1} \ , \ x_{\ 2} \ )
\ =
 \left (
 \hspace*{-0.1cm}
 \begin{array}{r}
 \left (
\ K^{\ +}
\ \right )_{\ 
\ \left \lbrack
\ \mu_{\ 1} \ \nu_{\ 1}
\ \right \rbrack
\ \left \lbrack
\ \mu_{\ 2} \ \nu_{\ 2}
\ \right \rbrack}
\ B^{\ (+)}
\vspace*{0.3cm} \\
+
\ \left (
\ K^{\ -}
\ \right )_{\ 
\ \left \lbrack
\ \mu_{\ 1} \ \nu_{\ 1}
\ \right \rbrack
\ \left \lbrack
\ \mu_{\ 2} \ \nu_{\ 2}
\ \right \rbrack}
\ B^{\ (-)}
\vspace*{0.3cm} \\
 +
\ B^{\ '}_{\ \left \lbrack \ \mu_{1} \ \nu_{1} \  \right \rbrack
\ , \ \ \left \lbrack  \ \mu_{2} \ \nu_{2}  \ \right \rbrack}
\end{array}
 \right )
\vspace*{0.5cm} \\
\ \left (
\ K^{\ +}
\ \right )_{\ 
\ \left \lbrack
\ \mu_{\ 1} \ \nu_{\ 1}
\ \right \rbrack
\ \left \lbrack
\ \mu_{\ 2} \ \nu_{\ 2}
\ \right \rbrack}
\ =
\ g_{\ \mu_{\ 1} \ \mu_{\ 2}}
\ g_{\ \nu_{\ 1} \ \nu_{\ 2}}
\ -
\ g_{\ \mu_{\ 1} \ \nu_{\ 2}}
\ g_{\ \mu_{\ 2} \ \nu_{\ 1}}
\vspace*{0.5cm} \\
\ \left (
\ K^{\ -}
\ \right )_{\ 
\ \left \lbrack
\ \mu_{\ 1} \ \nu_{\ 1}
\ \right \rbrack
\ \left \lbrack
\ \mu_{\ 2} \ \nu_{\ 2}
\ \right \rbrack}
\ =
\ \varepsilon_{\ \mu_{\ 1} \ \mu_{\ 2} \ \nu_{\ 1} \ \nu_{\ 2}}
\vspace*{0.2cm} 
\end{array}
\end{equation}

In eq. (\ref{eq:1063}) $K^{\ \pm}$ and the associated
bilinears and their induced amplitudes denoted
$B{\ (\pm})$ project on the spectral types $I^{\ \pm}$ according
to eq. (\ref{eq:1062}) , whereas 
$B^{\ '}_{\ \left \lbrack \ \mu_{1} \ \nu_{1} \  \right \rbrack
\ , \ \ \left \lbrack  \ \mu_{2} \ \nu_{2}  \ \right \rbrack}$
refers to the projection on the $II^{\ +}$ spectral type.
iThe unique projection of $B^{\ '}$ on the traceless part
of the Ricci bilinear identical to the
classical energy momentum tensor pertaining to gauge bosons
is derived in this appendix (A.5) .

The general decomposition of 
$B_{\ \left \lbrack \ \mu_{1} \ \nu_{1} \  \right \rbrack
\ , \ \left \lbrack  \ \mu_{2} \ \nu_{2}  \ \right \rbrack}$
in eq. (\ref{eq:1061}) is introduced in eq. (\ref{eq:1016})
reproduced below
\vspace*{-0.3cm} 

\begin{equation}
\label{eq:1064}
\begin{array}{l}
\hspace*{-1.0cm}
B_{\ \left \lbrack \ \mu_{1} \ \nu_{1} \  \right \rbrack
\ , \ \left \lbrack  \ \mu_{2} \ \nu_{2}  \ \right \rbrack}
\ = \ w_{\ \left \lbrack \ \mu_{1} \ \nu_{1} \  \right \rbrack
\ \left \lbrack  \ \mu_{2} \ \nu_{2}  \ \right \rbrack}
+ \Delta 
\ B_{\ \left \lbrack \ \mu_{1} \ \nu_{1} \  \right \rbrack
\ , \ \left \lbrack  \ \mu_{2} \ \nu_{2}  \ \right \rbrack}
\vspace*{0.3cm} \\
\hspace*{-1.0cm}
\Delta 
\ B_{\ \left \lbrack \ \mu_{1} \ \nu_{1} \  \right \rbrack
\ , \ \left \lbrack  \ \mu_{2} \ \nu_{2}  \ \right \rbrack}
\ = 
\begin{array}{l}
\ \frac{1}{2}
 \left (
\begin{array}{l}
 g_{\ \mu_{\ 1} \ \mu_{\ 2}} \ R_{\ \nu_{\ 1} \ \nu_{\ 2}}
- \  g_{\ \nu_{\ 1} \ \mu_{\ 2}} \ R_{\ \mu_{\ 1} \ \nu_{\ 2}}
\vspace*{0.3cm} \\
- \  g_{\ \mu_{\ 1} \ \nu_{\ 2}} \ R_{\ \nu_{\ 1} \ \mu_{\ 2}}
\ + \ g_{\ \nu_{\ 1} \ \nu_{\ 2}} \ R_{\ \mu_{\ 1} \ \mu_{\ 2}}
\end{array}
 \right )
\vspace*{0.3cm} \\
- 
\ \frac{1}{6}
\ K^{\ +}_{\ \left \lbrack \ \mu_{1} \ \nu_{1} \  \right \rbrack
  \ \left \lbrack  \ \mu_{2} \ \nu_{2}  \ \right \rbrack}
\ R
\end{array}
\vspace*{0.4cm} \\
\hline
\vspace*{-0.3cm} \\
\hspace*{-1.0cm}
\mbox{with :}
\hspace*{0.3cm}
g^{\ \mu_{\ 1} \ \mu_{\ 2}}
\ w_{\ \left \lbrack \ \mu_{1} \ \nu_{1} \  \right \rbrack
\ \left \lbrack  \ \mu_{2} \ \nu_{2}  \ \right \rbrack}
\ = \ 0
\vspace*{0.5cm} \\
\hspace*{-1.0cm}
g^{\ \mu_{\ 1} \ \mu_{\ 2}}
\ B_{\ \left \lbrack \ \mu_{1} \ \nu_{1} \  \right \rbrack
\ \ , \ \left \lbrack  \ \mu_{2} \ \nu_{2}  \ \right \rbrack}
\ =
\ g^{\ \mu_{\ 1} \ \mu_{\ 2}}
\ \Delta 
\ B_{\ \left \lbrack \ \mu_{1} \ \nu_{1} \  \right \rbrack
\ , \ \left \lbrack  \ \mu_{2} \ \nu_{2}  \ \right \rbrack}
\ = \ R_{\ \nu_{\ 1} \ \nu_{\ 2}}
\vspace*{0.5cm} \\
\hspace*{-1.0cm}
R \ =
\ g^{\ \nu_{\ 1} \ \nu_{\ 2}}
\ R_{\ \nu_{\ 1} \ \nu_{\ 2}}
\ =
\ 12 \ B^{\ (+)}
\end{array}
\end{equation}

In eq. (\ref{eq:1064}) we have denoted the still reducible parts
in the following way
\vspace*{-0.3cm} 

\begin{equation}
\label{eq:1065}
\begin{array}{l}
\begin{array}{lll}
w_{\ \left \lbrack \ \mu_{1} \ \nu_{1} \  \right \rbrack
\ \left \lbrack  \ \mu_{2} \ \nu_{2}  \ \right \rbrack}
& : &
\mbox{Weyl bilinear}
\vspace*{0.3cm} \\
R_{\ \mu \ \nu}
& : &
\mbox{Ricci bilinear}
\vspace*{0.3cm} \\
R
& : &
\mbox{Riemann scalar bilinear}
\end{array}
\end{array}
\end{equation}

The decomposition in eq. (\ref{eq:1064}) into positive parity
irreducible parts, leaving the Weyl bilinear reducible,
is introduced in eq. (\ref{eq:1020}) repeated below
\vspace*{-0.5cm} 

\begin{equation}
\label{eq:1066}
\begin{array}{l}
B_{\ \left \lbrack \ \mu_{1} \ \nu_{1} \  \right \rbrack
\ , \ \left \lbrack  \ \mu_{2} \ \nu_{2}  \ \right \rbrack}
\ = \ w_{\ \left \lbrack \ \mu_{1} \ \nu_{1} \  \right \rbrack
\ \left \lbrack  \ \mu_{2} \ \nu_{2}  \ \right \rbrack}
+ \Delta 
\ B_{\ \left \lbrack \ \mu_{1} \ \nu_{1} \  \right \rbrack
\ , \ \left \lbrack  \ \mu_{2} \ \nu_{2}  \ \right \rbrack}
\vspace*{0.3cm} \\
\hspace*{-1.0cm}
\Delta 
\ B_{\ \left \lbrack \ \mu_{1} \ \nu_{1} \  \right \rbrack
\ , \ \left \lbrack  \ \mu_{2} \ \nu_{2}  \ \right \rbrack}
\ = 
\begin{array}{l}
\ \frac{1}{2}
 \left (
\begin{array}{l}
 g_{\ \mu_{\ 1} \ \mu_{\ 2}} \ \varrho_{\ \nu_{\ 1} \ \nu_{\ 2}}
- \  g_{\ \nu_{\ 1} \ \mu_{\ 2}} \ \varrho_{\ \mu_{\ 1} \ \nu_{\ 2}}
\vspace*{0.3cm} \\
- \  g_{\ \mu_{\ 1} \ \nu_{\ 2}} \ \varrho_{\ \nu_{\ 1} \ \mu_{\ 2}}
\ + \ g_{\ \nu_{\ 1} \ \nu_{\ 2}} \ \varrho_{\ \mu_{\ 1} \ \mu_{\ 2}}
\end{array}
 \right )
\vspace*{0.3cm} \\
+ 
\ \frac{1}{12}
\ K^{\ +}_{\ \left \lbrack \ \mu_{1} \ \nu_{1} \  \right \rbrack
  \ \left \lbrack  \ \mu_{2} \ \nu_{2}  \ \right \rbrack}
\ R
\end{array}
\vspace*{0.4cm} \\
\hline
\vspace*{-0.3cm} \\
\mbox{with :}
\hspace*{0.3cm}
g^{\ \mu_{\ 1} \ \mu_{\ 2}}
\ w_{\ \left \lbrack \ \mu_{1} \ \nu_{1} \  \right \rbrack
\ \left \lbrack  \ \mu_{2} \ \nu_{2}  \ \right \rbrack}
\ = \ 0
\vspace*{0.5cm} \\
\hspace*{-1.0cm}
g^{\ \mu_{\ 1} \ \mu_{\ 2}}
\ B_{\ \left \lbrack \ \mu_{1} \ \nu_{1} \  \right \rbrack
\ , \ \left \lbrack  \ \mu_{2} \ \nu_{2}  \ \right \rbrack}
\ =
\ g^{\ \mu_{\ 1} \ \mu_{\ 2}}
\ \Delta 
\ B_{\ \left \lbrack \ \mu_{1} \ \nu_{1} \  \right \rbrack
\ , \ \left \lbrack  \ \mu_{2} \ \nu_{2}  \ \right \rbrack}
\ = \ R_{\ \nu_{\ 1} \ \nu_{\ 2}}
\vspace*{0.5cm} \\
R \ =
\ g^{\ \nu_{\ 1} \ \nu_{\ 2}}
\ R_{\ \nu_{\ 1} \ \nu_{\ 2}}
\ =
\ 12 \ B^{\ (+)}
\end{array}
\end{equation}

In eq. (\ref{eq:1066}) the irreducible positive parity parts introduce
the traceless part of the Ricci bilinear, i.e. the
classical traceless energy momentum bilinear pertaining to
gauge bosons. Thus the notations in eq. (\ref{eq:1065})
are extended, as shown in eq. (\ref{eq:1019})
\vspace*{-0.3cm} 

\begin{equation}
\label{eq:1067}
\begin{array}{l}
\begin{array}{lll}
w_{\ \left \lbrack \ \mu_{1} \ \nu_{1} \  \right \rbrack
\ \left \lbrack  \ \mu_{2} \ \nu_{2}  \ \right \rbrack}
& : &
\mbox{Weyl bilinear}
\vspace*{0.3cm} \\
R_{\ \mu \ \nu}
& : &
\mbox{Ricci bilinear}
\vspace*{0.3cm} \\
R
& : &
\mbox{Riemann scalar bilinear}
\vspace*{0.3cm} \\
\begin{array}{l}
- \ \varrho^{\ \mu \ \nu} \ = 
\ - \ R^{\ \mu \ \nu} \ + 
\ \frac{1}{4} 
\ g^{\ \mu \ \nu} \ R 
\vspace*{0.3cm} \\
\equiv
\ \vartheta_{\ cl}^{\ \mu \ \nu} 
\end{array}
& : &
\begin{array}{l}
\mbox{energy momentum}
\vspace*{0.0cm} \\
\mbox{bilinear}
\end{array}
\end{array}
\end{array}
\end{equation}

The structure of the energy momentum bilinear introduced in 
eq. (\ref{eq:1019}) ( and eqs. \ref{eq:1066} - \ref{eq:1067} ) 
with respect to chromoelectric and -magnetic
field strengths is reproduced in eq. (\ref{eq:1068}) below
\vspace*{-0.3cm} 

\begin{equation}
\label{eq:1068}
\begin{array}{l}
- \ \varrho^{\ \mu \ \nu} \ = 
\ - \ R^{\ \mu \ \nu} \ + 
\ \frac{1}{4} 
\ g^{\ \mu \ \nu} \ R \ =
\ \vartheta_{\ cl}^{\ \mu \ \nu} \ =
\vspace*{0.5cm} \\
\hspace*{-1.0cm} =
\ \left (
\begin{array}{cc}
\frac{1}{2} 
\ \left ( 
\begin{array}{l}
 \vec{E}^{\ A} \ \vec{E}^{\ D} 
\vspace*{0.3cm} \\
\hspace*{0.3cm} + \ \vec{B}^{\ A} \ \vec{B}^{\ D}
\end{array}
\ \right ) 
 & - \ \vec{S}^{\ k \ A \ D} 
 \vspace*{0.5cm} \\
- \ \vec{S}^{\ i \ A \ D} & 
\begin{array}{ll}
- \ \vec{E}^{\ i \ A} \ \vec{E}^{\ k \ D}
- \ \vec{B}^{\ i \ A} \ \vec{B}^{\ k \ D}
\vspace*{0.3cm} \\
\ + \ \frac{1}{2} \ \delta_{\ i k} 
\ \left (
\begin{array}{l}
\vec{E}^{\ A} \ \vec{E}^{\ D} 
\vspace*{0.3cm} \\
\hspace*{0.3cm} + \ \vec{B}^{\ A} \ \vec{B}^{\ D}
\end{array}
\ \right )
\end{array}
\end{array}
\ \right )
 \ U_{\ A \ D}
\end{array}
\end{equation}

The Weyl bilinear is decomposed into irreducible parts
according to eq. (\ref{eq:1031}) repeated in eq. (\ref{eq:1069})
below
\vspace*{-0.3cm} 

\begin{equation}
\label{eq:1069}
\begin{array}{l}
w_{\ \left \lbrack \ \mu_{1} \ \nu_{1} \  \right \rbrack
\ \left \lbrack  \ \mu_{2} \ \nu_{2}  \ \right \rbrack}
\ =
\ \left (
\begin{array}{r}
\left (
\ P^{\ RR} \ \left ( \ \vec{C} \ \otimes \ \vec{C} \ \right )
\ \right )_{\ \left \lbrack \ \mu_{1} \ \nu_{1} \  \right \rbrack
\ \left \lbrack  \ \mu_{2} \ \nu_{2}  \ \right \rbrack}
\vspace*{0.3cm} \\
+
\ \left (
\ P^{\ LL} \ \left ( \ \vec{G} \ \otimes \ \vec{G} \ \right )
\ \right )_{\ \left \lbrack \ \mu_{1} \ \nu_{1} \  \right \rbrack
\ \left \lbrack  \ \mu_{2} \ \nu_{2}  \ \right \rbrack}
\vspace*{0.3cm} \\
+
\ B^{\ (-)}
\ \left (
\ K^{\ -}
\ \right )_{\ 
\ \left \lbrack
\ \mu_{ 1} \ \nu_{ 1}
\ \right \rbrack
\ \left \lbrack
\ \mu_{ 2} \ \nu_{ 2}
\ \right \rbrack}
\end{array}
\ \right )
\vspace*{0.4cm} \\
 \left (
\ K^{\ -}
\ \right )_{\ 
\ \left \lbrack
\ \mu_{\ 1} \ \nu_{\ 1}
\ \right \rbrack
\ \left \lbrack
\ \mu_{\ 2} \ \nu_{\ 2}
\ \right \rbrack}
\ =
\ \varepsilon_{\ \mu_{\ 1} \ \mu_{\ 2} \ \nu_{\ 1} \ \nu_{\ 2}}
\end{array}
\end{equation}

The irreducible parts $\varrho$ , $P^{\ RR}$ and $P^{\ LL}$
representing the spin 2 parts of the energy momentum bilinear 
( $\varrho$ ) and the Weyl bilinear ($P^{\ RR}$ , $P^{\ LL}$)
are identified with the wave functions of the $II^{\ +}$
spectral series in eq. (\ref{eq:1038}) repeated in eq. (\ref{eq:1070})
below
\vspace*{-0.3cm} 

\begin{equation}
\label{eq:1070}
\begin{array}{l}
\begin{array}{lll ll}
\varrho^{\ \underline{.}}  
& \leftrightarrow & 
\ \widetilde{t}_{\ \underline{.}} 
\ ( \ \left \lbrace \ \varrho \ \right \rbrace
\ ; \ z \ , \ p \ , \  J^{\ P\ C} \ ; \ . \ ) 
& \rightarrow & 
 \widetilde{t} 
\ ( \ \left \lbrace \ \varrho \ \right \rbrace \ )
\vspace*{0.3cm} \\
w_{\ \underline{.}} 
& \leftrightarrow & 
\widetilde{t}_{\ \underline{.}} 
\ ( \ \left \lbrace \ w \ \right \rbrace
\ ; \ z \ , \ p \ , \  J^{\ P\ C} \ ; \ . \ ) 
& \rightarrow & 
 \widetilde{t} 
\ ( \ \left \lbrace \ w \ \right \rbrace \ )
\vspace*{0.3cm} \\
w^{\ RR}_{\ \underline{.}} 
& \leftrightarrow & 
\widetilde{t}_{\ \underline{.}} 
\ ( \ \left \lbrace \ w^{\ RR} \ \right \rbrace
\ ; \ z \ , \ p \ , \  J^{\ P\ C} \ ; \ . \ ) 
& \rightarrow & 
 \widetilde{t} 
\ ( \ \left \lbrace \ w^{\ RR} \ \right \rbrace \ )
\vspace*{0.3cm} \\
w^{\ LL}_{\ \underline{.}}
& \leftrightarrow & 
\widetilde{t}_{\ \underline{.}} 
\ ( \ \left \lbrace \ w^{\ LL} \ \right \rbrace
\ ; \ z \ , \ p \ , \  J^{\ P\ C} \ ; \ . \ ) 
& \rightarrow & 
 \widetilde{t} 
\ ( \ \left \lbrace \ w^{\ LL} \ \right \rbrace \ )
\end{array}
\end{array}
\end{equation}

The {\it new} result, worked out in this appendix (A.5) , shows,
that the wave functions pertaining to $P^{\ RR}$ and $P^{\ LL}$ ,
i.e. to the spin 2 irreducible parts of the Weyl bilinear vanish.

In this sense we identify here the bilinear with its gb wave functions,
keeping in mind that the full spin 2 Weyl bilinear operator 
does {\it not} vanish identically. This implies for the wave functions
defined in eq. (\ref{eq:1070})
\vspace*{-0.3cm} 

\begin{equation}
\label{eq:1071}
\begin{array}{l}
\widetilde{t} 
\ ( \ \left \lbrace \ w^{\ RR} \ \right \rbrace \ )
\ =
\ \widetilde{t} 
\ ( \ \left \lbrace \ w^{\ LL} \ \right \rbrace \ )
\ = \ 0
\end{array}
\end{equation}

Eq. (\ref{eq:1071}) is the main result of this appendix (A.5) .

With the above identification the full decomposition of the
adjoint string bilinear in eq. (\ref{eq:1066}) becomes
\vspace*{-0.3cm} 

\begin{equation}
\label{eq:1072}
\begin{array}{l}
\hspace*{-1.2cm}
 B_{\ \left \lbrack \ \mu_{1} \ \nu_{1} \  \right \rbrack
\ , \ \left \lbrack  \ \mu_{2} \ \nu_{2}  \ \right \rbrack}
\ = 
\ \left \lbrace
\begin{array}{c}
\ \frac{1}{2}
 \left (
\begin{array}{l}
 g_{\ \mu_{\ 1} \ \mu_{\ 2}} \ \varrho_{\ \nu_{\ 1} \ \nu_{\ 2}}
- \  g_{\ \nu_{\ 1} \ \mu_{\ 2}} \ \varrho_{\ \mu_{\ 1} \ \nu_{\ 2}}
\vspace*{0.3cm} \\
- \  g_{\ \mu_{\ 1} \ \nu_{\ 2}} \ \varrho_{\ \nu_{\ 1} \ \mu_{\ 2}}
\ + \ g_{\ \nu_{\ 1} \ \nu_{\ 2}} \ \varrho_{\ \mu_{\ 1} \ \mu_{\ 2}}
\end{array}
 \right )
\vspace*{0.3cm} \\
+ 
\ K^{\ +}_{\ \left \lbrack \ \mu_{1} \ \nu_{1} \  \right \rbrack
  \ \left \lbrack  \ \mu_{2} \ \nu_{2}  \ \right \rbrack}
\ B^{\ (+)}
\vspace*{0.3cm} \\
+ 
\ K^{\ -}_{\ \left \lbrack \ \mu_{1} \ \nu_{1} \  \right \rbrack
  \ \left \lbrack  \ \mu_{2} \ \nu_{2}  \ \right \rbrack}
\ B^{\ (-)}
\vspace*{0.3cm} \\
\end{array}
\right \rbrace
\vspace*{0.4cm} \\
\hline
\vspace*{-0.3cm} \\
\mbox{with :}
\ g^{\ \mu_{\ 1} \ \mu_{\ 2}}
\ B_{\ \left \lbrack \ \mu_{1} \ \nu_{1} \  \right \rbrack
\ , \ \left \lbrack  \ \mu_{2} \ \nu_{2}  \ \right \rbrack}
\ =
\ R_{\ \nu_{\ 1} \ \nu_{\ 2}}
\vspace*{0.5cm} \\
- \ \varrho^{\ \mu \ \nu} \ = 
\ - \ R^{\ \mu \ \nu} \ + 
\ \frac{1}{4} 
\ g^{\ \mu \ \nu} \ R \ =
\ \vartheta_{\ cl}^{\ \mu \ \nu} 
\vspace*{0.5cm} \\
R \ =
\ g^{\ \nu_{\ 1} \ \nu_{\ 2}}
\ R_{\ \nu_{\ 1} \ \nu_{\ 2}}
\ =
\ 12 \ B^{\ (+)}
\end{array}
\end{equation}

Comparing the form of the adjoint string components in eq. (\ref{eq:1072})
with eq. (\ref{eq:1063}) we find
\vspace*{-0.3cm} 

\begin{equation}
\label{eq:1073}
\begin{array}{l}
\hspace*{-0.6cm} 
B^{\ '}_{\ \left \lbrack \ \mu_{1} \ \nu_{1} \  \right \rbrack
\ , \ \ \left \lbrack  \ \mu_{2} \ \nu_{2}  \ \right \rbrack}
\ =
\ \frac{1}{2}
 \left (
\begin{array}{l}
 g_{\ \mu_{\ 1} \ \mu_{\ 2}} \ \varrho_{\ \nu_{\ 1} \ \nu_{\ 2}}
- \  g_{\ \nu_{\ 1} \ \mu_{\ 2}} \ \varrho_{\ \mu_{\ 1} \ \nu_{\ 2}}
\vspace*{0.3cm} \\
- \  g_{\ \mu_{\ 1} \ \nu_{\ 2}} \ \varrho_{\ \nu_{\ 1} \ \mu_{\ 2}}
\ + \ g_{\ \nu_{\ 1} \ \nu_{\ 2}} \ \varrho_{\ \mu_{\ 1} \ \mu_{\ 2}}
\end{array}
 \right )
\vspace*{0.3cm} \\
B^{\ '} \ \leftrightarrow \ \left \lbrace \ II^{\ +} \ \right \rbrace
\ \longleftrightarrow \ \vartheta_{\ cl}^{\ \mu \ \nu}
\end{array}
\end{equation}

\item ii) The parallel to abelian gauge fields

The relations of the nonabelian adjoint string variables,
contained in eqs. (\ref{eq:1072}) and (\ref{eq:1072}) ,
with the three spectral types, denoted $I^{\ +}$ ,  $I^{\ -}$ 
and $II^{\ +}$ here, equally apply to the corresponding 
string variables pertaining to electromagnetic fields 
\cite{Land} , \cite{Yang} . This has not been explicitely done
\cite{HFPM} ,
since the structure of the two isolated decay photons
yields a considerable simplification.

\end{description}


\newpage

\begin{thebibliography}{99}

\addcontentsline{toc}{section}{References}

\bibitem{KaKhoMaRy}  A.B. Kaidalov, V.A. Khoze, A.D. Martin and M.G. Ryskin,
Central exclusive diffractive production as a spin--parity analyser:
from hadrons to Higgs, hep-ph/0307064. 

\bibitem{LandauPom}  L.D. Landau  and E.M. Lifschitz,
Production of electrons and positrons by a collision
of two particles, Phys.Z.Sowjetunion 6 (1934) 244. 

L.D. Landau and I. Pomeranchuk (in Russian), 
Limits of applicability of the theory of Bremsstrahlung electrons
and pair production at high energies,
Dokl.Akad.Nauk Ser.Fiz.92 (1953) 535. 

L.D. Landau and I. Pomeranchuk (in Russian),
Electron cascade process at very high energies,
Dokl.Akad.Nauk Ser.Fiz.92 (1953) 735. 

L.D. Landau (in Russian),
On the multiparticle production in high energy collisions,
Izv.Akad.Nauk Ser.Fiz.17 (1953) 51. 

\bibitem{ChengWu} H. Cheng and  T. T. Wu, High energy elastic scattering in
quantum electrodynamics, Phys. Rev. Lett. 22, (1969) 666.  

H. Cheng and T. T. Wu,  
Impact picture and the eikonal approximation,
Phys. Lett. B34 (1971) 647. 

\bibitem{PMWO} P. Minkowski and W. Ochs, 
Identification of the glueballs and the scalar meson nonet of lowest
mass, Eur.Phys.J. C9 (1999) 283-312,
hep-ph/9811518. 

Scalar mesons and glueball in B-decays and gluon jets,
hep-ph/0304144. 

On the scalar nonet lowest in mass,
hep-ph/0209223. 

The glueball among the light scalar mesons,
hep-ph/0209225. 

\bibitem{HFPM} H. Fritzsch and P. Minkowski, 
Psi resonances, gluons and the Zweig rule,
Nuovo Cim.A30 (1975) 393. 

\bibitem{Land} H. Landau, On the angular momentum of a system
of two photons, Dokl. Akad. Nauk. SSSR, 60 (1948) 207,
see also in "Collected papers of L. D. Landau", D. ter Haar ed.,
Gordon and Breach, New York, 1965.

\bibitem{Yang} C. N. Yang,   
Selection rules for the dematerialization of a 
particle into two photons, 
Phys.Rev.77 (1950) 242. 

\bibitem{CaCoLeu} I. Caprini, G. Colangelo and H. Leutwyler,
work in progress.

H. Leutwyler, 
Electromagnetic form-factor of the pion,
Talk given at Continuous Advances in QCD 2002 / ARKADYFEST (honoring the 60th
birthday of Prof. Arkady Vainshtein), Minneapolis, Minnesota, 17-23 May 2002,
published in *Minneapolis 2002, Continuous advances in QCD* 23-40,
hep-ph/0212324. 

\bibitem{Kloe} The KLOE Collaboration, 
Recent results from the KLOE experiment at DA$\Phi$NE,
hep-ex/0308023. 

\bibitem{Schier} H. Ichie, V. Bornyakov, T. Streuer and G. Schierholz,
  The flux distribution of the three quark system in SU(3),
  Contributed to 20th International Symposium on Lattice Field Theory (LATTICE
  2002), Boston, Massachusetts, 24-29 Jun 2002,
  hep-lat/0212024. 

\bibitem{PMBar} P. Minkowski, On the Oscillatory Modes of Quarks in Baryons,
Nucl. Phys. B174 (1980) 258.

\bibitem{PDG} The Particle Data Group, K. Hagiwara et al., 
Phys. Rev. D 66, 010001 (2002) and 2003 off-year partial
update for the 2004 edition available on the PDG WWW pages (URL:
http://pdg.lbl.gov/). 

\bibitem{Ruefe}  
F. Niedermayer, P. R\"{u}fenacht and U. Wenger,
Fixed point SU(3) gauge actions : scaling properties
and glueballs, 
Submitted to 18th International Symposium on Lattice Field Theory (Lattice
2000), Bangalore, India, 17-22 Aug 2000,
Nucl.Phys.Proc.Suppl.94 (2001) 636, hep-lat/0011041. 

\bibitem{Teper} H. B. Meyer and M. J. Teper,
Glueballs and the Pomeron, hep-lat/0308035. 

\bibitem{Michael} C. Michael, Hybrid Mesons from the Lattice,
hep-ph/0308293. 

\bibitem{Kuni} T. Kunihiro, S. Muroya, A. Nakamura, C.  Nonaka, 
M. Sekiguchi and H. Wada,
Scalar Particles in Lattice QCD,
to be published in Proceedings of `International Symposium
on Hadron Spectroscopy, Chiral Symmetry and Relativistic Description of Bound
Systems' (in a series of KEK proceedings),
hep-ph/0308291. 

\bibitem{Chung}  S.U. Chung, E. Klempt and J.G. K\"{o}rner,
$ SU(3) $ Classification of $ p $-wave $ \eta\pi $ and $ \eta'\pi $
Systems, Eur.Phys.J. A15 (2002) 539, hep-ph/0211100. 

D. R. Thompson et al., Phys. Rev. Lett 79 (1997) 1630.

S. U. Chung et al., Phys. Rev. D60 (1999) 092001.

E. Ivanov et al., Phys. Rev. Lett. 86 (2001) 3977.

\bibitem{CLG} G. Colangelo, J. Gasser  and H. Leutwyler, \\
  Pi Pi scattering,
  Nucl.Phys.B603 (2001) 125, hep-ph/0103088. 

\bibitem{Jfunc} R. Jost, \"{U}ber die falschen Nullstellen der
Eigenwerte der S-Matrix, Helv. Phys. Acta 20 (1947) 256.

\bibitem{Lesniak} R. Kaminski, L. Le\'{s}niak and B. Loiseau,
Elimination of ambiguities in pion-pion phase shifts using crossing
symmetry, Phys.Lett. B551 (2003) 241, hep-ph/0210334. 

R. Kaminski, L. Le\'{s}niak and K. Rybicki,
A joint analysis of the S-wave in the $\pi^{+} \ \pi^{-}$ and 
$\pi^{0} \ \pi^{0}$ data,
Eur.Phys.J.direct C4 (2002) 4, hep-ph/0109268,

and references cited therein.

\bibitem{Les1} R. Kaminski, L. Le\'{s}niak and K. Rybicki,
Separation of S- wave pseudoscalar and pseudovector amplitudes in
   $\pi^- p \ \rightarrow \pi^+ \pi^- n$ reaction on polarized target,
Z.Phys. C74 (1997) 79, hep-ph/9606362. 

\bibitem{Amsler} The crystal barrel collaboration, 
CERN-PS-197 Experiment,
C. Amsler et al., 
High statistics study of $f_{\ 0} \ ( \ 1500 \ )$ decay
into $\pi^{\ 0} \pi^{\ 0}$,
Phys.Lett.B342 (1995) 433. 

\bibitem{spanier} S. Spanier and N.A. Tornqvist, 
 Scalar mesons,
 page 450 of the Review of Particle Properties in ref. \cite{PDG},

and references cited therein.

\bibitem{amslerrev} C. Amsler,
Non $q \overline{q}$ candidates,
in ref. \cite{PDG},

and references cited therein.

\bibitem{aitala} E.M. Aitala et al.,  E791 Collaboration,
Experimental evidence for a light and broad scalar resonance
in $D^{\ +} \ \rightarrow \ \pi^{\ -} \ \pi^{\ +} \ \pi^{\ +}$
decay, Phys.Rev.Lett.86 (2001) 770, hep-ex/0007028. 

E.M. Aitala et al.,  E791 Collaboration,
Study of the 
$D^{\ +}_{\ s} \ \rightarrow \ \pi^{\ -} \ \pi^{\ +} \ \pi^{\ +}$
decay and measurement of $f_{\ 0}$ masses and widths,
Phys.Rev.Lett.86 (2001) 765, hep-ex/0007027. 

\bibitem{bediaga} P.L. Frabetti et al.,  E687 Collaboration,
Analysis of the $D^{\ +} \ , \ D^{\ +}_{\ s} \ \rightarrow
\ \pi^{\ -} \ \pi^{\ +} \ \pi^{\ +}$ Dalitz plots,
Phys.Lett.B407 (1997) 79, 

and Phys.Lett.B351 (1995) 591. 

\bibitem{garmash} A. Garmash et al.,  Belle Collaboration,  
Study of B meson decays to three body charmless hadronic
final states, hep-ex/0307082,

A. Garmash et al.,  Belle Collaboration,
Three body charmless $B \ \rightarrow \ K \ h \ h$ decays at Belle,
hep-ex/0207003,

K. Abe et al., Belle Collaboration,
Study of three body charmless B decays,
Phys.Rev.D65 (2002) 092005, hep-ex/0201007,

K. Abe et al., Belle Collaboration,
Study of charmless B decays to three kaon final states,
Contributed to 31st International Conference on High Energy Physics (ICHEP
 2002), Amsterdam, The Netherlands, 24-31 Jul 2002,
hep-ex/0208030. 

\bibitem{babar} B. Aubert et al., BABAR Collaboration,
Measurements of the branching fractions of charged B decays to
$K^{\ \pm} \ \pi^{\ \mp} \ \pi^{\ \pm}$ final states,
hep-ex/0308065,

B. Aubert et al., BABAR Collaboration,
Measurements of the branching fractions and charge asymmetries
of charmless three body charged B decays,
Phys.Rev.Lett.91 (2003) 051801, hep-ex/0304006,

B. Aubert et al., BABAR Collaboration,
Measurements of the branching fractions of charged B decays
to $K^{\ +} \ \pi^{\ -} \ \pi^{\ +}$ final states,
hep-ex/0303022,

P. C. Bloom for the BABAR Collaboration, 
Measurements of rare B decays at BABAR,
eConf C020805 (2002) TTH06, hep-ex/0302030,

B. Aubert et al., BABAR Collaboration,
Measurements of the branching fractions of charmless
three body charged B decays, hep-ex/0206004. 

\bibitem{amslerrev2} C. Amsler,
Proton - antiproton annihilation and meson spectroscopy
with the crystal barrel,
Rev.Mod.Phys.70 (1998) 1293, hep-ex/9708025. 

\bibitem{akesson} T. Akesson et al., AFS Collaboration,
A search for glueballs and a study of doouble pomeron exchange at the CERN
intersecting storage rings, Nucl. Phys. B264 (1986) 154.

\bibitem{WA102} A. Singovsky for the WA102 collaboration, 
New results from the WA102 experiment,
Nucl.Phys.A675 (2000) 47c, 

A. Kirk,
Resonance production in central p p collisions at the CERN OMEGA spectrometer,
Phys.Lett.B489 (2000) 29, hep-ph/0008053, 

and references cited therein.

\bibitem{Close} F. Close , 
Light hadron spectroscopy and experiment,
Int.J.Mod.Phys.A17 (2002) 3239, hep-ph/0110081,

F. Close and A. Kirk,
Scalar glueball $q \ \overline{q}$ mixing above 1 GeV and
implications for lattice QCD,
Eur.Phys.J.C21 (2001) 531, hep-ph/0103173,

F. Close,
Glueballs: a central mystery,
Acta Phys.Polon.B31 (2000) 2557, hep-ph/0006288, 

F.Close, A. Kirk and G. Schuler,
Dynamics of glueball and $q \ \overline{q}$ production
in the central region of p p collisions,
Phys.Lett.B477 (2000) 13, hep-ph/0001158, 

and references cited therein.


\bibitem{RJ} R. Jost, The general theory of quantized fields,
American Mathematical Society, Providence R.I., 1965,

R. Jost, CTP-Invarianz der Streumatrix und interpolierende Felder,
Helv. Phys. Acta 36 (1963) 77,

R. Jost, Einiges \"{u}ber die Lorentzgruppe und das ein\"{a}ugige
Sehen 
(engl. transl. : Some things on the Lorentz group and monocular vision) ,
Verein Schweizerischer Mathematik- und Physiklehrer,
Bulletin Nr. 2, 1966.

\bibitem{latt} for a review see M. Creutz,
The early days of lattice gauge theory, hep-lat/0306024, 

J. Greensite, The confinement problem in lattice gauge theory,
hep-lat/0301023, 

M. Luescher, S.Sint, R. Sommer and P. Weisz,
Chiral symmetry and O(a) improvement in lattice QCD,
Nucl.Phys. B478 (1996) 365 , hep-lat/9605038.

\bibitem{alvarez} for a review see E. Alvarez,
Loops versus strings, gr-qc/0307090,

A. M. Polyakov and V. S. Rychkov, 
Gauge fields-strings duality and the loop equation,
Nucl.Phys.B581 (2000) 116, hep-th/0002106, 

Y. M. Makeenko and A.A. Migdal, Exact equation for the loop
average in multicolor QCD, Phys. Lett. B88 (1979) 135 and
Erratum Phys. Lett. B89 (1980) 437.

\bibitem{Nachtmann} O. Nachtmann,  Perturbative and nonperturbative
aspects of QCD, H. Latal and W. Schweiger edts. Springer Verlag,
Berlin, Heidelberg 1997, hep-ph/9609365. 

\bibitem{encyc} Spider, The Encyclopedia Americana, Int. edition,
Vol. 25, p. 494, Americana Corporation 1976.

\bibitem{spiderfot} Lui Bernard, www.onlinekunst.de/ bernard/04.html .

\bibitem{MF} M. Fierz, Die unit\"{a}ren Darstellungen der
homogenen Lorentzgruppe, in Preludes in theoretical physics, in
honor of V. F. Weisskopf, A. De-Shalit, H. Feshbach and L. van Hove
ets., North Holland, Amsterdam 1966, p.1.


\end{thebibliography}
\end{document}